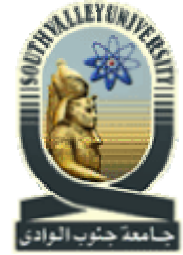

**Aswan Faculty of Engineering**
**Electrical Engineering Department**

# Efficient Web-Based SCADA System

## M. Sc. Thesis

Submitted by

## Eng. Hosny Ahmed Abbas Ahmed

An Automation Engineer in Qena Paper Company

**Electrical Engineering Department**
**Faculty of Engineering**
**South Valley University**
**Aswan, Egypt.**

June 2011

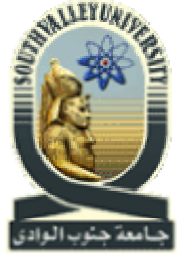

**Aswan Faculty of Engineering**
**Electrical Engineering Department**

# Efficient Web-Based SCADA System

By

## Eng. Hosny Ahmed Abbas Ahmed

B. Sc., Electrical Engineering Assiut University, 1998.
An Automation Engineer in Qena Paper Company

## A THESIS
Submitted in Partial Fulfillment of the
Requirements for the Degree of

## MASTER OF SCIENCE

Electrical Engineering Department
Faculty of Engineering
South Valley University
Aswan, Egypt.

Supervised by:                                Examined by:
**Prof. Dr. Abd EL-Magid Mohamed Ali**        **Prof. Dr. Mohammed Hussein Amin**
Faculty of Engineering- South Valley University   Faculty of Engineering- Assiut University
**Dr. Ahmed Mostafa Abdel-Rahman**            **Prof. Dr. Youssef Basioni Mahdy**
Faculty of Engineering- South Valley University   Vice Dean for Graduate Studies and Research-Faculty
                                              of Computers and Information Science- Assiut University
                                              **Prof. Dr. Abd EL-Magid Mohamed Ali**
                                              Faculty of Engineering- South Valley University

June 2011

بِسْمِ اللَّهِ الرَّحْمَنِ الرَّحِيمِ

رَبِّ أَوْزِعْنِي أَنْ أَشْكُرَ نِعْمَتَكَ الَّتِي أَنْعَمْتَ عَلَيَّ وَعَلَى وَالِدَيَّ وَأَنْ أَعْمَلَ صَالِحاً تَرْضَاهُ وَأَدْخِلْنِي بِرَحْمَتِكَ فِي عِبَادِكَ الصَّالِحِينَ

**صدق الله العظيم**

النمل ١٩

ACKNOWLEDGEMENTS

# ACKNOWLEDGEMENTS

After more than 10 years from graduation in Assiut University (1998) I decided to come back to research in computer science and I think that Allah supports me by guiding me to work with **Dr. Ahmed Mostafa,** I found that his way of thinking is very close to my way, he could understand me easily and he helps me psychologically to continue in the research area. His guidance is a main reason to complete this thesis.

I would like to thank **Prof. Dr. Abd EL-Magid Mohamed Ali** for his guidance, his support and continuous encouragement.

I wish to thank the Deputy Chief of factories in Qena Paper Mill **Eng. Abd El-Hameid Omar;** he always pushes me in the research direction and encourages me to do great projects, thanks for him.

Most importantly, I'd like to dedicate this thesis to my family: my parents, whose love, support and encouragement, are more deserving of my thanks than anyone; my wife, whose unconditional love, support, patience and caring were and always will be a source of inspiration; my kids *Rasha*, *Rania* and *Mohammed*, who are the blessings of my life; my brothers and sister, who offered support and encouragement. This thesis would not have been possible without the help of my family. I could never be as happy as I am without each of them. They are all very precious to me.

*Hosny A. Abbas*

# ABSTRACT

# ABSTRACT

Computer-based supervisory control and data acquisition (SCADA) systems have evolved over the past 50 years, from standalone, compartmentalized operations into networked architectures that communicate across large distances. In addition, their implementations have migrated from custom hardware and software to standard hardware and software platforms. These changes have led to reduced development, operational, and maintenance costs as well as providing executive management with real-time information that can be used to support planning, supervision, and decision making. For reasons of efficiency, maintenance, economics, data acquisition and control platforms have migrated from isolated in-plant networks using proprietary hardware and software to PC-based systems using standard software, network protocols, and the Internet. There is an emerging trend in many organizations comprising SCADA and conventional IT units toward consolidating some overlapping activities. For example, control engineering might be absorbed or closely integrated with the corporate IT department. This trend is motivated by cost savings achieved by consolidating disparate platforms, networks, software, and maintenance tools. In addition, integrating SCADA data collection and monitoring with corporate financial and customer data provides management with an increased ability to run the organization more efficiently and effectively. In this thesis we design an approach for web based SCADA systems by accessing an OPC DA server through the Internet. To do that we will aggregate some of the modern IT technologies like web services, ASP.NET and AJAX with standard process protocols such as OPC DA protocol. Our goal is to design and develop an efficient and feasible web based SCADA system which consumes as little as possible of the available resources, but to achieve this goal we will have to face and solve many challenges which related to the Internet limitations.

# TABLE OF CONTENTS

# TABLE OF CONTENTS

## APPENDICES

# LIST OF FIGURES

# LIST OF FIGURES

**Figure No.**             **Title**                              **Page**

# LIST OF TABLES

# LIST OF TABLES

# CHAPTER 1

# INTRODUCTION

# CHAPTER 1
# INTRODUCTION

## 1.1 Automation

Automation is the use of control systems (such as numerical control, programmable logic control, and other industrial control systems), in addition to other applications of information technology (such as computer-aided technologies [CAD,CAM,  CAx]), to control industrial machinery and processes, reducing the need for human intervention. In the scope of industrialization, automation is a step beyond mechanization. Whereas mechanization provided human operators with machinery to assist them with the muscular requirements of work, automation greatly reduces the need for human sensory and mental requirements as well [40]. If we imagine automation as a stack consists of three layers; it will be as shown in Figure 1.1.

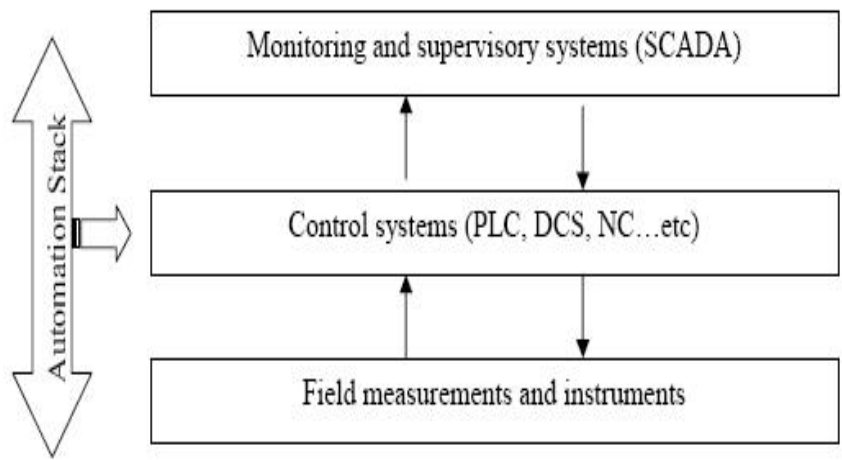

**Figure 1.1: Automation stack**

As we see in Figure 1.1 that the first (bottom) layer in the automation stack is field measurements from which we collect all the field data like temperatures, levels, flows, speed, etc. Second (middle) layer is control systems layer in which the control of the process takes place. The third (top) layer in automation stack is real time monitoring and supervisory systems which can be hardware or software and in most of cases its software.

## 1.2 SCADA Systems

As the technical capabilities of computers, operating systems, and networks improved, organizational management pushed for increased knowledge of the real-time status of remote plant operations. In addition, in organizations with a number of geographically separated operations, remote data acquisition, control, and maintenance became increasingly attractive from management and cost standpoints. These capabilities are known collectively as Supervisory Control and Data Acquisition or SCADA. As the name indicates, it is not a full control system, but rather focuses on the supervisory level. As such, it is a purely software package that is positioned on top of hardware to which it is interfaced, in general via Programmable Logic Controllers (PLCs), or other commercial hardware modules. SCADA systems analyze real-time data to monitor and control the proper operation of control processes.

SCADA systems are vital components of most nations' critical infrastructures. They control pipelines, water and transportation systems, utilities, refineries, chemical plants, and a wide variety of manufacturing operations. SCADA provides management with real-time data on production operations; implements more efficient control paradigms; improves plant and personnel safety and reduces costs of operation. These benefits are made possible by the use of standard hardware and software in SCADA systems combined with improved communication protocols and increased connectivity to outside networks, including the Internet. However, these benefits are acquired at the price of increased vulnerability to attacks or erroneous actions from a variety of external and internal sources [18]. SCADA systems use mostly open loop control due to the less reliable communication they use [16]. It enables remote monitoring and control of a variety of industrial devices as diverse as water and gas pumps, track switches and traffic signals.

Automation technology produces large amounts of data; this typically represents physical variables such as temperature, current, pressure or other production data such as number of units, error information and so on. The data comes from equally diverse sources: controllers, level measuring devices, weighing machines, scanners and similar. These data end up being crunched or otherwise retained by all sorts of software applications. Some of it must be stored in databases or used directly. It may need to be placed straight into visualization software for HMI (Human Machine Interface). In short, there are widely differing data, devices, manufacturers and applications which somehow all have to work in the same space. The drive towards networked industrial control systems is due to several factors. Integration of geographically distributed assets through centralized control improves agility in responding to supply and demand fluctuations, reduces cost of operations and enables process efficiencies unachievable in the past [6]. A typical SCADA system is shown in Figure 1.2.

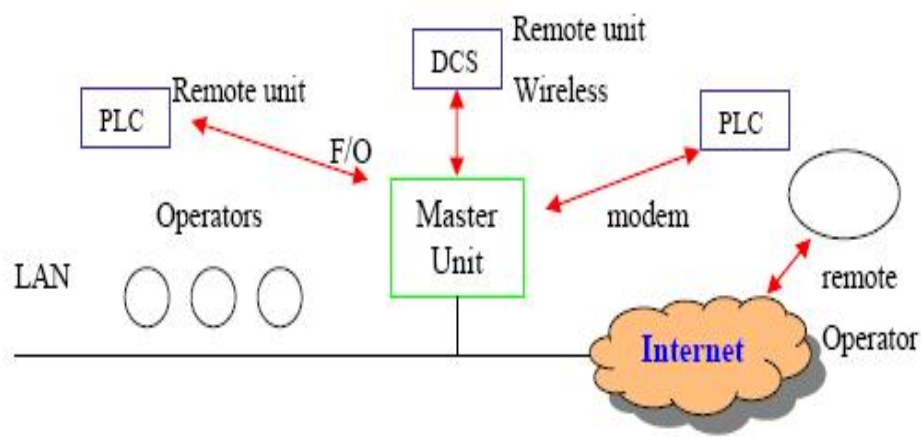

**Figure 1.2: Typical SCADA system**

## 1.1  OPC Protocol

As mentioned earlier that SCADA systems transferred from using proprietary communication protocols to using standard, open protocols. Now, vendors compete based upon their computing equipment's features, costs and quality factors instead of who has the best protocol. Customers are not bound to one manufacturer after the initial sale [7].

The integration of Ethernet into manufacturing and process control would have counted for nothing if wide-ranging and different applications were not able to communicate with each other. It was not until a new approach to achieve commonality of software interface was developed, a method of treating frequently encountered application tasks such as interface, communication protocol and data exchange in a standardized way. That method was Objects for Process Control and it has changed the face of industrial software [9]. One of the most familiar application protocols now is OPC protocol.

OPC stands for OLE (Object linking and Embedding) for Process Control and now its name is Open Process Control or openness, productivity and connectivity draws a line between hardware providers and software developers. It provides a mechanism to provide data from a data source and communicate the data to any client application in a standard way. A vendor can now develop a reusable, highly optimized server to communicate to the data source, and maintain the mechanism to access data from the data source/device efficiently. Providing the server with an OPC interface allows any client to access their devices. The utility of OPC has now reached the point where automation without OPC is unthinkable. This interface is supported by almost all SCADA, visualization, and process control systems [12]. OPC delivers connectivity and interoperability benefits to measurement and automation systems in the same way that standard printer drivers deliver connectivity and interoperability benefits to word processing. OPC-enabled applications access data the same way, whether data is coming from an OPC server connected to a programmable logic controller; an industrial network such as Foundation fieldbus, Profibus, or DeviceNet; SCADA; a laboratory information management system; or a production management system. Therefore, it gives production and business applications across the manufacturing enterprise access to real-time plant floor information in a consistent manner, making multivendor interoperability and "plug and play" connectivity a reality [34]. Interoperability is assured through the creation and maintenance of open standards specifications.

There are currently seven standards specifications completed or in development. Based on fundamental standards and technology of the general computing market, the OPC Foundation adapts and creates specifications that fill industry-specific needs. OPC will continue to create new standards as needs arise and to adapt existing standards to utilize new technology. OPC is a series of standards specifications. The first standard (originally called simply the OPC Specification and now called the Data Access Specification) resulted from the collaboration of a number of leading worldwide automation suppliers working in cooperation with Microsoft. Originally based on Microsoft's OLE COM (component object model) and DCOM (distributed component object model) technologies, the specification defined a standard set of objects, interfaces and methods for use in process control and manufacturing automation applications to facilitate interoperability. The COM/DCOM technologies provided the framework for software products to be developed. There are now hundreds of OPC Data Access servers and clients [31].

OPC is based on the DCOM/COM component-object programming model developed by Microsoft in which software is divided into smaller, independent units – the objects (components). An object can execute certain actions (i.e., supply certain methods). While the available methods are visible from the outside, the inner life of the object remains hidden. For this reason, objects are enclosed and one or more interface(s) are defined for the outside. The interface contains the description of which functions are available and how these are to be used. The functions are called with the help of so-called virtual function tables, which point to different functions.

OPC simplifies the link to applications by providing a standard interface, which is not dependent on the concrete application. This allows the user to choose freely between the products of different software and hardware manufacturers. Both OPC server and OPC clients are COM objects. It makes no difference who developed these objects, when they were developed, or in what programming language and on what Windows operating system they were written. As COM objects, they can all communicate with each another. Network capability was added to COM with DCOM (distributed COM). In other words, not only the services of the OPC servers on the same computer are available but also those of all servers, which can be accessed over the network; the client remains unaware of the local or remote server status [12].

Process visualization is currently the main application area for OPC, which is not surprising since it was primarily the HMI manufacturers who demanded standardized access to the different controllers. The pointer or object-oriented approach of COM offers user's capabilities, which far exceed the basic connectivity between finished OPC servers and clients. COM objects can be implemented in a wide variety of different applications. It is possible to program a separate client which uses the methods of an existing OPC server provided that the programming language is object or pointer-oriented as in C, C++, Delphi, VB6, C# or VB.NET. The OPC client accesses an OPC server via one of two types of interface: the custom or automation interface. While function-oriented languages such as C++ use the custom interface, one can communicate effectively with the automation interface by using script languages such as Visual Basic. The advantage of doing your own client programming is of course that the application can be tailored to precise local requirements. The OPC interface architecture is shown in Figure 1.3.

Since the beginning of OPC, there have been those people skeptical that a standard generic interface could be specified, designed, implemented and adopted by a number of vendors, and at the same time provide the level of performance and through-put of vendor specific proprietary data exchange mechanisms. The architecture of OPC was designed to address the following:

1. Clients and servers residing on different physical computers(distributed)
2. Multiple clients communicating to a single server
3. A Single client communicating to multiple servers
4. Multiple clients simultaneously communicating to multiple servers

The OPC Foundation and the OPC Foundation member companies will focus on making sure that products are interoperable together, and by definition; vendors push the OPC Foundation to make sure that performance and throughput expectations are not compromised to achieve interoperability. Therefore, the Objects, Interfaces and functionality defined by OPC will continue to evolve and change as the technology continues to grow [25].

Finally, by adopting the OPC technology, differences of the local plant communication protocols to different PLCs can be masked, and a uniform data communication manner to remote clients can be provided [30].

## 1.3.1 OPC Protocol history

- 1996 OPC DA (Data Access)
    1. OPC classic based on Microsoft COM/DCOM Technology
    2. DCOM status frozen by Microsoft
    3. Operates on tags
- Various Other OPC Specifications based on COM/DCOM

    4.  OPC HDA (historical data access)

    5.  OPC AE (alarms & events)

    6.  OPC Complex Data

- 2001 OPC-XML DA

    7.  Web based access, SOAP/HTTP

    8.  Platform-neutral

    9.  In practice too slow (even much slower than OPC DA)

- 2003 Start Specification of OPC UA (Unified Architecture)

    a.  Avoid DCOM

    b.  Unified Architecture

    c.  Intended to run on PLC (Embedded)

    d.  Platform neutral

    e.  custom binary protocol, security

- 2006 Release of OPC UA Specifications

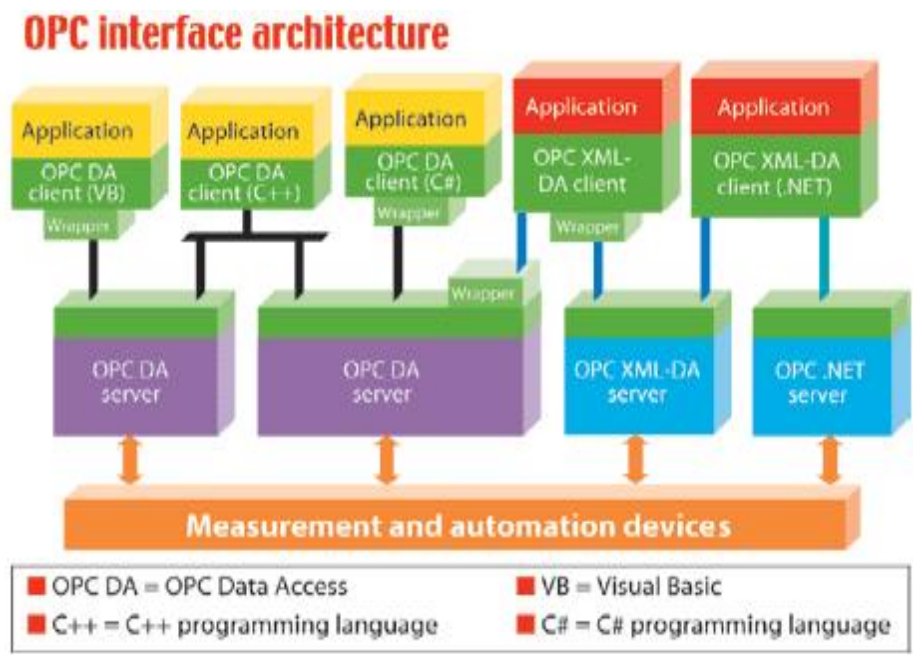

**Figure 1.3: OPC interface architecture**

## 1.3.2 OPC DA Internal Structure

Regarding to Figure 1.4 and Figure 1.5, the server provides management of the OPC groups and controls mediates and optimizes access to the physical devices by multiple clients.

The SERVER is essentially an I/O Driver which 'understands' how to talk to some vendor specific data provider (hardware or  software) and which Exposes the data from this data provider via the standard OPC Interfaces.

The GROUP provides a convenient way for an application to organize the data it needs. Different Groups can be used by different parts of the application; can have different refresh rates; can be polled or Event based. Operator Displays, Recipes, Reports might each use one or more groups.

The ITEM provides a connection point to a value in the physical device. Typically this is a single variable such as a process value or setpoint. The ITEM provides information to the client; Value, Quality, Timestamp, and Data Type. As mentioned before, the OPC item is a transient object, which exists within the OPC server. It is used to create a connection between the server and the real data. More information about OPC protocol can be found at "*www.opcfoundation.org*".

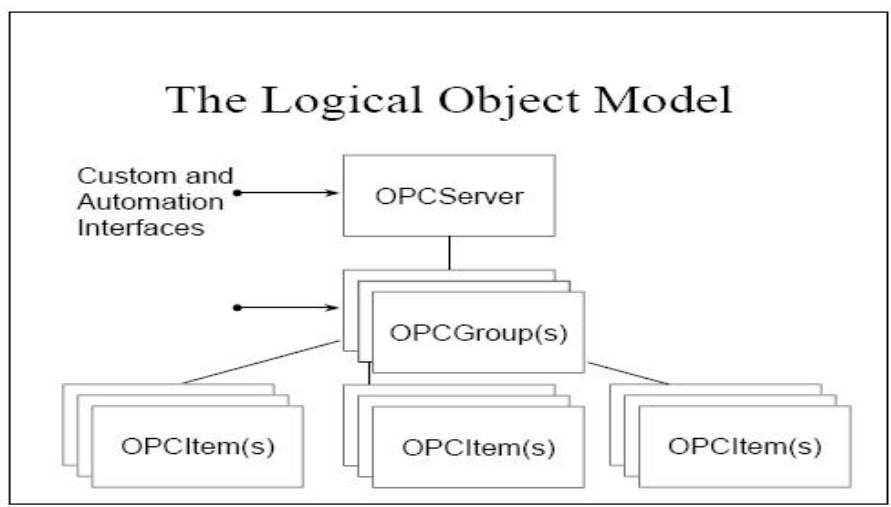

**Figure 1.4: OPC DA logical object model**

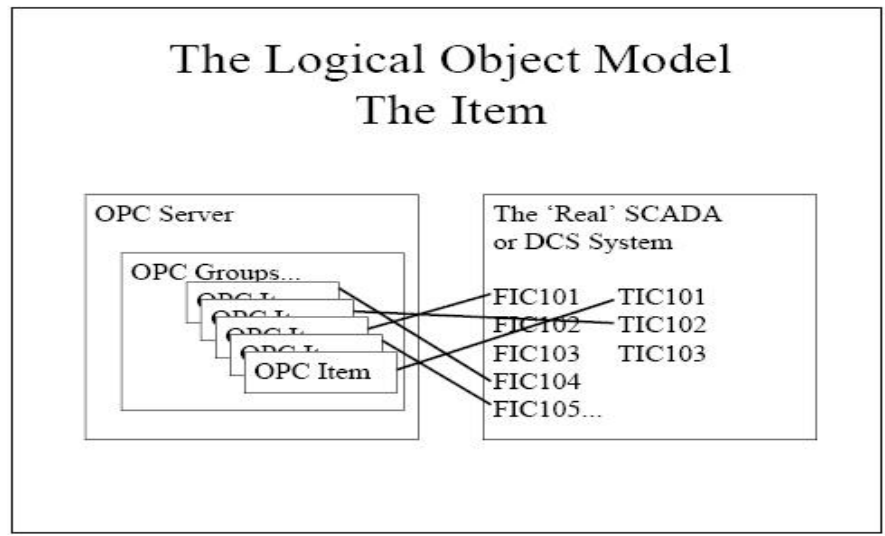

**Figure 1.5: OPC DA mapping to DCS or PLC.**

## 1.4 Real-Time Systems

In computer science, real-time computing (RTC), or reactive computing, is the study of hardware and software systems that are subject to a "real-time constraint"—i.e., operational deadlines from event to system response. By contrast, a non-real-time system is one for which there is no deadline, even if fast response or high performance is desired or preferred.

The needs of real-time software are often addressed in the context of real-time operating systems, and synchronous programming languages, which provide frameworks on which to build real-time application software. A real time system may be one where its application can be considered (within context) to be mission critical. The anti-lock brakes on a car are a simple example of a real-time computing system — the real-time constraint in this system is the short time in which the brakes must be released to prevent the wheel from locking. Real-time computations can be said to have failed if they are not completed before their deadline, where their deadline is relative to an event.

A real-time deadline must be met, regardless of system load. So we can define a real-time system as a system that must satisfy explicit (bounded) response-time constraints or risk severe consequences, including failure. Also we have do define the response time which is the time between the presentation of a set of inputs to a system (stimulus) and the realization of the required behavior (response),including the availability of all associated outputs, this is called the response time of the system [15].

Defining Real-Time was very important to understand the meaning of real time monitoring which is very important in automation now. An example of the response time deadline in a real time system is shown in Figure 1.6. As shown for a system to be real time system, its total response time = T1+T2+T3 must be less than the deadline. Also in a real time monitoring system, the changed values (process outputs) should be monitored immediately after they changed by an interval which can be considered as a deadline.

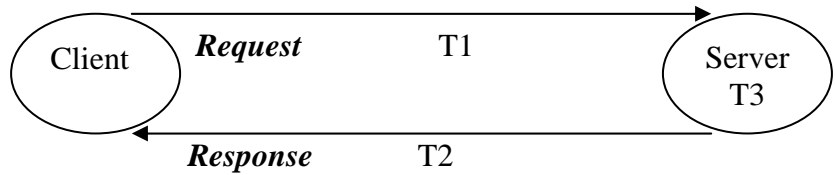

**Figure 1.6: System response time.**

## 1.5 Web-based SCADA systems

Nowadays, Internet is playing an important role in different domains. It had gained a lot of researches and investments. Hundreds of billions of dollars had been spent on the infrastructure and the backbone of the Internet; as a result, Internet now is covering almost the entire planet. This made the Internet an excellent choice to transfer any kind of data between any two or more points. Other hundreds of billions had been spent on developing tools, framework, platforms, protocols and computer languages to make the development of Internet applications easier and less expensive. Average programmers could now develop Internet applications easily. These reasons are strong enough to adopt the Internet and its technologies in any type of distributed applications. SCADA makes no exception on this [10].

Web-based SCADA System makes use of the Internet and Hypertext Transfer Protocol (HTTP) and other Web technologies as a communication layer of the system. It also uses development tools, framework, platforms and computer languages which are used by regular Internet applications as development environment of SCADA application.

Also as stated in [3] that Web-based SCADA system uses the Internet to transfer data between the RTUs (Remote Terminal Units) and the MTU (Master terminal Unit) and/or between the operators' workstations and the MTU. This will reduce the cost of the installation of the SCADA network if compared with installing a dedicated network for it. It also uses the Internet browser programs such as Mozilla Firefox, Netscape Navigator or Microsoft Internet Explorer as Graphical User Interface (GUI) for the operators HMI. This would give all the benefits of browser-based systems, such as simplifying the installation process of the client side of the SCADA systems and also enable the users to access the system using wide range of platforms, as the browsers is now available in most of the modern operating systems. Web-based SCADA system has many advantages including:

- Using the Client/Server n-tier platforms and development tools to develop the Web-based SCADA system will get the development cost and time to the minimum.

- Using the Infrastructure of the Internet or the corporate intranet will get the deployment cost to the minimum.

- Increase distance, data sharing and data provision for monitoring and control systems.

- Enabling collaboration between skilled plant managers situated in geographically diverse locations.

- Enabling the business to relocate the physical location of plant management staff easily in response to business needs.
- For the educational and researching purposes, the risk involving in real laboratory may be avoided by doing the dangerous experiments remotely.

## 1.6 Dissertation layout

This thesis consists of six chapters:

**Chapter 1:** This chapter serves as an introduction to give a brief account of the problem that the author is trying to address in this dissertation. The importance and significance of the work are discussed. Also this chapter contains detailed definitions for the main topics in this thesis.

**Chapter 2:** The purpose of this chapter is to provide a background about the latest IT technologies. Advantages and disadvantages of each one are mentioned.

**Chapter 3:** In this chapter we shall discuss the previous and related work used to design and implement web based SCADA systems with OPC protocol technology. The pros and cons of each approach are discussed. Actually this chapter will help us to determine from where we will start to create our own approach.

**Chapter 4:** This chapter contains our proposed approach to solve the problems of the previous work in the area of web based SCADA systems. The challenges we will face to achieve our goals are discussed and solved. Also this chapter contains an evaluation section to test our designed approach and compare it with other approaches. Simulation software created for testing purposes and finally there will be a case study to apply our proposed approach practically. The case study carried out in Qena Paper mill.

**Chapter 5:** When we mention Internet we have to think about security. This chapter discusses the attacks which affect web based SCADA systems. Also in this chapter we will design a security policy to protect our designed SCADA system from cyber attacks

**Chapter 6:** In this chapter, we conclude the proposed work and suggest certain points of research to improve our designed web based SCADA system.

# CHAPTER 2

# MODERN IT TECHNOLOGIES

# CHAPTER 2
# MODERN IT TECHNOLOGIES

In this chapter we introduce some of the modern IT technologies used in web applications from which we will choose the best to be able to achieve our desired goals; also we will explain the advantages and disadvantages of each one.

## 2.1 Static Web Pages

A static page is a page that gives exactly the same response over and over again. It is the basic HTML page. The static page consists of URLs that are free of any special characters, remains in its final form and has no server-side interaction. An example of a static URL is (xxx equals the domain name): xxxxxxx.com/index.html. Since static web pages are html driven, they offer the webmaster the following quick advantages:

- For those who have little or no experience with design, the static URL allows for an easy and quick solution to getting a website up and running.

- There are no surprises--what one person sees, is what all will see.

- The pages are easier for the search engine spiders to crawl and index.

- No database downtime or uptime.

The major downside of a static web site is when the site begins to grow. 10 to 20 pages is pretty manageable to update, but when it gets to 200+ static pages the daily maintenance and cost can grow exponentially -- in time and money.

## 2.2 Dynamic Web Pages

A dynamic web page is a hypertext document rendered to a World_Wide Web user presenting content that has been customized or actualized for each individual viewing or rendition or that continually updates information as the page is displayed to the user. In dynamic sites page content and page layout are created separately. The content is retrieved from a database and is placed on a web page only when needed or asked for. The benefit of this is that it allows for quicker page loading and it allows just about anyone, with limited or no web design experience, to update their own website via an administrative tool. This set-up is ideal for those who wish to make frequent changes to their websites including text and image updates. Dynamic sites are also great for image galleries, online calendars or e-commerce, etc. However, a web page can also provide a live user experience. Content (text, images, form fields, etc.) on a web page can change, in response to different contexts or conditions. There are two ways to create this kind of effect:

- Using client-side scripting to change interface behaviors within a specific web page, in response to mouse or keyboard actions or at specified timing events. In this case the dynamic behavior occurs within the presentation.

- Using server-side scripting to change the supplied page source between pages, adjusting the sequence or reload of the web pages or web content supplied to the browser. Server responses may be determined by such conditions as data in a posted HTML form, parameters in the URL, the type of browser being used, the passage of time, or a database or server state.

Advantages of Dynamic Pages

A dynamic page can be customized by a response on a server to help personalize your site to meet your customer's need. All page content will come from a database connected to the Web site. Since the dynamic template is maintained separately from the content, it means that content changes can be made when needed. In addition, the web site can be updated without major maintenance, editing and reviewing, which translates into lower maintenance costs and time.

Disadvantages of Dynamic Pages

A downside to dynamic pages is that, in days gone by, many of the web spiders could not read the URL parameters to the right of the question mark in the dynamic URL. Today, webmasters have seen some improvement. Google and some of the other search engines can handle simple dynamic URLs, but if the query parameters gets too long or complicated (having more than one ? for example) their crawler will ignore the link completely. Yahoo, on the other hand, suggests that you use dynamic links only in directories that are not intended to be crawled or indexed [35].

## 2.3 Java Applets/Servlets

A Java applet is an applet delivered to the users in the form of Java bytecode. Java applets can run in a Web browser using a Java Virtual Machine (JVM), or in Sun's AppletViewer which is a stand-alone tool for testing applets. Java applets were introduced in the first version of the Java language in 1995. Java applets are usually written in the Java programming language but they can also be written in other languages that compile to Java byte code such as Jython, Ruby or Eiffel. Applets are used to provide interactive features to web applications that cannot be provided by HTML alone. They can capture mouse input (like rotating 3D object) and also have controls like buttons or check boxes. In response to the user action an applet can change the provided graphic content. This makes applets well suitable for demonstration, visualization and teaching purposes. An applet can also be text area only, providing, for instance, cross platform command-line interface to some remote system. If needed, applet can leave the dedicated area and run as separate window. However applets have very little control on web page content outside the applet dedicated area, so they are less useful for improving the site appearance in general. Applet can also play media in formats that are not natively supported by the browser. Java applets run at a speed that is comparable to (but generally slower than) other compiled languages such as C++, but many times faster than JavaScript. In addition they can use 3D hardware acceleration that is available from Java. This makes applets well suitable for non trivial, computation intensive visualizations. HTML page may embed parameters that are passed to the applet. Hence the same applet may appear differently depending on that parameters were passed. First implementations were downloading an applet class by class. While classes are small files, there are frequently a lot of them, so applets got a reputation of slow loading components. However since jars were introduced an applet is usually delivered as a single file that has a size of the bigger image (hundreds of kilobytes to several megabytes). Since Java's byte code is platform independent, Java applets can be executed by browsers for many platforms, including Windows, Unix, Mac OS and Linux. It is also trivial to run Java applet as an application with very little extra code. This has the advantage of running a Java applet in offline mode without the need for internet browser software and also directly from the development IDE. Java applets are executed in a *sandbox* by most web browsers, preventing them from accessing local data like clipboard or file system. The code of the applet is downloaded from a web server and the browser either embeds the applet into a web page or opens a new window showing the applet's user interface.

A Java applet may have any of the following disadvantages:

- It requires the Java plug-in which may not be available on some less popular web browsers or operating systems.

- Some organizations only allow software installed by the administrators. As a result, some users can only view applets that are important enough to contact the administrator asking to install Java plug-in.
- As with any client side scripting, security restrictions may make difficult or even impossible for untrusted applet to achieve the desired goals.
- Some more badly designed code may require a specific JRE.

Java Servlets are Java programming language objects that dynamically process requests and construct responses. Servlets can be considered as "server-side applet". The Java Servlets API allows a software developer to add dynamic content to a Web server using the Java platform. The generated content is commonly HTML, but may be other data such as XML. Servlets are the Java counterpart to non-Java dynamic Web content technologies such as PHP, CGI and ASP.NET, and as such some find it easier to think of them as 'Java scripts' (Not to be confused with JavaScript). Servlets can maintain state across many server transactions by using HTTP cookies, session variables or URL rewriting. Servlets also have some disadvantages:

- Developers MUST know JAVA.
- Web Administrator will need to learn how to install and maintain Java Servlets.
- Tedious uses of out.println () statements.
- Can be remedied by using Java Server Page (JSP).

## 2.4 Webservices

A web service (also Webservice) is traditionally defined by the W3C as "a software system designed to support interoperable machine-to-machine interaction over a network. It has an interface described in a machine-processable format (specifically Web Services Description Language WSDL).

Other systems interact with the web service in a manner prescribed by its description using SOAP (Simple Object Access protocol) messages, typically conveyed using HTTP with an XML serialization in conjunction with other web-related standards. Web services today are frequently just Application Programming Interfaces (API) or web APIs which can be accessed over a network, such as the Internet, and executed on a remote system hosting the requested services.

Advantages of web services:

- Web services provide interoperability between various software applications running on disparate platforms.
- Web services use open standards and protocols. Protocols and data formats are text-based where possible, making it easy for developers to comprehend.
- By utilizing HTTP, web services can work through many common firewall security measures without requiring changes to the firewall filtering roles.
- Web services easily allow software and services from different companies and locations to be combined easily to provide an integrated service.
- Web services allow the reuse of services and components within an infrastructure.

On the other hand Webservices have some disadvantages:

- Webservices standards for features such as transactions are currently nonexistent or still in their infancy compared to more mature distributed computing open standards such as CORBA.

- Webservices may suffer from poor performance compared to other distributed computing approaches such as RMI, CORBA, or DCOM. This is a common trade-off when choosing text-based formats. XML explicitly does not count among its design goals either conciseness of encoding or efficiency of parsing.

- By utilizing HTTP, web services can evade existing firewall security measures whose roles are intended to block or audit communication between programs on either side of the firewall [36].

## 2.5 Client Side Scripting

Client-side scripting generally refers to the class of computer programs on the web that are executed by client-side, by the user's web browser, instead of server-side (on the web server). This type of computer programming is an important part of the Dynamic HTML (DHTML) concept, enabling web pages to be scripted; that is, to have different and changing content depending on user input, environmental conditions (such as the time of day), or other variables. Web authors write client-side scripts in languages such as JavaScript and VBScript. Client-side scripts are often embedded within an HTML document (hence known as an "embedded script"), but they may also be contained in a separate file, which is referenced by the document (or documents) that use it (hence known as an "external script"). Upon request, the necessary files are sent to the user's computer by the web server on which they reside. The user's web browser executes the script, and then displays the document, including any visible output from the script. Client-side scripts may also contain instructions for the browser to follow if the user interacts with the document in a certain way, e.g., clicks a certain button. These instructions can be followed without further communication with the server, though they may require such communication. By viewing the file that contains the script, users may be able to see its source code. In contrast, server-side scripts, written in languages such as Perl, PHP, and server-side VBScript, are executed by the web server when the user requests a document. They produce output in a format understandable by web browsers (usually HTML), which is then sent to the user's computer. The user cannot see the script's source code (unless the author publishes the code separately), and may not even be aware that a script was executed. The documents produced by server-side scripts may, of course, contain client-side scripts. Client-side scripts have greater access to the information and functions available on the user's browser, whereas server-side scripts have greater access to the information and functions available on the server. Server-side scripts require that their language's interpreter be installed on the server, and produce the same output regardless of the client's browser, operating system, or other system details. Client-side scripts do not require additional software on the server (making them popular with authors who lack administrative access to their servers); however, they do require that the user's web browser understands the scripting language in which they are written. It is therefore impractical for an author to write scripts in a language that is not supported by the web browsers used by a majority of his or her audience. Due to security restrictions, client-side scripts may not be allowed to access the user's computer beyond the web browser application. Techniques like ActiveX controls can be used to sidestep this restriction. Unfortunately, even languages that are supported by a wide variety of browsers may not be implemented in precisely the same way across all browsers and operating systems. Authors are well-advised to review the behavior of their client-side scripts on a variety of platforms before they put them into use.

## 2.6 Server Side Scripting

Server-side scripting is a web server technology in which a user's request is fulfilled by running a script directly on the web server to generate dynamic web pages. It is usually used to provide interactive web sites that interface to databases or other data stores. This is different from  Client side scripting where scripts are run by the viewing web browser, usually in JavaScript. The primary advantage to server-side scripting is the ability to highly customize the response based on the user's requirements, access rights, or queries into data stores. When the server serves data in a commonly used manner, for example according to the HTTP or FTP protocols, users may have their choice of a number of client programs (most modern web browsers can request and receive data using both of those protocols).  In the case of more specialized applications, programmers may write their own server, client, and communications protocol that can only be used with one another. Programs that run on a user's local computer without ever sending or receiving data over a network are not considered clients, and so the operations of such programs would not be considered client-side operations. In the earlier days of the web, server-side scripting was almost exclusively performed by using a combination of C programs, Perl scripts and shell scripts using the Common Gateway Interface (CGI). Those scripts were executed by the operating system, mnemonic coding and the results simply served back by the web server. Nowadays, these and other on-line scripting languages such as ASP and PHP can often be executed directly by the web server itself or by extension modules -e.g. Mod_perl or Mod_php- to the web server. Either form of scripting (i.e., CGI or direct execution) can be used to build up complex multi-page sites, but direct execution usually results in lower overhead due to the lack of calls to external interpreters. Dynamic websites are also sometimes powered by custom web application servers, for example the Python "Base HTTP Server" library, although some may not consider this to be server-side scripting. Some server-side scripting languages include:

- ASP/ASP.NET (*.asp/*.aspx)
- ColdFusion Markup Language (*.cfm)
- ANSI C Server Scripts TrustLeap G-WAN ANSI C Scripts (*.c)

- Java via JavaServer Pages (*.jsp)
- Javascript using Server Side Javascript (*.ssjs)
- PHP (*.php)
- Perl (*.pl)
- SMX (*.smx)
- Python (*.py)
- Ruby (*.rb)
- Lasso (*.lasso)
- WebDNA (*.dna,*.tpl)

## 2.7 AJAX

Ajax (shorthand for Asynchronous JavaScript And XML) is a group of interrelated web development techniques used on the client-side to create interactive web applications. With Ajax, web applications can retrieve data from the server asynchronously in the background without interfering with the display and behavior of the existing page. The use of Ajax techniques has led to an increase in interactive or dynamic interfaces on web pages. Data is usually retrieved using the *XMLHttpRequest* object. Despite the name, the use of JavaScript and XML is not actually required, nor do the requests need to be asynchronous. Like DHTML and LAMP, Ajax is not a technology in itself, but a group of technologies. Ajax uses a combination of:

- HTML and CSS for marking up and styling information.
- The DOM accessed with JavaScript to dynamically display and interact with the information presented.
- A method for exchanging data asynchronously between browser and server, thereby avoiding page reloads. The XMLHttpRequest (XHR) object is usually used, but sometimes an IFrame object or a dynamically added <script> tag is used instead.

- A format for the data sent to the browser. Common formats include XML, pre-formatted HTML, plain text, and JavaScript Object Notation (JSON). This data could be created dynamically by some form of server-side scripting.

- The term *Ajax* has come to represent a broad group of web technologies that can be used to implement a web application that communicates with a server in the background, without interfering with the current state of the page.

Since then, however, there have been a number of developments in the technologies used in an Ajax application, and the definition of the term Ajax. In particular, it has been noted that:

- JavaScript is not the only client-side scripting language that can be used for implementing an Ajax application. Other languages such as VBScript are also capable of the required functionality. However JavaScript is the most popular language for Ajax
- Programming due to its inclusion in and compatibility with the majority of modern web browsers.
- XML is not required for data interchange and therefore XSLT is not required for the manipulation of data. JavaScript Object Notation (JSON) is often used as an alternative format for data interchange, although other formats such as preformatted HTML or plain text can also be used.

Advantages of AJAX:

*Asynchronous*: Ajax allows for the ability to make asynchronous calls to a web server, this allows the client browser to avoid waiting for all data to arrive before allowing the user to act once more.

*Minimal data transfer*: By not performing a full post back and sending all form data to the server, the network utilization is minimized and quicker operations occur. In sites and locations with restricted pipes for data transfer, this can greatly improve network performance.

*Limited processing on the server*: With the fact that only the necessary data is sent to the server, the server is not required to process all form elements. By sending only the necessary data, there is limited processing on the server, There is no need to process all form elements, process the Viewstate, send images back to the client, and no need to send a full page back to the client.

Disadvantages of using AJAX:

One disadvantage of using AJAX is security. Sometimes developers do not put checks on the data coming into the server - they assume that it's coming from their own website. Unfortunately, this is subject to injection attacks. Furthermore, there is several ways to fake how the data is coming in, making detection all but impossible. Another disadvantage is that AJAX does not play well in encrypted environments. AJAX relies on plain text transmission (nothing but text can be transmitted through AJAX anyways), and so encrypting this stream and having the server-side program deal with it presents large problems [37].

# CHAPTER 3

# RELATED WORK

# CHAPTER 3
# RELATED WORK

## Introduction

As our main problem will be accessing an OPC DA server, which is a COM/DCOM server through the Internet, hence we have two options, using DCOM or using modern IT technologies. DCOM is best suitable for LANs where there are less number of nodes and small delay times. However, when used through Internet there will be some limitations related to its nature. DCOM is windows platform dependent, difficult to configure, has very long and non-configurable timeouts, and cannot be used for Internet communication. On the other hand, if modern IT technologies used, we will have to face Web problems. Internet is based on HTTP protocol that is a stateless protocol because it has no built-in state mechanism. Our goal is to make the web application behaves similar to a desktop application, which has higher user interactivity and availability. To do that, we need to integrate some modern web technologies to achieve our goals. In a classic client/server application, there is a static communication between client and server but with web application this static communication does not exist. Thus the need for a persistent communication mechanism by which a client can communicate with the server independent of any specific action taken in the user interface by the user, so we want that communication takes place even if the user is not clicking or using any of the controls in the user interface. There are many research efforts in this area, which used both DCOM and modern IT technologies like XML, Webservices, Java and AJAX. In the following sections, we will take an overview of the previous work in this area.

## 3.1 DCOM

Most of the previous researches used DCOM for communication with OPC DA server through LANs. For example Xiaofeng Lee et al. [29] uses DCOM communication between an OPC DA client and OPC DA server, then they transfer the OPC DA client data to XML format to be able to access these data through Internet with an XML-DA client which communicate to an XML server (Web server) to get data as shown in Figure 3.1. Also Truong Chau et al. [26] uses DCOM to enable the C# server script to access OPC DA through LAN and because the OPC DA client is a .NET client, they used an OPC .NET wrapper to make the transformation from .NET to COM and COM to .NET as shown in Figure 3.2. Zhang Lieping et al. [31] uses the OPC DA Toolbox which is integrated in MATLAB 7 and above editions, this Toolbox enables MATLAB applications to communicate with OPC DA servers through DCOM then the user can simply and conveniently realize the operation to the OPC objects. As they claimed that these features could simplify the process of development and provide an effective method to realize the remote real-time communication between MATLAB and process devices as shown in Figure 3.3.

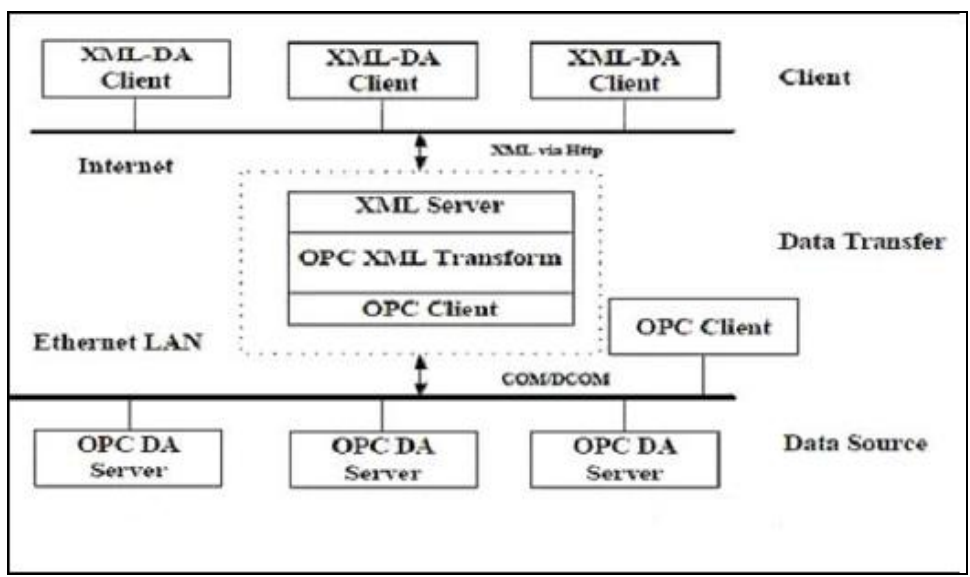

**Figure 3.1: Xiaofeng Lee et al. [29] design**

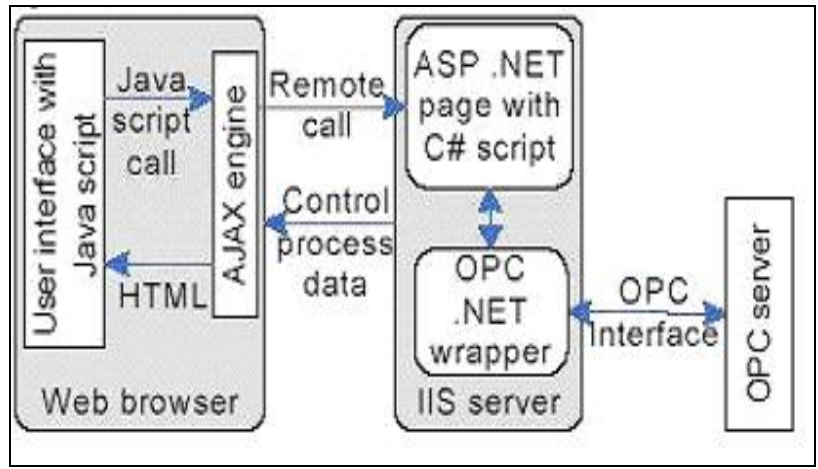

**Figure 3.2: Truong Chau et al. [26] design**

.

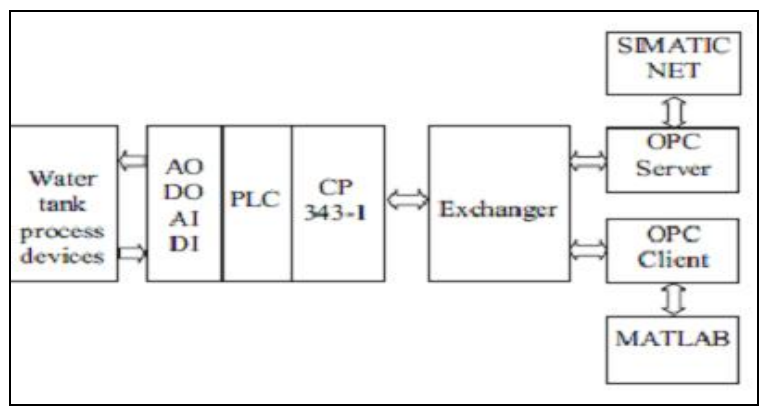

**Figure 3.3: Zhang Lieping et al. [31] design**

From the above discussion, we conclude that the only way to communicate directly to the OPC DA server is DCOM (or COM if the client and server are on the same machine). Therefore, the first step to access an OPC DA through web is to use COM or DCOM through LAN, and then we have to do a suitable transformation to enable accessing the OPC data through Internet. In our approach, we also use DCOM for communication between the web server and the OPC DA server, which connected to the same plant LAN.

## 3.2 XML

XML is a platform-independent, which is an important feature to achieve interoperability between different applications, which are running on different platforms. This is the reason, which forced the OPC FOUNDATION to release the OPC XML-DA specifications to allow the XML applications to access OPC data in a standard way. The other advantage of OPC XML-DA is the simple administration as it based on SOAP and XML. On the other hand, it has some disadvantages such as:

- Not suitable for transferring large data volumes

- XML technology is generally slower than COM

- The interaction parameters coded using XML, which leads to an overhead.

- An OPC XML-DA Service is stateless.

During browsing, no information about the position of the client in the namespace is stored in the OPC XML-DA Service, but all information about the namespace (or a defined part of it) transferred to the client at the same time. The client can poll for values at the server, but it should also be possible, to receive changed values automatically. XML-DA servers may stand alone, or may be developed to wrap COM based 3.0, and even 2.0x servers [14] [3].

Xiaofeng Lee et al. [29] suggested designing an information integration system which will adopt OPC DA to OPC XML-DA; the design includes three layers structure, data source layer, data transfer layer and client layer as shown in Figure 3.1. The authors claim that because of adopting industry standard OPC interface and webservice transfer interface, the remote monitoring system based on OPC XML-DA technology makes it convenient to update and expand system. If we analyze this approach, we will find that there is an overhead in layer2 because of the COM-XML transformer. In addition, the authors did not expose to the problem of client data update, is there a data polling mechanism that enables the client to get the new data- if there is –in an efficient way that consumes as little as possible of available resources. For similar work, that uses XML and/or OPC XML-DA techniques see [27, 20, and 24]. Due to possible performance limitations, OPC XML-DA is unlikely used for real time applications, although it commonly used as a bridge between the enterprise and control network. Furthermore, only OPC-DA functionality is provided in XML-DA. So it can be best seen as a transitional path to a true Webservice architecture that is just released by the OPC Foundation, which is OPC-UA (Unified Architecture) project [1]. For more information about OPC-UA, see [28].

## 3.3 Webservices

Webservices as defined by the W3C as "a software system designed to support interoperable machine-to-machine interaction over a network. As Shekhar M. Kelapure et al. [21] mentioned about the Webservices features, which are:

- Communicate via open protocols (HTTP, SMTP, etc.)
- Processes XML messages framed using SOAP
- Describes its messages using XML Schema
- Provides an endpoint description using WSDL
- Can be discovered using UDDI

The authors found that they could achieve a good web based SCADA system using Webservices as shown in Figure 3.4. They concluded that the advantages obtained when using Webservices in SCADA systems are:

- Web services provide data publishing on the internet through HTTP protocol, thereby eliminating the need to compromise on security of SCADA servers.

- Web services provide the means to publish the data through various devices like thin clients, PDAs and mobile phones.

- Web services facilitate interfacing of multiple control centers, like ICCP, once the services are standardized.

- Webservices supported by .NET as well as JAVA technologies.

- Data can be fetched from Hard Disc as well as RAM using web services

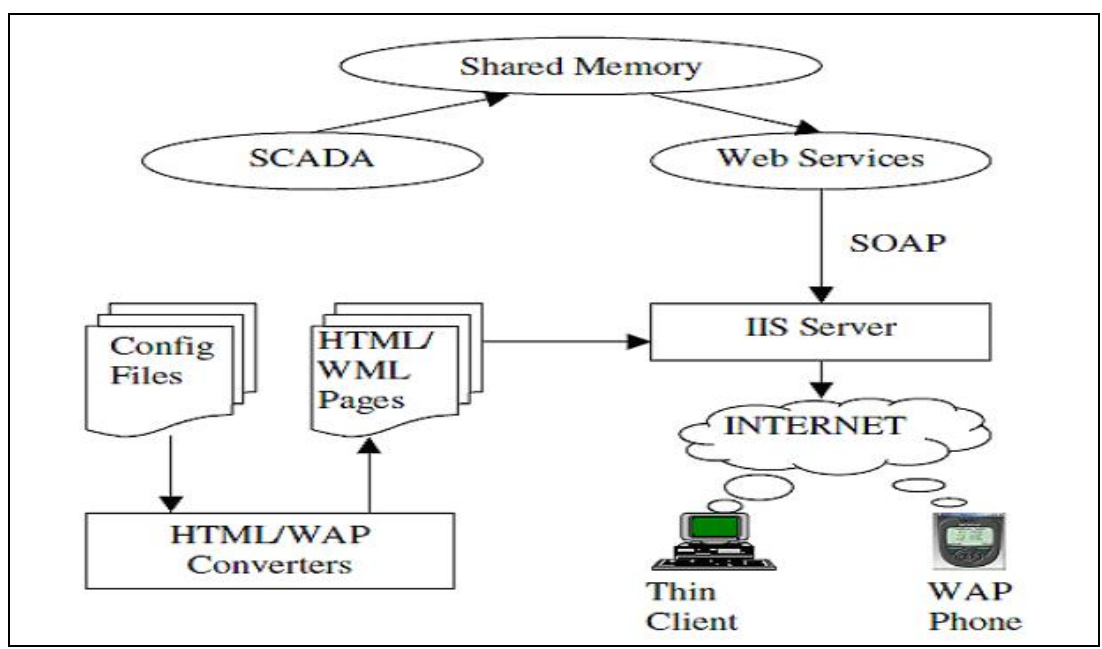

**Figure 3.4: Shekhar M. Kelapure et al. [21] design**

Because of all these advantages of Webservices, we will use it in our approach. On the other hand, Webservices have some disadvantages like:

- Webservices standards for features such as transactions are currently nonexistent or still in their infancy compared to more mature distributed computing open standards such as CORBA.

- Webservices may suffer from poor performance compared to other distributed computing approaches such as RMI, CORBA, or DCOM. This is a common trade-off when choosing text-based formats. XML explicitly does not count among its design goals either conciseness of encoding or efficiency of parsing.

- In addition, as mentioned in [31] that by utilizing HTTP, webservices can evade existing firewall security measures whose roles are intended to block or audit communication between programs on either side of the firewall.

Nunzio M. Torrisi et al. [11] proposed what they called CyberOPC which is a communication system anticipates the use of a gateway station called CyberOPC gateway, which will process messages sent to the OPC towards the public network and vice versa. Their proposed communication system targeted to best effort network with minimum bandwidth reserved for periodic traffic as shown in Figure 3.5. As they claimed that the necessity to satisfy time-critical and security requirements for remote control has stimulated the study of a new protocol for process control. Moreover, to obtain maximum interoperability with existing factory floor technologies, they built their communication project over the OPC technology. Unlike many OPC gateways for the Internet that use Webservices with HTTP and SOAP, the CyberOPC gateway does not use Webservices because the loss in performance will not be balanced by the advantage of a high-level programming language offered by Webservices. However, in order to use Webservices that transport OPC data, OPC libraries for processing OPC messages are required. Therefore, assuming that the use of libraries is necessary for processing non-open source and usually not free OPC data, they believe that it is useful to develop a set of free and open source libraries in order to implement the OPC communication over the Internet with the best possible performance.

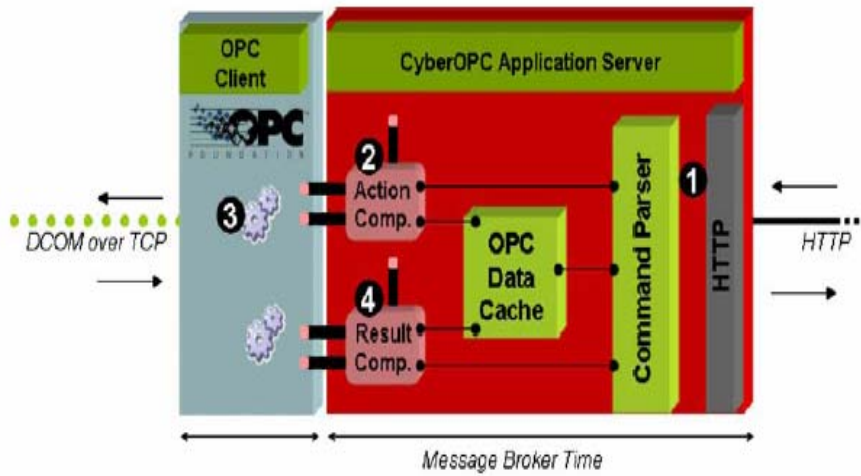

**Figure 3.5: Nunzio M. Torrisi et al. [11] design**

This approach does not solve the problem of periodic data update of the remote client to get new data. Moreover, when the remote client makes a request, the CyberOPC will not check if there is a change in this data from the last sent data or not, so if the control process has a high frequency data change rate, the remote client has to increase its periodic data-requesting rate, which can affect the server efficiency and network bandwidth.

## 3.4 Classic approaches

Thomas Bauer et al. [23] mentioned that to integrate OPC technology with Internet, its necessary to have an asynchronous data transmission method that is time-independent bidirectional communication, which is not possible because of the architecture of Internet. Therefore, they suggested an approach to solve this problem as shown in Figure 3.6. As they mentioned, first the Client (browser) will initiate a connection request for establishing at least one transmission channel that should be permanently open to send data at any desired time (asynchronously) and independently of the action of the user. In order to keep this data connection permanently open, the web server should continuously send data to the client, if there is no useful data it will send dummy data or send information to tell the client that a useful data coming. The need of dummy data is to maintain the data connection. Keeping this channel permanently open will enable the server to initiate the communication to the client, in the same time the client is still can initiate a new request to the server using another channel. In other words, the permanent channel will be for the server and the other channels are for the client. The main disadvantage of this approach is that the server and network will suffer because of the continuous transmit of useful and unuseful data. What will happen if the data change rate is slow? The web server will send a dummy data for long time wasting the network bandwidth with no feasibility. Also what will happen if for some reason the permanently open channel broken? The client will have to reinitiate the channel again. In addition to that, the client must keep checking the permanent open channel for new useful data the action that can affect the user interactivity.

Duo Li et al. [3] suggested an approach to implement the function of real time monitoring which provides periodically updated data to operators. The target data, which saved in a database, is marked and mapped to an HTML file-, say file2- in a web server; the database server automatically refreshes file2 with the latest data whenever the target data in the server updated. To show target data in an HTML file a Java applet developed that connects to file2 to get the latest data periodically and display them in the required format, another HTML file -say file1- in which the applet is included, developed to establish a basic human-machine interface (HMI) and complete some initiations. Data monitor functions commence from a web browser sending an HTTP GET command to the web server asking for file1, which then fetched to the browser and displayed and the applet is downloaded to run to show the target data as shown in Figure 3.7.

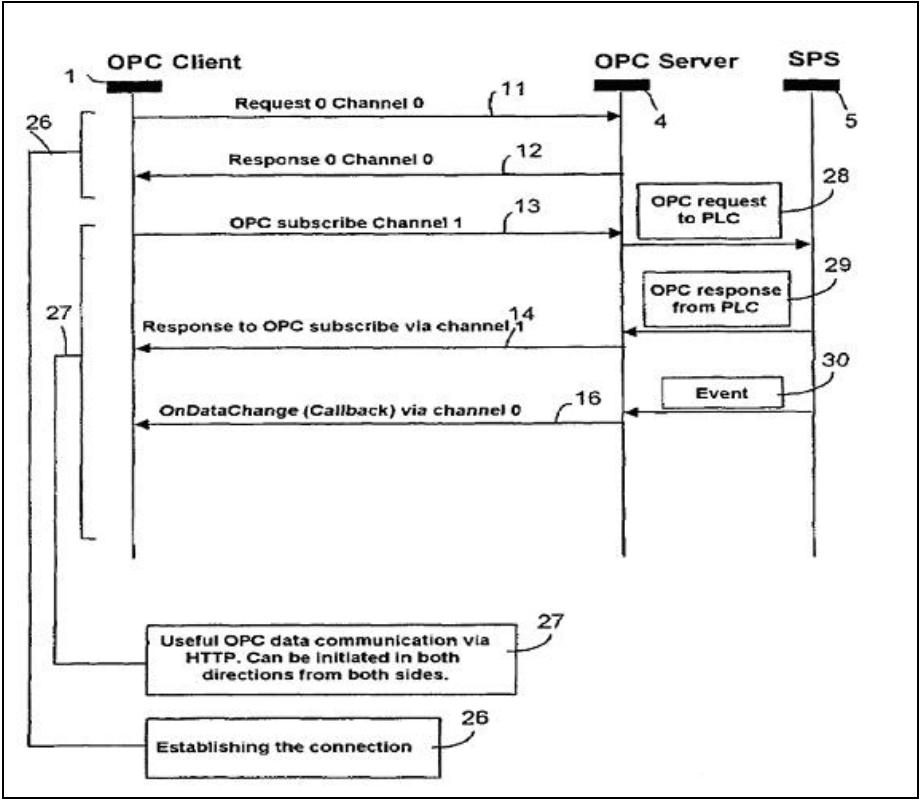

**Figure 3.6: Thomas Bauer et al. [23]**

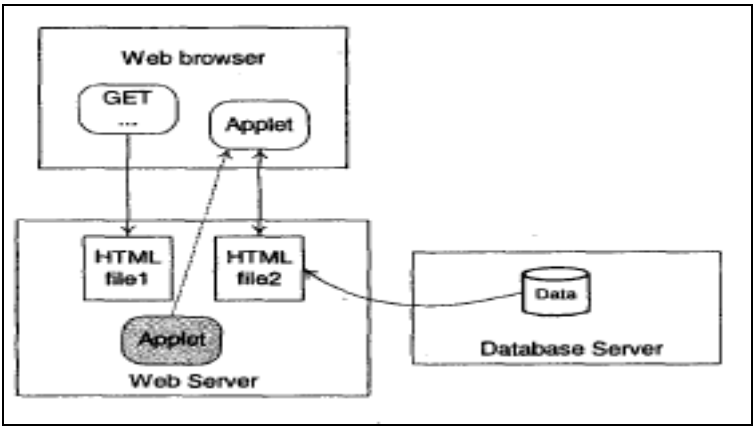

**Figure 3.7: Duo Li et al. [3] approach**

Actually, Java is a perfect programming language especially for web application development because its programs after compiled transferred to what known as byte code format, which can be executed by Java virtual machine (JVM) that is now supported by most platforms so a good interoperability can be achieved by Java. In addition, Java has many other features like Java applets and Servlets. Applets run on client (browser) and Servlets run on a web server. Applets and Servlets can communicate with each other in an efficient and persistent way, so they used to develop web based SCADA systems. On the other hand, they have some disadvantages as mentioned in Chapter 2.

Mike Clayton et al. [10] designed a TCP client-server Java architecture based on socket support to develop applications to integrate SCADA systems and Web applications. As he mentioned, that he achieved a solution to have a bridge between two complex applications (SCADA software and web servers), which evolve independently because they belong to different worlds. The proposed design is shown in Figure 3.8, which illustrates the TCP socket communication between the web server and the SCADA system. As shown in the figure that there is a Java application on the SCADA system machine, this Java application is a server, which allows the TCP connections from the remote clients. The Java application or the server communicate with the SCADA system by JNI (Java native interface) which enables Java classes to communicate with standard programs using custom librarries in C/C++. For each client TCP connection request from the web server, the Java server will create a new thread to handle that connection.

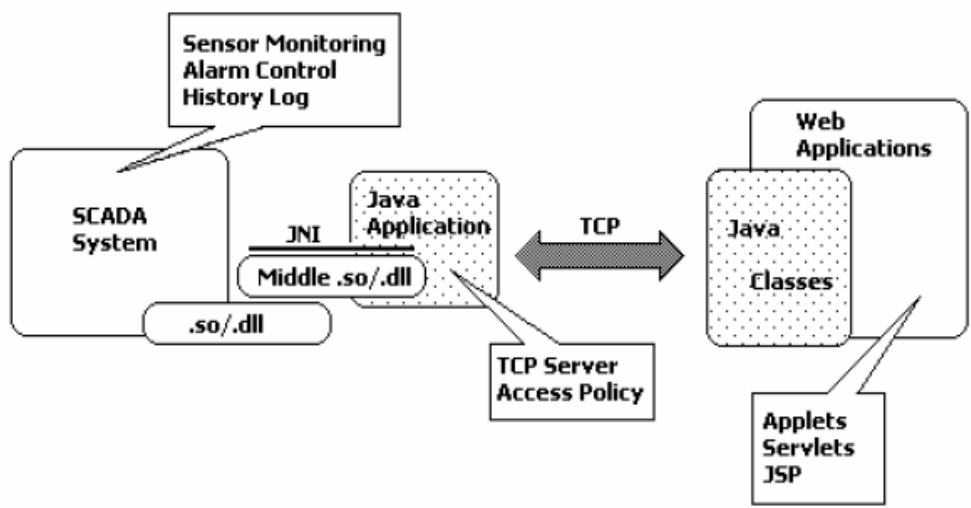

**Figure 3.8: Mike Clayton et al. [10] approach**

This approach shares the same disadvanteges like Duo Li et al. [3] approach, in addition to the complexity and non-flexibility in the Java application (server), because if the SCADA system changed or updated the Java application will need to be modified to use a new custom library. Also with large number of clients, the SCADA machine (Java server machine) will be very loaded with threads that may impact the SCADA system performance. In addition, the author of this approach concentrated on the communication between the web server and the JAVA server on the SCADA machine and ignored the client to the web server (browser), how the browser updates its data with new data. There should be a data polling mechanism, which enable the browser to ask for data update.

S.H. YANG et al. [19] mentioned in their guidance for web based Process control systems that Internet can be linked with the local computer system at any level in the information architecture, or even at the sensor/actuator level. These links result in a range of 4Rs (response time, resolution, reliability, and reparability). For example, if a fast response time is required a link to the control loop level should be made. If only abstracted information needed, the Internet should linked with a higher level in the information architecture such as the management level or the optimization level as shown in Figure 3.8. In addition, they mentioned that the Internet is a public transmission media, which is fundamentally different from other private transmission media used by many end-users for different purposes. The Internet transmission performance is associated with time delay and packet loss and possesses large temporal and spatial variation. In detail, the Internet time delay is characterized by the processing speed of nodes, the load of nodes, the connection bandwidth, the amount of data, the transmission speed,…etc. Therefore, it is somewhat unreasonable to model the Internet time delay for accurate prediction at every instant. In addition, they exposed to the problem of concurrent users, as they said that the special feature of the Internet-based SCADA is multiple-users and the uncertainty about users. The number and location of the users keep changing and the operators cannot see each other or may never have met. It is likely that multiple-users may try to control concurrently a particular process variable in which case some problems may arise. So, coordination among multiple-users becomes very important.

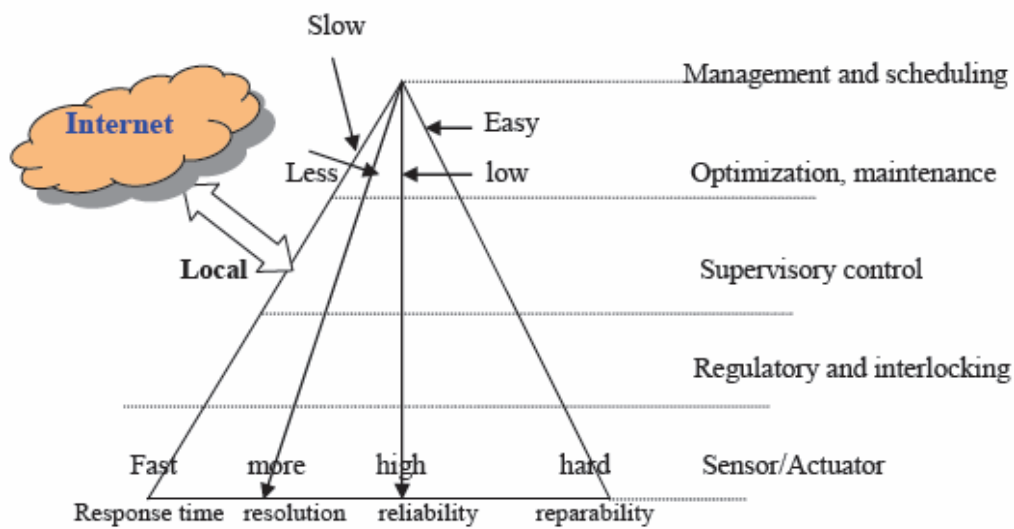

**Figure 3.9: S.H. YANG et al. [19] guidance (information architecture levels)**

## 3.5 AJAX

Actually, AJAX (Asynchronous JavaScript and XML) started a new line in the design of web based SCADA systems because it enables us to create more interactive web pages, which are our way to replace traditional desktop SCADA application. AJAX considered as the future of the Internet because of its ability to simulate desktop applications by the new features it offers like asynchronous client-server calls and partial-page updates. Ajax is a group of interrelated web development techniques used on the client-side to create interactive web applications. With Ajax, web applications can retrieve data from the server asynchronously in the background without interfering with the display and behavior of the existing page. The use of Ajax techniques has led to an increase in interactive or dynamic interfaces on web pages. Truong Chau et al. [26] used AJAX in their approach to get the OPC DA data to the client web page with high user interactivity as shown in Figure 3.2. But because of the problems of the HTTP protocol as a stateless protocol they had to use a simple data polling mechanism by using an AJAX Timer to poll for new OPC DA data set from the OPC DA server through the web server. When using an AJAX style of programming the old, classic programming approach must be given up. There is no form submit any more that posts all the client state to the server and requests for a complete new page description using HTML. Instead of loading several pages until the functionality done, only one page loaded and stay in the browser until the end of functionality. With the Ajax model, the execution of the web page processed in three different phases:

1. Loading the page
2. Loading data from the server using AJAX techniques
3. Interaction with the page using AJAX techniques

The web standards used in AJAX are well defined, and supported by all major browsers. AJAX Applications are browser and platform independent.

The main disadvantage of AJAX is security. Sometimes developers do not put checks on the data coming into the server - they assume that it is coming from their own website. Unfortunately, this is subject to injection attacks. Furthermore, there are several ways to "fake" how the data is coming in and making detection to all is impossible. Another disadvantage is that AJAX does not play well in encrypted environments. AJAX relies on plain text transmission (nothing but text can be transmitted through AJAX anyways), and so encrypting this stream and having the server-side program deals with it presents large problems [37].

## 3.6 Motivation

None of the mentioned approaches considered the feasibility of the realization of their work. If we look closely to [29, 27, 20, 24] we will find that they use XML and/or OPC XML-DA technologies, which have the disadvantages mentioned in Chapter 2 such as:

- Not suitable for transferring large data volumes
- XML interactions are generally slower than COM interactions.
- The interaction parameters coded using XML, which leads to an overhead.
- An OPC XML-DA Service is stateless.

Actually, OPC XML-DA designed for Internet access and enterprise integration and based on its platform-independence; it mainly implemented in embedded systems and on non-Microsoft platforms. However, due to its high resource consumption and limited performance, it was not as successful as expected for this type of applications.

In [21] the authors used the features offered by Webservices to design and implement a web based SCADA system. Moreover, they could solve the challenges, which they faced such as thin client support, refresh of GUI window, and firewalls restrictions.   However, they could not solve real time data collection challenge in an efficient and effective way because they used a constant refresh frequency of a few seconds, which will lead to non-synchronized process data transfer.

In [23], to allow each node (client or server) to start sending data to the other node independently at any time (bidirectional communication), the authors had to maintain a connection in a permanent open state by making the server continuously sends useful or unuseful data to the client. In addition, this will affect the performance of the system especially with large number of clients and low network bandwidth, and this may lead to server crash.

In [3] the authors will need to find a way to periodic update of file2, which may be a loop or any other way. Also the java applet, which embedded in the HTML file1, periodically will need to use a Timer to get the new data from file2 and then they will face the problem of synchronization between the applet Timer interval and the process of updating file2 with new data and this will be difficult. In addition, they ignored the heavy network load and the need for large server memory to handle the large number of requests and the heavy load on the server CPU. It is very difficult (nearly impossible) to guarantee a real time behavior with such limitations. The only solution is to spend money to increase the network bandwidth and the server CPU speed and memory capacity (The hardware resources). In our work, we try to find alternative, much cheaper solutions to these problems. In [11], the designed CyberOPC does not solve the problem of periodic data update of client to get new data. Moreover, when the remote client makes a request, the CyberOPC will not check if there is a change in these data from the last sent data or not so, if the control process has a high frequency data change, the remote client has to increase its periodic data requests, which affects the server efficiency and network bandwidth.

In [26], the authors did not take any attention to the efficiency and performance of their approach. The Timer, which they used, will get the data without any care if there is a significant data change or not. In addition, how they can specify the Timer interval, there are two cases for that, first, when the interval is small i.e. one second (high frequency data polling), where the server and network will suffer. Second, if the interval is large they will loss some data change events that maybe very important and the system will not considered as a Real-Time monitoring system.

So this approach still needs some modifications which enable us to get optimal efficiency and performance and to achieve a better real time behavior, for this reason we choose this approach to be our reference approach and because we believe that AJAX can be the way to make web applications compete with desktop applications by the interesting features it offers. We will have to replace the simple data polling mechanism the authors used by a more efficient mechanism which enables us to achieve real time system behavior and to get optimal usage of the available resources.

Finally, still the impression that "the Internet" naturally entails a uniform way of remote access to automation networks is deceptive. The user interface is only one aspect, the underlying mechanisms and data structures are another [17].

# CHAPTER 4

# THE PROPOSED APPROACH

# CHAPTER 4
# THE PROPOSED APPROACH

## 4.1 Problem description

How to design and implement an adaptive web-based SCADA system that has the following features:

- Efficient, that makes the best consumption of the available system resources i.e. CPU load and network bandwidth for the web server.

- Have a reasonable real time behavior that is adaptive to the real application.

- Have a good performance.

- Have a responsive user interface.

As the OPC DA server –which is a COM/DCOM server- still used and the world will still depend on it for many years. In addition, there are more and more applications implemented to work with it in all over the world, so we will design and implement a new methodology for providing a direct web access to controllers using OPC DA server.

Actually, we have two options to achieve this target, first using DCOM, second using a web server and modern IT technologies. If we used the first option, we will face the DCOM problems, which are:

1. Windows platform dependency

2. DCOM is difficult to configure

3. Has very long and non-configurable timeouts

4. Can be disabled by firewalls

That is why DCOM is avoided to be used for Internet. Therefore, we will use the second option, which is web server and modern IT technologies. However, with this option we are going to face many design and implementation challenges such as:

- <u>HTTP limitations</u>
    1. Stateless Protocol
    2. Request/Response architecture
    3. Refreshing GUI
    4. Client data update

    HTTP is a stateless protocol because it has no built in state mechanism, the web server maintains no information about the past client requests and each request is considered as a new request. In addition, the web server will not send any data to clients without requests from them. In addition, normally each request of HTML fetches new HTML page on the browser. To update the time varying data, the MMI screen needs to be updated regularly (Refresh rate). This causes flickering of the screen, which is undesirable. In addition, since the real time data is continuously varying, it is always desirable to have a quicker data update on the MMI. This update is limited on the browser by data volume and the network bandwidth.

- <u>Responsiveness</u> (user interactivity) which is lower in web applications compared to desktop applications because of the communication overhead with web applications and the architecture of the Internet which depends on request-response approach. The responsiveness of an interactive system describes how quickly it responds to user input (i.e. the rate of communication with the system). The user in the client machine has to wait for the response of the previous request before going ahead with a new request. In addition, the web page will freeze if the response delayed because of long time processing in the server machine or network congestion.

- <u>Real Time constraints</u>, which are very important for SCADA systems functionality, any SCADA system should maintain real time behavior. In critical SCADA systems, there should be a deadline for the response time, for example as the control process data changes, it must immediately sent to the client (Operator) who can take action according to these changes in the control process data.

- <u>Efficiency, which</u> means obtaining the optimal consumption of the available resources, or making the best use of the available resources to run the system. In our approach, we have a system with certain resources such as CPU ,RAM and network bandwidth, our mission will be how to run the system efficiently and effectively even with high load (large number of clients requests).

- <u>Performance</u>, which means how well, is the system doing the work it is supposed to do, or the extent to which the system is reliable and available. A system with high performance should have the following features:

  1. Short response time for a given piece of work
  2. High throughput (rate of processing work)
  3. Low utilization of computing resource(s)
  4. High availability of the computing system or application
  5. Fast (or highly compact) data compression and decompression
  6. High bandwidth / short data transmission time

- <u>Security</u>, for several years, security risks have been reported in control systems, upon which many of the nation's critical infrastructures rely to monitor and control sensitive processes and physical functions. In addition to general cyber threats, which have been steadily increasing, several factors have contributed to the escalation of risks specific to control systems, including the:

1. Adoption of standardized technologies with known vulnerabilities

2. Connectivity of control systems to other networks

3. Constraints on the use of existing security technologies and practices

4. Insecure remote connections

5. Widespread availability of technical information about control systems

- OPC DA protocol constraints, there are two constraints belong to OPC DA protocol, .NET-COM transformation and DCOM security problems. The first one appears because we will need a Webservice which implemented by VB.NET languages to access a legacy COM interface which belong to the OPC DA protocol. The second constraint is due to DCOM issues, When Windows XP Service Pack 2 is installed with its default configuration settings, OPC communication via DCOM will cease to work. SP2 includes many changes and security enhancements, two of which directly affect OPC via DCOM. First new DCOM limit settings have been added. Secondly, the software firewall included with XP has been greatly enhanced and is turned on by default.

- LANs local (virtual) IP addresses, usually LANs (Local Area Networks) use local IP (Internet protocol) addresses i.e. 192.168.16.1, this class of IP addresses is not visible through Internet and the only real and visible IP address will be the router international IP which given to the router by the ISP (Internet Service Provider). So, how we could access the plant web server, which has a local IP address, from the Internet.

- ADSL router dynamic real IP address, ADSL (Asymmetric Digital Subscriber Line) Internet connections assign addresses dynamically with DHCP, giving the router a different IP address on each connection. This will be a problem because we will access the plant LAN using the router public IP address by writing it in the Internet explorer address bar to open our SCADA web site. Each time this IP address changes the client should get the new one to be able to login again to the system.

Therefore, we will have to find some tools to deal with these problems. Our goal is to make the web application behaves similar to a desktop application, which has high user interactivity, and to do that we will look for some modern web technologies which enable us to achieve our targets. In a classic client/server application, there is a static communication between client and server but with web applications, this static communication does not exist thus the need for a persistent communication mechanism by which a client can communicate with the server independent of any specific action taken in the user interface by the user. Therefore, we want that communication take place even if the user is not clicking or using any of the controls in the user interface.

To do that we will use AJAX (Asynchronous JavaScript and XML) technology using JavaScript client scripting to communicate with a server side VB.NET Webservice located in a web server, which in turn is located in the plant. In addition, we called our approach an efficient approach to distinguish it from the other previous approaches, which concentrated on realizing the communication and data transfer without any attention to the consumption of server resources like CPU load, RAM capacity and Network bandwidth. Actually, our approach can be considered as an improvement to Truong Chau et al. [26] approach which is an example of the approaches that used a client side fixed data polling and with which we will compare our obtained results.

## 4.2 System Overview

As we have mentioned earlier that our main target will be accessing an OPC DA server which is a COM/DCOM server through Internet efficiently. So our target system consists of a web server, an OPC server and a PLC units which will be connected in the same plant network, which may be industrial Ethernet network or Fieldbus network (Profibus, MPI... etc), in our case it will be an Ethernet LAN because it is now a standard and familiar protocol in industry. This LAN connected to internet through an ADSL router which has a local (virtual) IP address and an international (Real) IP address. The browser can connect to the Webservice residing in the web server though internet using HTTP protocol, then the Webservice will connect to the OPC server in the same LAN using DCOM to get or set the new Data as shown in Figure 4.1. The used Technologies in our design will be:

1- IIS6.0 web server

2- OPC DA server "OPC.SimaticNet" for Siemens

3- AJAX and Web services

4- VB.NET 2008 programming language (server side scripting)

5- JavaScript programming (client side scripting)

6- Microsoft visual web developer 2008 (for active web pages design)

7- Simulated OPC data based on Siemens OPC DA server

8- PCs operating systems will be windows XP SP2

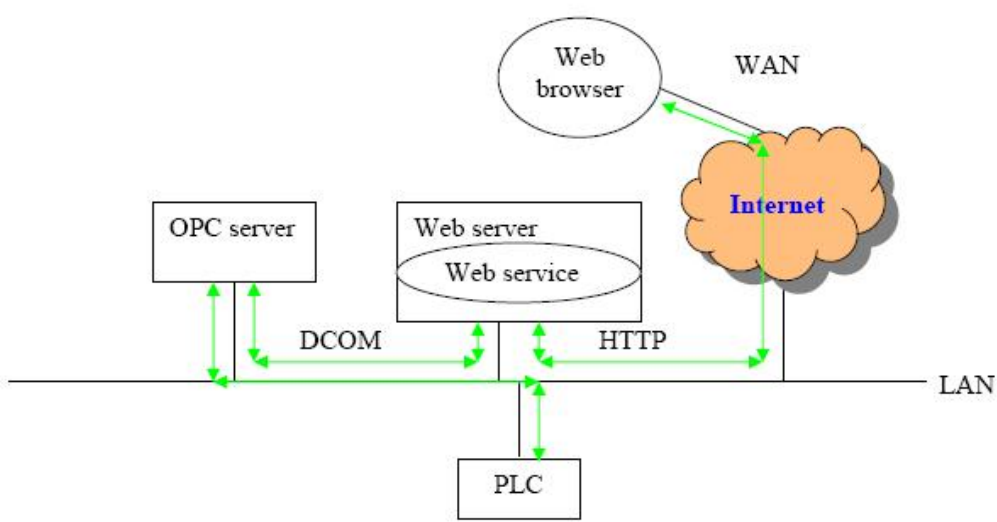

**Figure 4.1: System Overview**

## 4.3 System modules

To achieve our goals we will use a modular architecture in the design of our web based SCADA system. As shown in Figure 4.2 our system consists of following modules:

1. The Client, which is a web browser that will be used to run and display the designed SCADA web page and will send the required requests to update the web page with the new data. The requests will be done asynchronously (in background) by the AJAX engine and the web page will still responsive to the user. When the request's response comes the AJAX engine will forward the returned data to the web page, then the web page will send another request and so on. A new request will be sent only if the response to its predecessor request comes. Another one-direction channel can be used to write data (setpoints) directly to the control process.

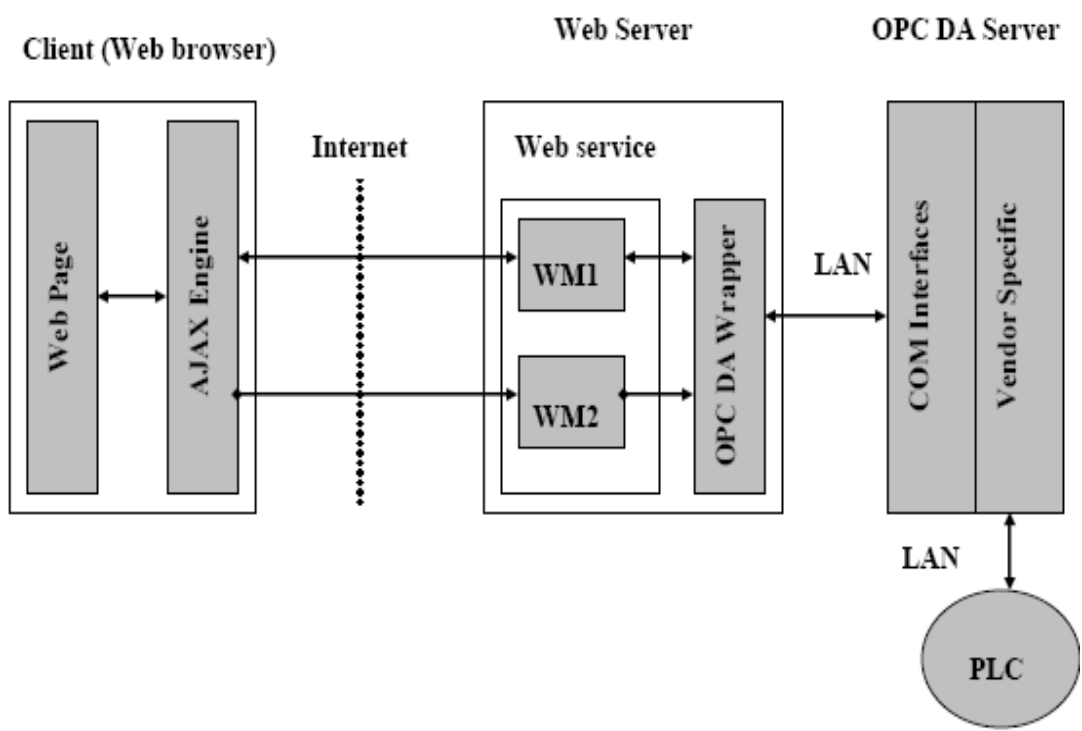

**Figure 4.2: System Modules overview**

2. The web server which will wait for the HTTP client's requests and forward these requests to the webservice which in turn will run the proper web method (WM1 if the request asks for new data and WM2 if the request contains setpoints to the OPC DA server). The WM1 will wait for a data change in the OPC DA server cache and then return the new data to the client. WM2 will write the coming client's setpoints to the OPC DA server and return to the client without waiting.

3. The OPC DA server, which is the interface to the physical control process (PLC), the OPC DA server has two layers, one to provide the standard COM interface to its clients and another one, which is a vendor specific driver to communicate with the PLC.

4. OPC DA Wrapper, as we mentioned that OPC DA server is based on Microsoft COM technology. However, the webservice, which we will use to access this server, will be implemented in the modern .NET technology, which is a new Microsoft technology and is different from the legacy COM technology. So we will have to find a way to access a legacy technology from a modern technology, the OPC FOUNDATION solved this problem by creating wrappers to enable .NET clients to access COM servers.

5. The PLC, which contains the program that controls the physical control process

The web server, OPC DA server and PLC units will be connected in the same LAN, and the client (browser) and the web server will communicate through the Internet (HTTP). The flow chart for each module can be found in appendix A, and the implementation for each module can be found in appendix B through D.

## 4.4 Solving the Challenges

### 4.4.1 HTTP Limitations

The first challenge that we will have to solve and will have a direct effect on solving other challenges belongs to the Internet infrastructure which is based on HTTP protocol which -as we have mentioned- is a stateless protocol. By stateless we mean that there is no persistent connection between the client and server, it is a request-response relation between them.

First, the client initiates a request for which a TCP connection established to the web server, and then the server sends back its response and closes the connection, the server will not send data without a request for it from clients. By this way, we cannot achieve a real time behavior, which is needed for any SCADA system. From the discussion of previous work and because of the Internet infrastructure, we conclude that data polling is the only solution, which is robust and viable; our challenge will be implementing a data polling that is effective and wastes as little as possible of the available resources.

In Truong Chau et al. [26] approach they used an AJAX update panel with AJAX Timer to make data polling every Timer interval, there will be two cases, first when the Timer interval is large the data polling will be done slowly and in this case, they may miss many data changes, which take place on the OPC server. Second case when they use high frequency data polling (small Timer interval) the client will generate large number of requests to the web server. These requests may queue up either at the network queue or at the web server queue. Consequently, the response time for a request will increase; also, the server and the network may reach a bad utilization. In addition, if the client requests rate is large compared to the control process data change rate, the client will get some data that it already has.

As a result, the system will suffer from up using its resources (server and network). This is a sort of damned-if-you-do and damned-if-you-don't situation [2]. As shown in Figure 4.3 there are three types of data polling, the first is simple refresh polling, second one is advanced refresh polling with no data change events and the third is advanced refresh polling with changed data available. A traditional data poll such as the one used by Truong Chau et al. [27] will query the server, get an answer, and then wait until the next poll. The waiting period until the next poll is a dead time during which neither the client nor the server can communicate with each other. By converting the dead time into a wait created by the web server, the client is waiting for the potential of data change happened, to achieve a better response time and enhance the resources utilization we will use a modified version of the third type data polling approach. Now how we can achieve our goals with HTTP limitations? To answer this question, first we have to specify our needs as follows:

1. We need web server (Web service) to know if there is any data change reported by the PLC system.

2. We need the server to inform its client about this data change without a request from the client.

3. In addition, we need that if there is no data change the web server should hold up the connection until data changes.

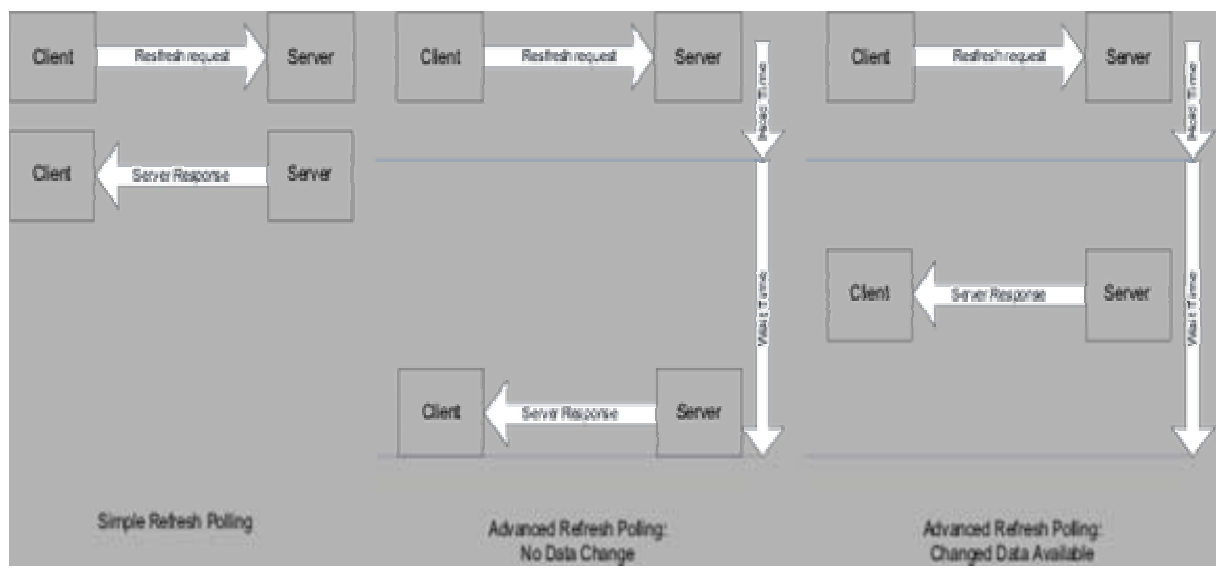

**Figure 4.3: Data polling types**

The OPC DA server will help us to achieve our needs because it has a feature, which is the subscription-based architecture of OPC, where the OPC server notifies its clients when there is a data change in a specified group of data.

The proposed architecture is shown in Figure 4.4. We are going to use AJAX technology in our system implementation. With AJAX, we will achieve two important features in a web application, partial page update and asynchronous communication with the web server. With partial page update, we can update a certain area in the web page without interfering with the rest of the page this technique is used now in the Google suggestions and Gmail. Asynchronous communication means that the client-server communication will be in background without the user feels that, and the page will still be interactive and responsive to user actions. In our approach, the AJAX enabled client will call a Webservice method continuously (as explained later) depending on the data change in the OPC server because of its change in PLC. We will implement the Webservice in VB.NET, and actually, it can be implemented in C # too. There will be two methods in the Webservice one for real time monitoring (From server to client) and the other for data writing from client to server as shown in Figure 4.4.

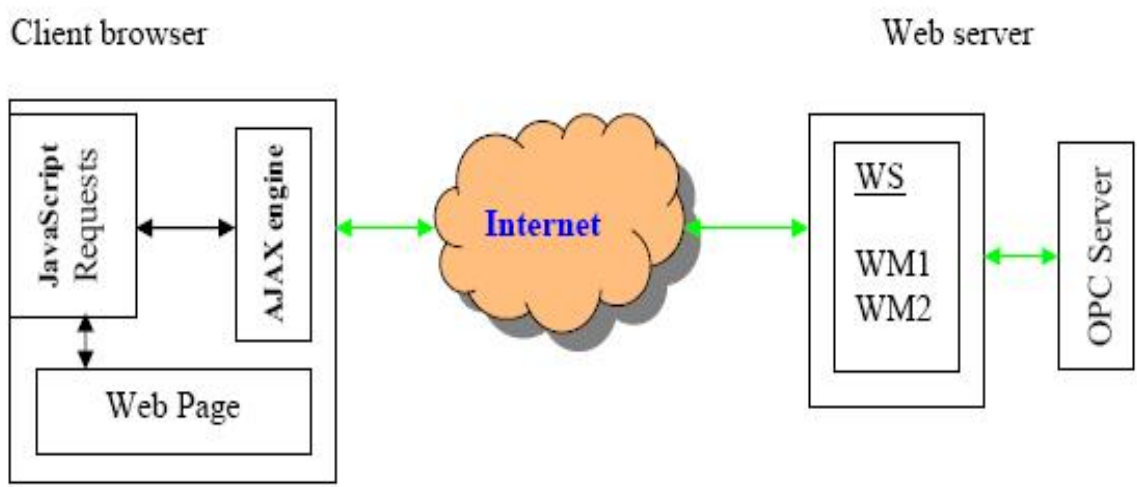

**Figure 4.4: AJAX-Webservice Communication**

To implement the PCP (Persistent Communication Pattern) of AJAX, we will use JavaScript *setTimeout ()* method which allows us to specify that a piece of JavaScript code (called an expression) will be run a specified number of milliseconds from when the *setTimeout ()* method was called without repeating. The general syntax of the method is:

*setTimeout (expression, Delay);*

Where *expression* is the JavaScript code to run after *Delay* milliseconds have elapsed. It is worth pointing out that *setTimeout ()* does not halt the execution of the script during the timeout period; it merely schedules the specified expression to be run at the specified time. After the call to *setTimeout ()* the script continues normally, with the timer running in the background. We will create a function called *InitConstantCall()* as follows:

```
<script type="text/javascript" language="JavaScript" >
var Delay=50; // time is in miliSeconds
Function InitConstantCall()
{
setTimeout('ConstantCall',Delay); // Run ConstantCall after 50 ms
}
</script>
```

Then we will put it in the *onload* event of the HTML *body* tag to be called when the page loaded. The function *ConstantCall()* is another JavaScript function, which implemented in this way:

```
Function ConstantCall()
{
MyWebService.GetUpdate(OnComplete);}// Calling the webservice web method
```

*MyWebService* is the Webservice which resides on the web server and it has two methods *GetUpdate()* which is used for real time monitoring channel and *senddata()* which used by the client (browser) to write data directly to the OPC server. *OnComplete()* is the callback function which will be called when the response comes from the web server (Webservice web method return) this function will contain the necessary scripts for parsing the received data and update the web page controls, then it will call the *ConstantCall()* function again as follows :

*Function onComplete(data)* *//data is the returned data*

*{*

*//code for parsing the received data*

*//code for updating the web page*

*.*

*.*

*.*

*.*

*//*

*setTimout('ConstantCall',Delay); //make a new rquest again*

*}*

This means that the function *ConstantCall()* will be called if the response from the Webservice web method comes which will come only if there is a data change in the OPC server cache according to a change in the PLC. The user will be able to send data to the server while the real time monitoring is running using another channel for sending data (a one-way channel) as shown in the following JavaScript code segment:

```
Function senddata()
{
        var mystr=new String("");
        mystr=$get('TextBox2').value + ";" + $get('TextBox3').value + ";" +
        $get('TextBox4').value;
        MyWebService.senddata(mystr,onReturn); //sending three values

}
```

where *mystr* is a string containing all the data which the user need to send to the OPC server - in our example there are three variables- and *onReturn()* is the feedback function which can be used to do any thing for example informing the user that the data arrived to the server. Finally, to call a Web service from the Internet, the Web service client needs to know the location of the Web service and the input and output information required for accessing the Web service. Therefore, as it's an AJAX enabled web page we have to add an AJAX *ScriptManager* from the VWD (Visual Web Developer 2008) Toolbox and we will add a reference to the Webservice (URL or path) in the following way:

```
<asp:ScriptManager ID="ScriptManager1" runat="server">
  <Services>
   <asp:ServiceReference Path="MyWebService.asmx" />
  </Services>
</asp:ScriptManager>
```

Flow charts of client side scripting can be found in appendix A and the complete JavaScript code for client side implementation can be found in appendix B.

The final user interface may look like Figure 4.5. The important part of this research is server side and network. We aim to make an efficient web based SCADA system without wasting more server resources like CPU usage and Network traffic, that is why we divide the system to client side scripting and server side scripting to free the server from managing the presentation of the web page and to let this part to the client machine.

As we mentioned before we will use OPC DA server to communicate with the controller (PLC), OPC DA server is a COM server so we need a way to transfer. NET calls –from the Webservice- to COM calls and vice versa so, we will use OPC.NET wrapper. We will apply our project on Siemens OPC DA server "OPC.SimaticNET".

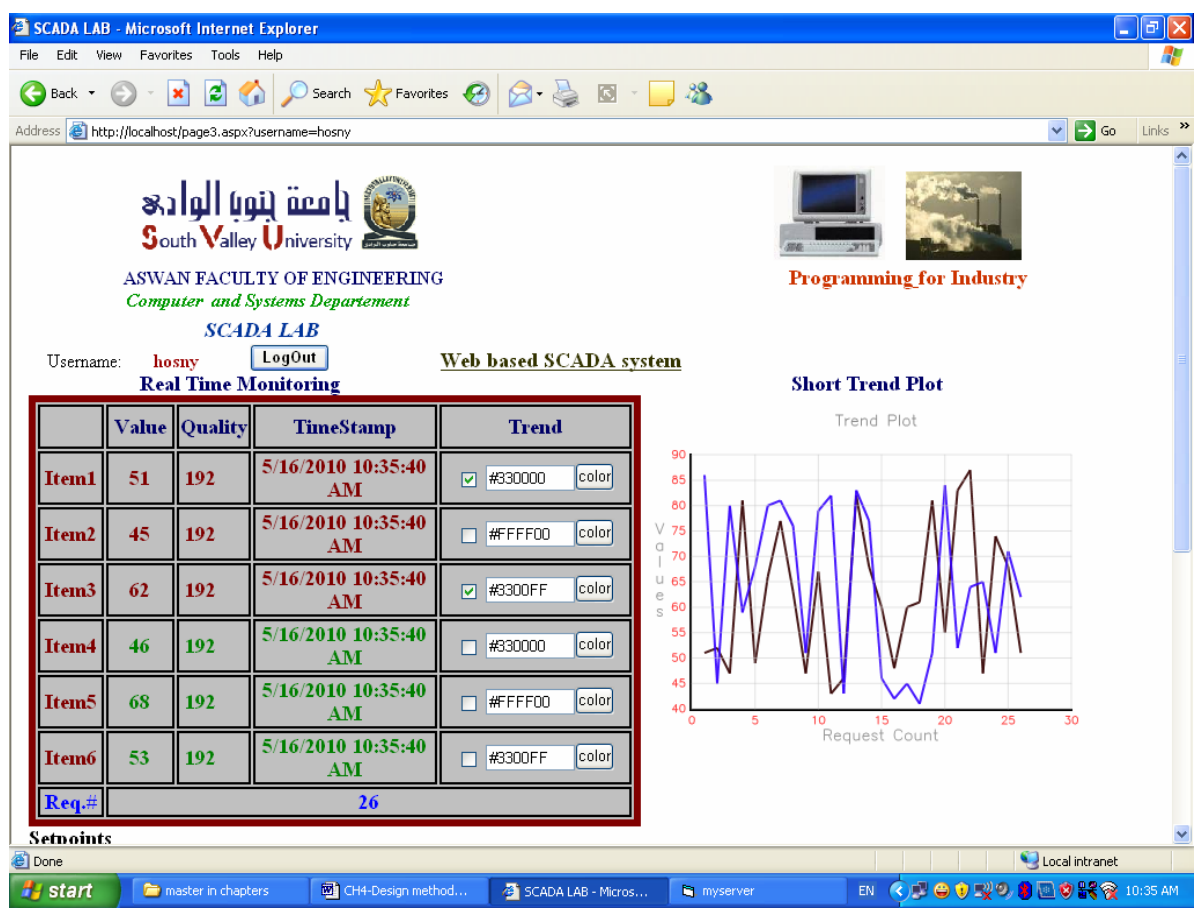

**Figure 4.5: Final Presentation sample**

As we mentioned earlier the Webservice method will only return if there is a change in OPC data so we will use the OPC API *datachange ()* event which triggered by the OPC server if there is data change in the group of items which we define to contain the needed items in our project. The global declarations in the Webservice will be similar to this:

```
Imports System.Web
Imports System.Web.Services
Imports System.Web.Services.Protocols
Imports System
Imports System.Threading
Dim OPCMyServer As OPCAutomation.OPCServer
Dim OPCMyGroups As OPCAutomation.OPCGroups
Dim WithEvents OPCMyGroup As        OPCAutomation.OPCGroup
Dim OPCMyItems As OPCAutomation.OPCItems
Dim OPCMyItem As OPCAutomation.OPCItem
Dim sItemName(8) As String
Dim cH(8) As Integer
Dim sH(8) As Integer
Dim oVal(8) As Object
Dim dTime(8) As Date
Dim wQuality(8) As Short
Dim ServerHandles As Array
Dim Errors As Array
Dim mystr As String
Private m_Thread As Thread
```

As shown we declare an OPC server and OPC group objects also we declare the required arrays for communication with OPC server. In addition, we declare a thread called *m_thread* this thread will be the way by which the web method will know if there is a data change in the OPC server. In computer science, a thread of execution results from a fork of a computer program into two or more concurrently running tasks. The implementation of threads and processes differs from one operating system to another, but in most cases, a thread is contained inside a process. Multiple threads can exist within the same process and share resources such as memory, while different processes do not share these resources [33]. To use threading in our project we have to import "*System. Threading*" class to our Webservice. The idea will be as shown in the following code segment.

```vbnet
   <WebMethod()> _
  Public Function GetUpdate() As String
    opcconnect()                ' connecting to the OPC server
    Thread.sleep(50)            'for stability
     'Instantiation of the new thread
    m_Thread = New Thread(AddressOf mythread)
     ' m_thread will work in background
    m_Thread.IsBackground = True
     ' starting m_thread
    m_Thread.Start()
     'main thread will suspended untill m_thread terminates
    m_Thread.Join()
     'm_thread terminated then main thread resumed
    getdata()
    m_Thread = Nothing
    'disconnecting from the OPC server
    opcdisconnect()
     'returning to the client (AJAX)
    Return mystr
  End Function

  Sub mythread()
      'm_thread suspend it self until data change event triggered
      m_Thread.Suspend()
  End Sub
    'the datachange event procedure
Private Sub OPCMyGroup_DataChange(Arguments) Handles
OPCMyGroup.DataChange
     ' data change event trigered and resume m_thread
    m_Thread.Resume()
  End Sub
```

As shown in the above VB.Net code segment, we first call the function *opcconnect()* which contains the required instructions to connect to the OPC server. Then instantiate a thread *m_thread* and assign it the function *mythread ()* then we choose to let it works in background, as we see this thread function has only one statement *m_thread.Suspend()* and this is the point, this thread started to suspend itself. After assigning the thread it's function, the web method starts it, and then the web method will call *m_thread.join()* this is another important statement by which the main thread (web method) will suspend itself too at this point of execution and will not resume until its child thread terminates, which will happen only if there is a data change as shown in the *datachange()* event handling function which triggered by the OPC server when there is a data change and when this happened the web method will resume and call the function *getdata()* to read the new data from the OPC server and to compact it in a string or XMLDocument object then returns it as a response to the client. We intended to make the server side scripting as simple as possible to achieve our targets, which are optimal efficiency and performance. If there is no data change, all the running threads in the web server will sleep until *datachange ()* event triggered by the OPC server. This is our way to enable the web method to know if there is a data change or not. Also, this is our way to make the communication between the client and server seems to be persistent, in the same time the client still can send data to the server using the other web method *senddata ()* which is just a direct data writing to the OPC server. Flow charts for the server side-scripting can found in appendix A and the complete VB.NET code for server side implementation is found in appendix C.

**4.4.2 Responsiveness** (user interactivity)

Using AJAX in client side enables us to obtain higher user interactivity in spite of HTTP limitations; actually, our user interface is simple because we concentrate on realizing the main approach ideas, but it can be complicated as needed. Using AJAX asynchronous communication between client (browser) and the web server is the main reason for this, now the user can use any control in the web page at any time without exposing to page freezing, all the requests will be managed in background by the AJAX engine. Another responsiveness problem can happen on the web server, which related to 100% CPU usage problem which cause the system to become unresponsive either intermittently or continuously, so the web server cannot handle any requests from the clients. The probability of this problem to happen is too high with Truong Chau et al. [26] approach especially with high frequency data polling and large number of clients. But in our approach, the rate of requests depends on data change rate, which means that in cases that there is no data change the web server will be more effective and responsive to new clients' requests because in this case the server CPU load will be a minimum and only the server RAM will be used to store the suspended connections parameters for short time until a data change takes place in which case all the used RAM will be released. Here is the client side AJAX instruction to achieve the desired responsiveness:

*Function ConstantCall()*
*{*
*MyWebService.GetUpdate(OnComplete);//Calling the webservice web method*
*}*

As shown, the client will call the server side Webservice and will assign *onComplete ()* function to wait for the response of this request. The JavaScript code for the *onComplete ()* will be as follows:

*Function onComplete(data) //data is the returned data*

*{*

*//code for parsing the received data*

*//code for updating the web page*

*.....*

*setTimout('ConstantCall',Delay); //make a new rquest again*

*}*

As shown, this function has two missions, first get the received new data, and second update the web page with this new data. This processing happens in background without the user knows and he still can send user data (set points) to the web server, which in turn will forward this data to the OPC server then to the PLC, this will happen using another communication channel for writing to OPC server. Actually, AJAX is designed to achieve higher user interactivity that was difficult to be achieved with traditional web technologies.

### 4.4.3 Real Time constraints

A real time system can be defined as a system that must satisfy explicit (bounded) response-time constraints or risk severe consequences, including failure. In our approach, as the data changes in the OPC (PLC) the user will get it immediately and without asking for, this is a type of real time behavior which achieved by the using of the threading mechanism in the Webservice web method in server side by which we enabled the Webservice web method to know if there is a change in the OPC DA server data, on the other hand there is no any processing happens to the data, just-in case of a change it is red from the OPC server then compacted into a string and sent to the client immediately. However, with Truong Chau et al. [26] the worst case is to miss the just changed data or get it after one Timer tick or in average half of the Timer tick. In the evaluation section, we will show the extent to which we achieved a better real time behavior compared to Truong Chau et al. [26] approach.

### 4.4.4 Efficiency

We can define efficiency as a measure of how much effort is required to achieve a required outcome, or making the best use of the available resources. Resource utilization is the percentage of time that a component is actually occupied, as compared with the total time that the component is available for use. For example, if a CPU processes transactions for a total of 40 seconds during a single minute, its utilization during that interval is 67 percent. A resource is said to be critical to performance when it becomes overused or when its utilization is disproportionate to that of other components. Our approach enables us to obtain higher efficiency because in the server side if there is no data change all processes will be suspended until there is a data change this will save the CPU time for important events. Also only the useful data will be sent to the client so consumption of the network bandwidth will be better. Also, the rate of clients' requests using our approach will be lower than Truong Chau et al. [26] approach and accordingly with our approach there will be no need to increase the capacity of the resources like CPU speed, memory or network bandwidth. Figure 4.6 shows how the client and server communicate together efficiently. As we see in Figure 4.6 that the client (browser) initiates the first request to the web server asking for the new data and waits for the response of this request before sending a second request. In addition, as we see that the web server (webservice) will send back the response only if there is a data change, if there is no data change the web server will hold the connection with the client, waiting for data change in the OPC DA server cache. If we can suppose that, the OPC data change rate is (Z) and the web server holding (waiting) time is (Tw) then the relation between the two values will be

$Tw = 1 / Z$

This means that the waiting time in the web server depends on the OPC data change rate or the control process data change rate. In the evaluation section, we will show the extent to which we achieved a better efficiency than Truong Chau et al. [26] approach by monitoring the web server CPU load, calculating the number of client requests and calculating the amount of data transferred through the network during a specified time interval. These calculations will be done for our approach and Truong Chau et al. [26] approach.

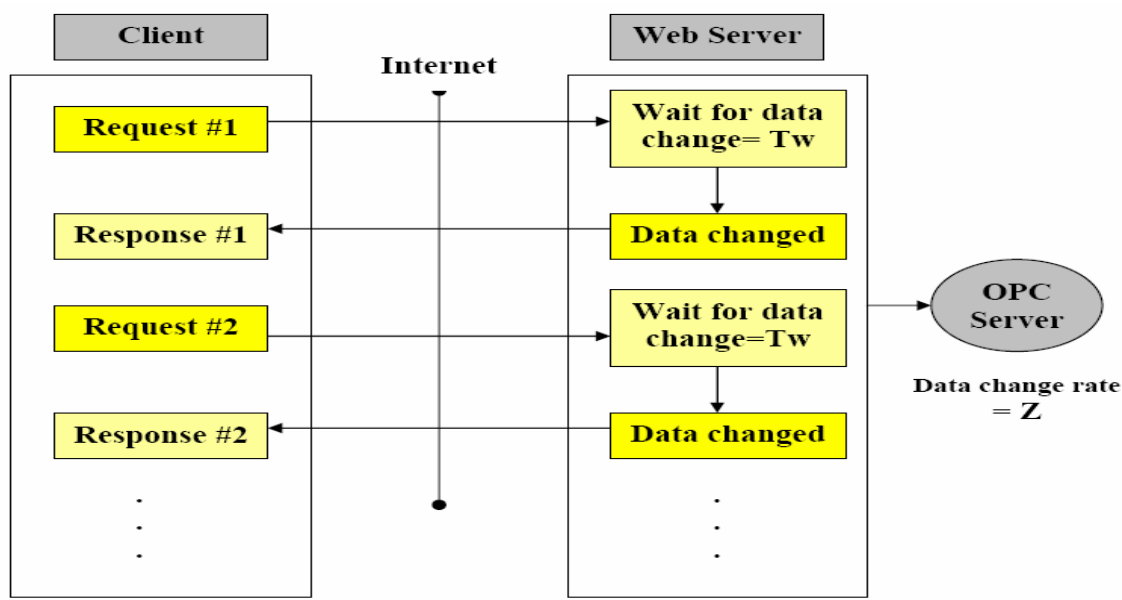

**Figure 4.6: Advanced data polling between client and server**

### 4.4.5 Performance

Most of software developers often use the terms "speed" and "performance" interchangeably. So response time to clients' requests can be considered as an indication to system performance. Web latency comes from several sources. Firstly, web servers can take a long time to process a request, especially if they are overloaded or have slow disks. Secondly, Web clients can add a delay if they do not quickly parse the retrieved data and display it for the user. Finally, the main contributor to Web latency, however, is network communication. Latency caused by client or server slowness - in principle - can be solved simply by buying a faster computer, or faster disks, or more memory. However, for network delays, some of them depend on network bandwidth so reducing this delay will be by buying a higher bandwidth link. However, much of the latency seen by web users comes from propagation delay that cannot be improved (past a certain point) no matter how much money we have.

For client side delays, instead of spending money, AJAX solved the problem. With AJAX the user will no longer wait for the response of his requests, the web page will not freeze and he will always feel that web page is responsive that is because all data retrieving and page updating will be carried out in the background by the AJAX engine. For server side delays, no need for spending more money to improve server performance because in our approach we take care about this. In our design, there will be processing only if there are data changes, if the data change rate is slow then the system scalability will be high and the system can handle more and more clients' requests. We think that with our approach we achieved a reasonable performance compared to Truong Chau et al. [26] approach because their client side requests rate will be very large especially if they use a high frequecny data polling. With our approach we bounded with the data changes on the server side. Also the unnessesary high number of clients' requests can cause network congestion in which case the response time to the client request will increase. In the evaluation section we will show that with our approach the inter-arival time between client requests to the web server will depend on the OPC DA server data change rate which will put the web server in a light load compared with Truong Chau et al.[26] approach which uses a fixed rate client requests that will put the web server always under high load handling high frequency client requests rate.

### 4.4.6 Security

Security is a very important issue in the web based SCADA systems. In the past SCADA systems were isolated in the plant so the security was by "obscurity", but today's SCADA which is in most cases connected to enterprise networks and the Internet is vulnerable to cyber attacks and penetration by unauthorized persons or terrorists. In the next chapter (5), we will cover the security in details and will include a design of a security policy.

### 4.4.7 OPC DA protocol Constraints

To enable the modern .NET Webservice to communicate with the legacy OPC DA COM technology we need a software wrapper to wrap the COM technology to a .NET interface. The OPC Foundation developed many wrappers for this purposes and we will use it in our system. For the DCOM security problems and since the callback mechanism used by OPC essentially turns the OPC Client into a DCOM Server and the OPC Server into a DCOM Client, the instructions provided by OPC Foundation in [13] must be followed on all nodes that contain either OPC Servers or OPC Clients to enable communication between them.

### 4.4.8 LANs local (virtual) IP numbers

We have to find a way to access the local plant web server, which contains the designed SCADA web site. The router itself can solve this problem. Most modern ADSL routers have a utility called NAT (Network Address Protocol) which enables us to configure the router as a virtual server to direct any external traffic to port 80 (for HTTP) to the local plant web server machine. As shown in Figure 4.7.

### 4.4.9 ADSL router dynamic real IP number

One solution to solve the ADSL router dynamic IP address problem is to ask the ISP to deliver a static real IP address, which will be somewhat expensive compared to the dynamic one. Then we will be sure that the router IP address will not change dynamically. Another solution is to use a Dynamic DNS (DDNS) and redirection service to map a dynamic IP address to an easy to remember sub-domain. One of the DDNS service providers is "No-IP" company (www.no-ip.com) which enables us to use text format web site name instead of IP addresses. We will need to download and install a small client program called

DUC (Dynamic Update Client) on the web server machine. This client program will periodically check for router IP changes. In addition, we will need to choose the domain name i.e. (QenaPaperCo.no-ip.org), and then if there is an IP change the DUC will send the new IP address to the DDNS (Dynamic Domain Name Server) to be mapped with the specified domain name. This process will be transparent to us.

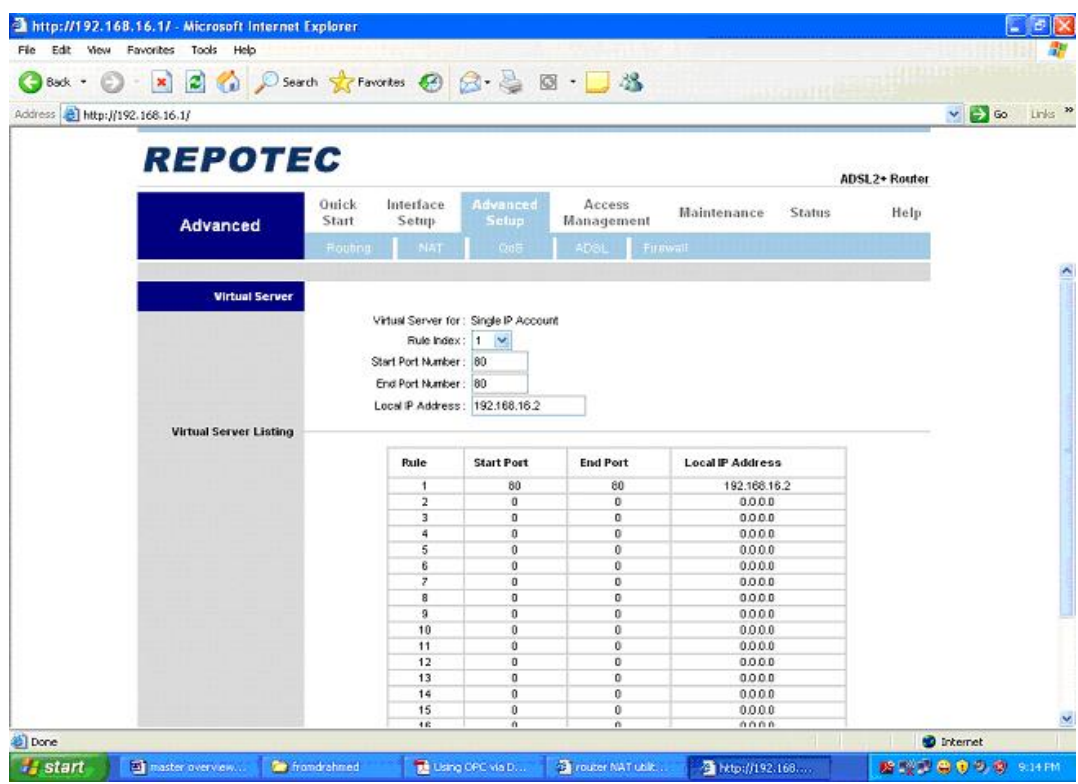

**Figure 4.7: Router NAT utility**

## 4.5 Evaluation

To evaluate our approach we have used two different approaches. First, we used simulation to compare our methodology to Truong Chau et al. [26]. As an example of the approaches that used client side fixed data polling, then we applied our new methodology to a case study, which was implemented in Qena Paper mill.

### 4.5.1 Simulation

Any control process has a number of measurable physical quantities, which are measured by different sensors, and then the raw signals transferred to the PLC to be converted by the A/D converters to digital form to be manipulated by the PLC processing unit. The OPC DA server updates its cache by the new values available in the PLC according to a specified update rate. So a change in the control process quantities reflected in a change in the OPC DA server cache.

Hence, to simulate a typical control process we will create a software application, which connects to the OPC DA server, and continuously change its cache data. To do the continuous change we will use a Timer control and we will configure its Interval as needed so the rate at which the software simulation changes the OPC DA server data can be configured to be fixed rate or variable (random) rate. The simulation application created using VB.NET 2008 and designed to offer many options such as the data change rule (Fixed or Random), the number of data changes (Limited or Unlimited), and the range for Timer Interval in case of random data change can be specified, and the changed data range can be specified too. In addition, there is a facility to start and stop data changing at any time to check the behaviors of our proposed approach and other approaches with/without data change. The exact time of each data change can be stored in a database for analysis purposes. For simplicity the number of items, which we will use in our simulation application, is six items and they will simulate any of the control process variables. In the case study section, we will replace the simulation application with a real control process, which is the Winder station in Qena Paper mill. The complete implementation of the simulation software application can be found in appendix E. The simulation application in run mode will be similar to Figure 4.9.

Now to show the extent to which we achieved our goals we will compare our approach results to Truong Chau et al. [26] approach results, we will use windows Task Manager - shown in Figure 4.10 - to monitor the web server's performance measures and network bandwidth utilization. As shown in Figure 4.8, we designed a networked system to evaluate our approach using the simulation software instead of a real control process. As shown the OPC DA server, which belongs to Siemens, and a web server, which is IIS6.0, which belongs to Microsoft, are connected in the same LAN, this LAN connected to the Internet using an ADSL router (REOPOTEC). The client machine -from which we will run the web browser to access the web server which contains the VB.NET webservice which in turn will connect to the OPC DA server to get or set a new data- will be connected to the internet through another router in another place. The simulation software will be installed and run on the OPC DA server machine to start changing the OPC DA server cache data locally according to the specified Timer Interval which maybe fixed or randomly chosen. In the case study section, we will connect a real PLC to the LAN instead of the simulation software to test our approach with a real control process. Applying Truong Chau et al. [26] approach and using simulated OPC data, which as we mentioned above is six items, this data will change according to the Timer Interval, which will be fixed with 1 second, and we got the CPU usage shown in Figure 4.11. As we see, even if there is no data change –which is indicated by the ovals where we stopped the data change in our simulation software application- the server still working approximately with the same CPU usage rate.

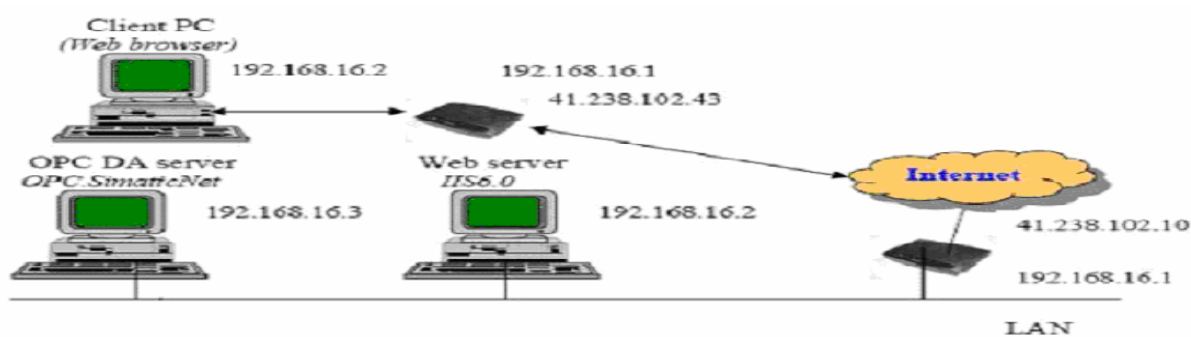

**Figure 4.8: Designed Simulation System**

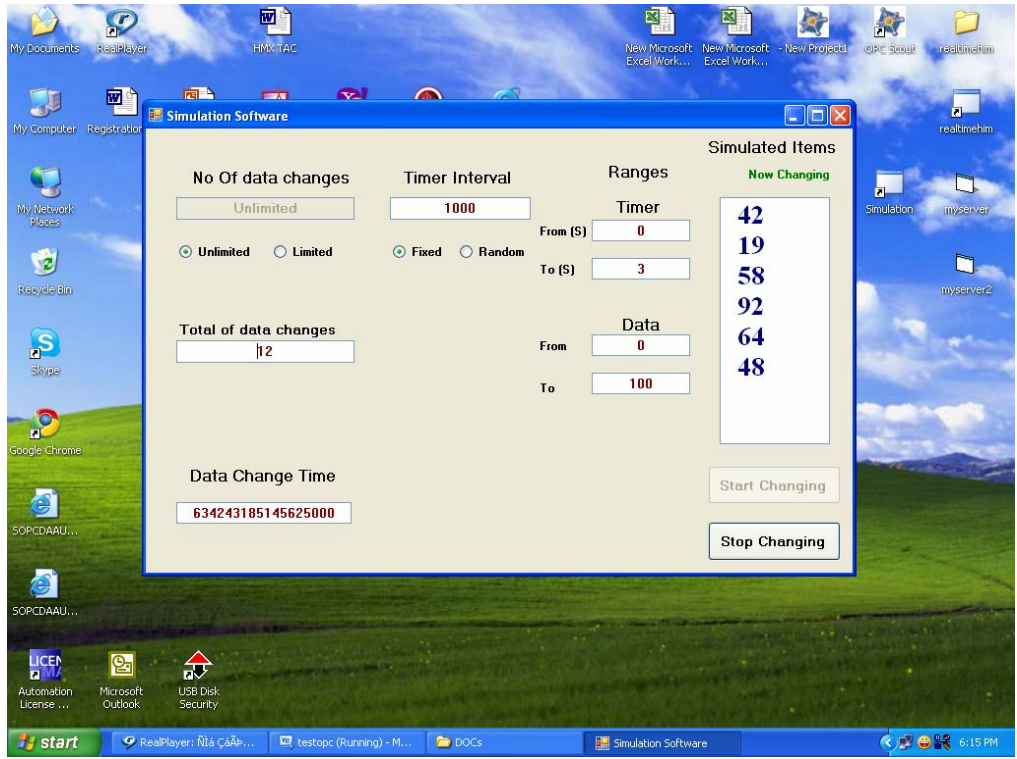

**Figure 4.9: Simulation application**

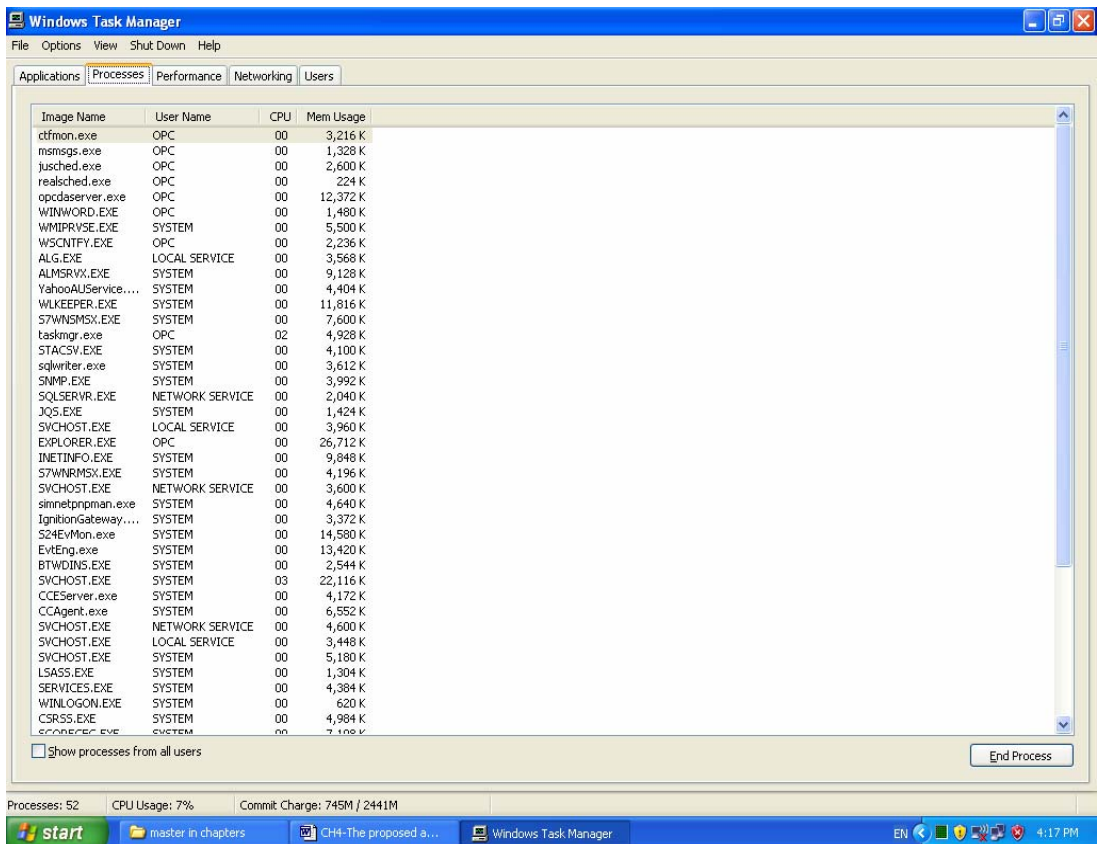

**Figure 4.10: Windows XP Task Manager**

However, with our approach the CPU usage is approximately zero when there is no data change as shown in Figure 4.12. Also the data transferred through network with Truong Chau et al. [26] approach is shown in Figure 4.13, as shown his approach transfers data continuously through network (suppose his Timer interval is 1 second) regardless if the data changed or not. However, with our approach the data transferred only if there is a data change as indicated by the ovals in Figure 4.14 where we stopped the data change in our simulation.

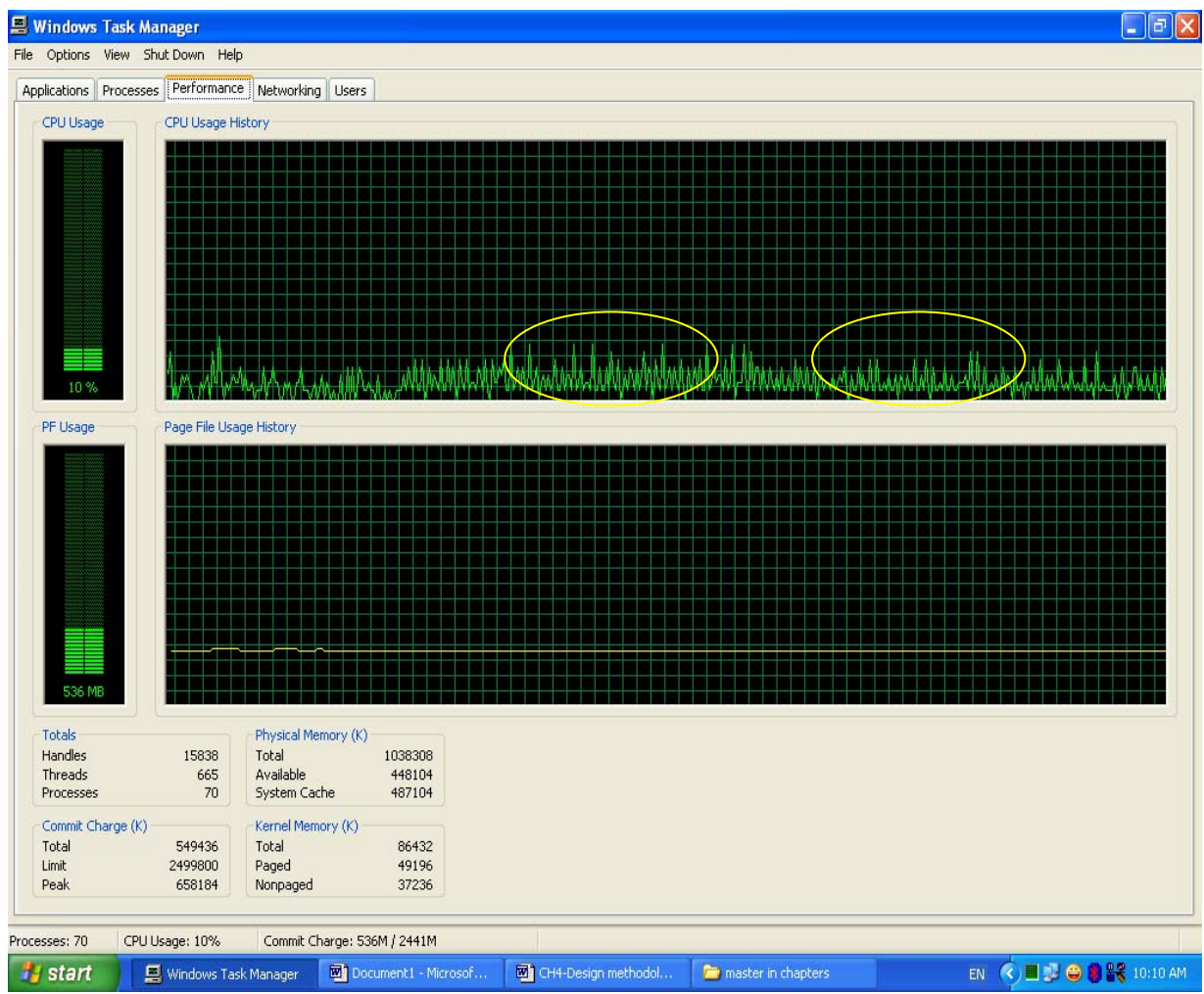

**Figure 4.11: Truong Chau et al. [26] test results (CPU load) with Timer interval=1 seconds**

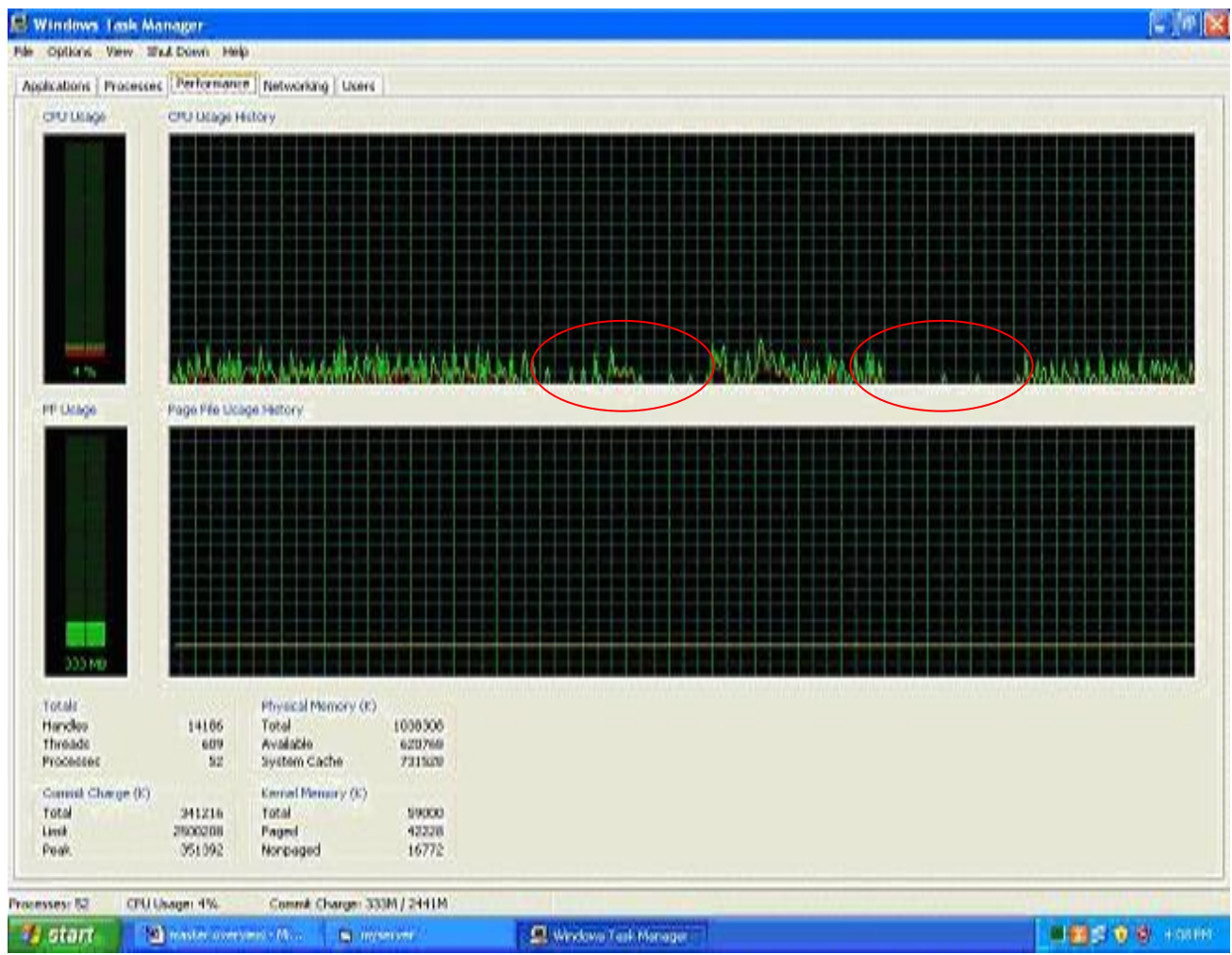

**Figure 4.12: Our approach test results (CPU Load).**

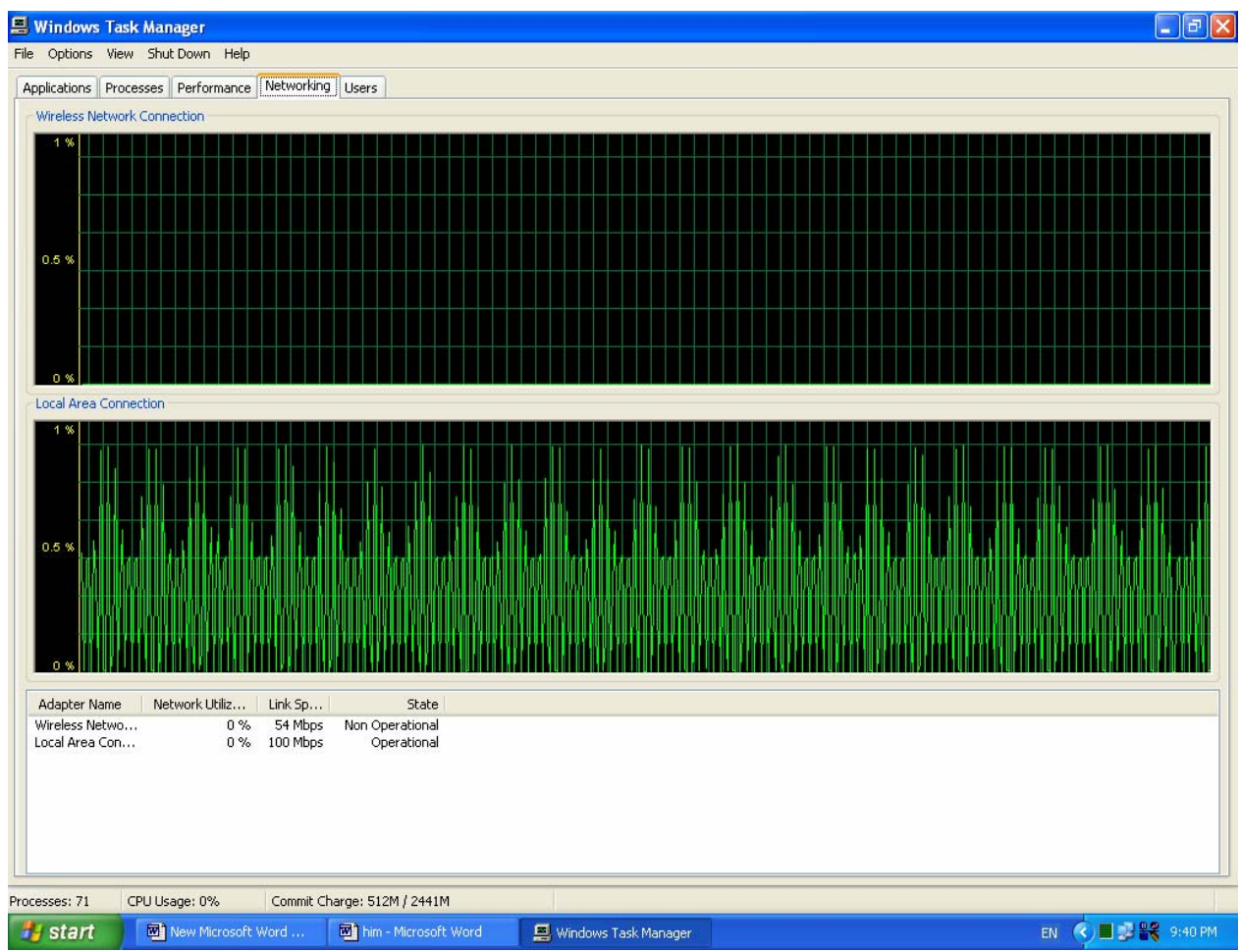

**Figure 4.13: Truong Chau et al. [26] results (Network traffic) with Timer interval=1 seconds**

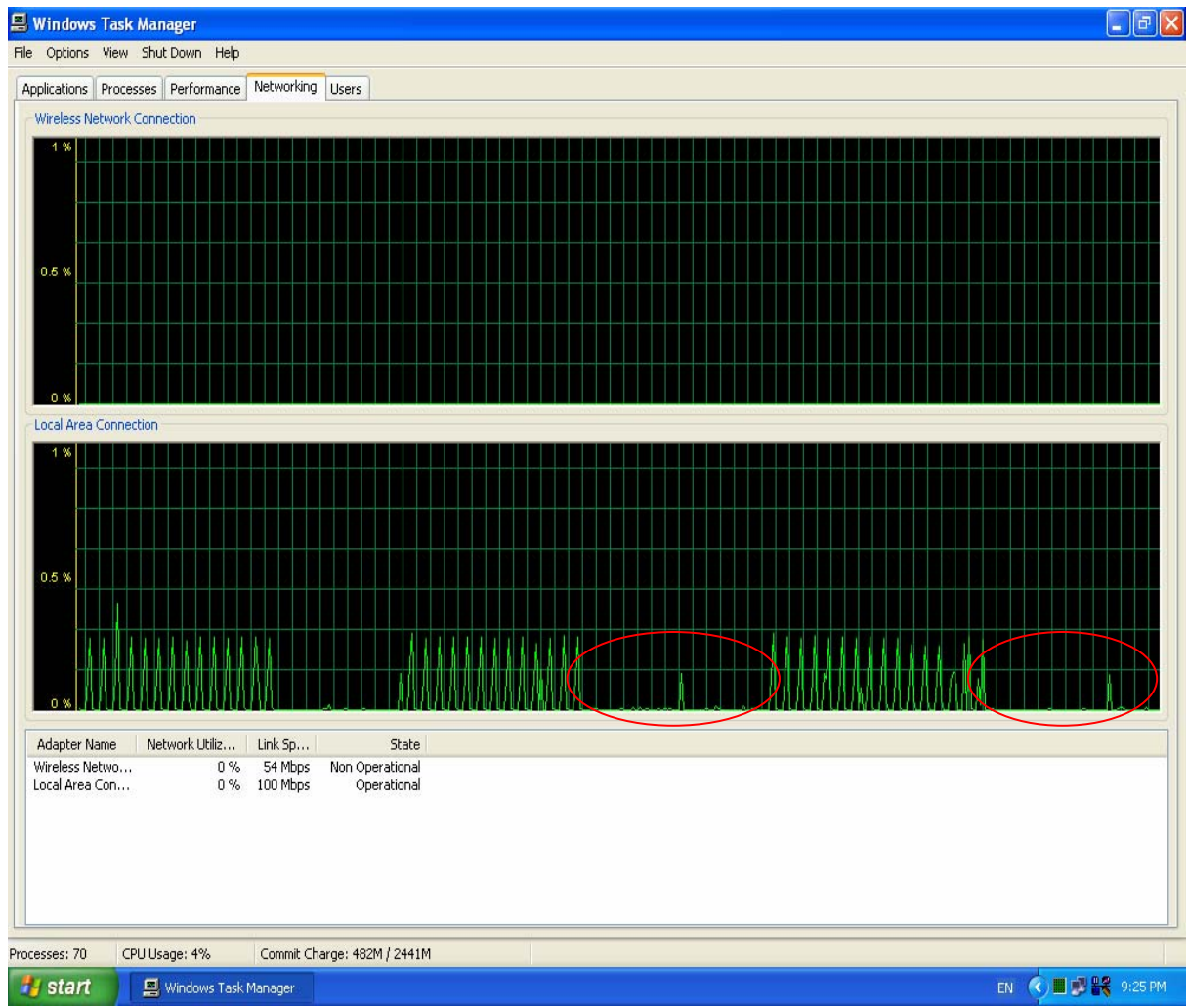

**Figure 4.14: Our approach test results (Network traffic).**

Another situation in which our approach wins, if we configure the OPC data change rate to be fixed with one change per second and applying Truong Chau et al. [26] approach with Timer interval 5 seconds then the network traffic will be as shown in Figure 4.15. However, with our approach the Network traffic is shown in Figure 4.16.

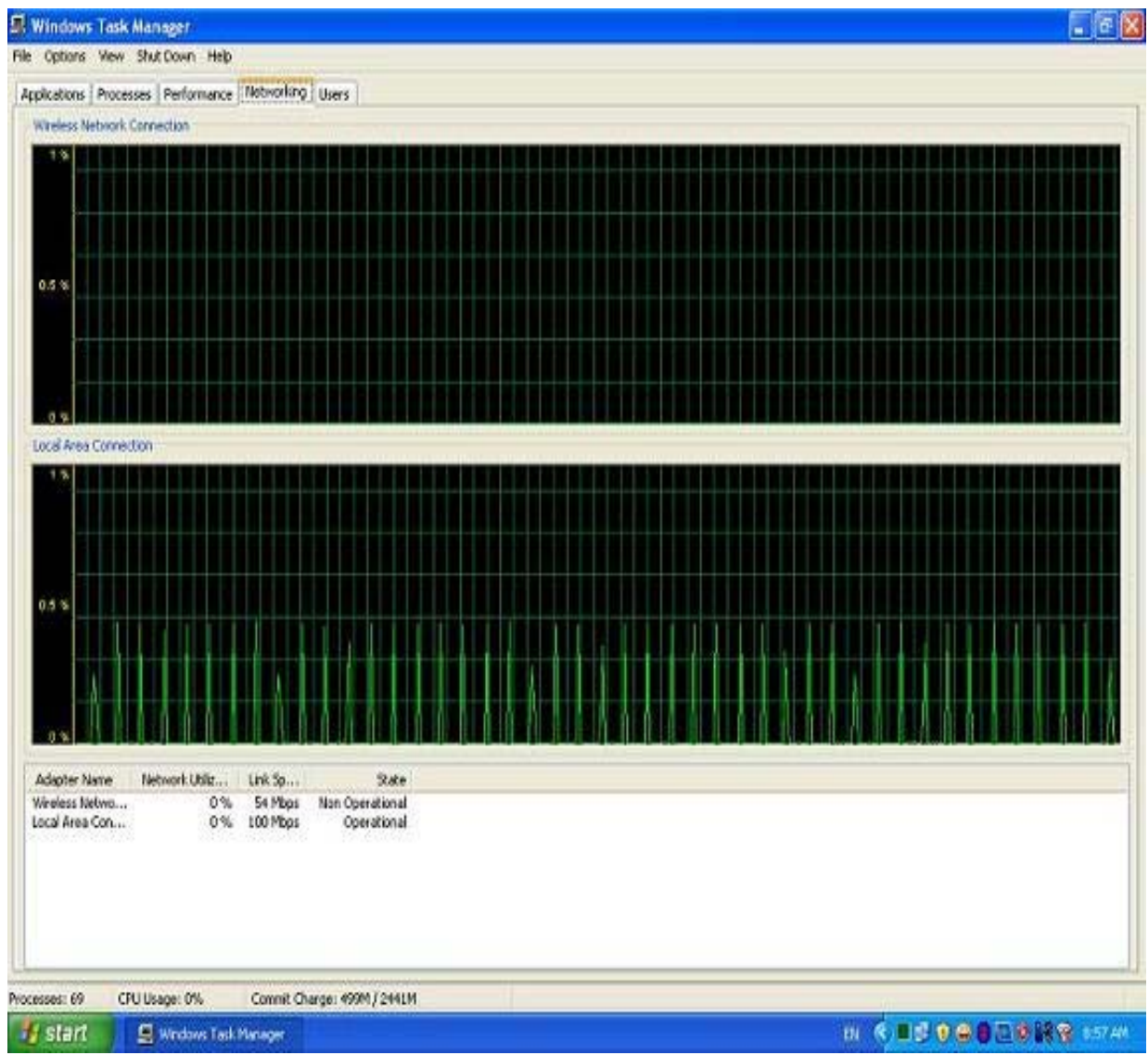

**Figure 4.15: Truong et al. [26] test results Network traffic with Timer interval=5 seconds**

As shown in Figure 4.15 that Truong Chau et al. [26] approach gets data on intervals of 5 seconds only, even there are about four data change events between Timer ticks, which are not visible to their approach, but with our approach, we could observe and get most of the data changes events as shown in Figure 4.16.

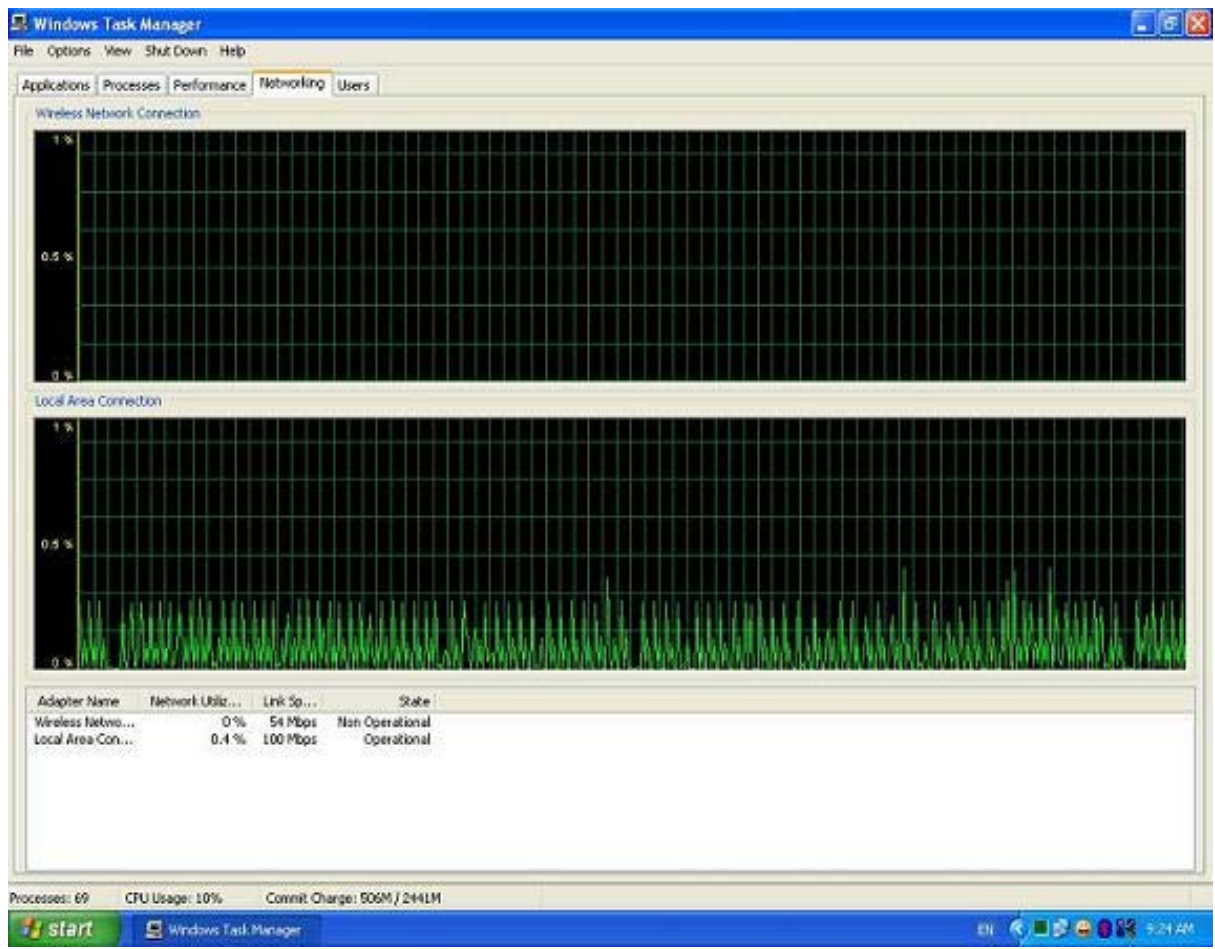

**Figure 4.16: Our approach test results (Network traffic)**

## Conclusions

From the above experimental results with the simulation application, we conclude that with our approach the web server CPU usage and the network bandwidth utilization are better than with Truong Chau et al. [26] approach. That is achieved by dividing the code between the client and the server and the magic threading mechanism used in server side Webservice that enabled us to add a time delay, which is proportional to the data change rate of the OPC DA server data (process data). As shown in the previous section, there will be processing in the web server only if there is a data change events, this is in addition to time needed to handle the clients requests, which will be given new Webservice instances then suspended until a data change event happens. In addition, the utilization of network bandwidth will be more efficient with our approach because only new and effective data will be transferred to the client, also if there is no new data the network will not be used for the current client.

Another indirect benefit of our approach is that as we depend on data change events to make a new request then we can calculate the frequency of requests per second, which is proportional to the number of data change events per second, this frequency maybe used as an indication of the control process stability. In Truong Chau et al. [26] work this frequency is fixed according to the Timer interval. Also in our approach we do not miss any information (data change) but in [26] they may miss some important information if the data change rate is faster than the request rate.

**Numerical Calculations**

Now we can show numerically the extent to which we achieved our goals by calculating the following values:

1. Number of client requests during a certain period.
2. Effective process data transferred through the network.
3. The delay for a just changed process value to be delivered to client side (The response time)

First, we will make a relation between the client requests number and the number of data changes in the OPC DA cache. We will make the calculations during one minute (60 seconds). Using our simulation application, we will change the rate of data change to be no changes at all, 1 change/ 4 seconds, 1 Change/ 2 Seconds, 1 Change/Second. Applying Truong Chau et al. [26] and our approach we got the following numerical data in Table 4.1 and this data is graphed in Figure 4.17.

**Table 4.1: No. of requests with data change rates**

| Number of Actual data changes /Minute | Number of client requests/Minute | |
|---|---|---|
| | *Truong Chau et. Al.* with *T=1 Second* | *Our approach* |
| 0 | 60 | 0 |
| 15 | 60 | 15 |
| 30 | 60 | 30 |
| 60 | 60 | 60 |

As shown in Table 4.1 that with Truong Chau et al. [26] approach the number of requests is constant for all data change rates and it depends on the Timer interval, which is 1 second in this case, But with our approach it depends on the data change rates.

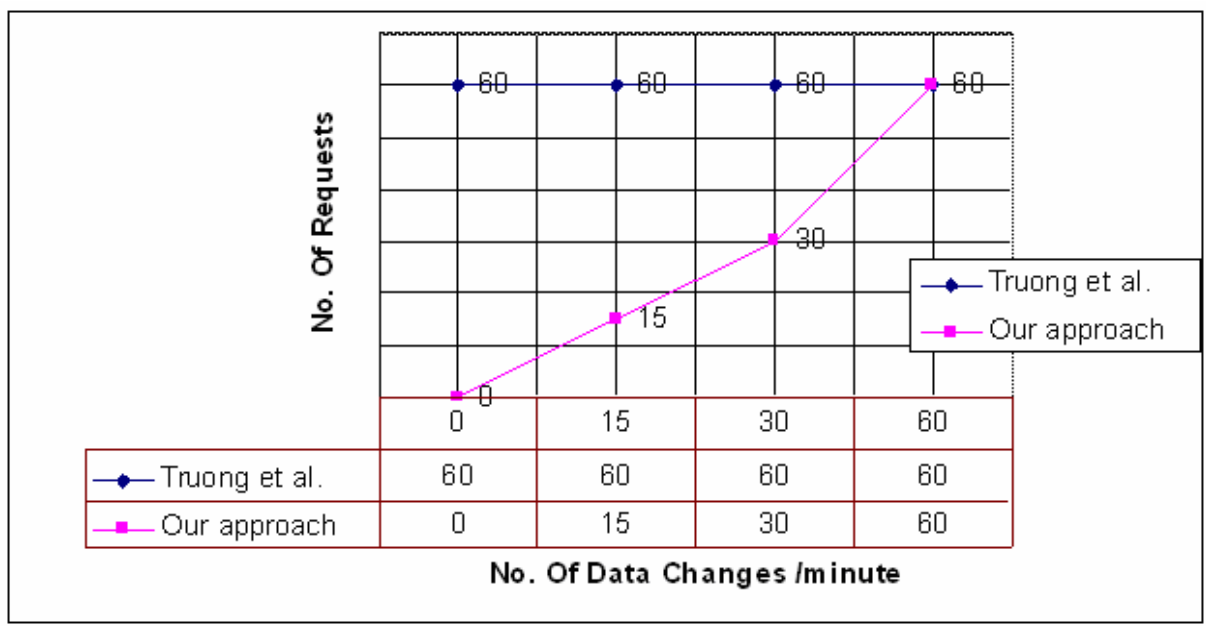

**Figure 4.17: No. of requests with data change rates plot**

By the same way, we can find the amount of data transferred through the network with different data changes rates as shown in Table 4.2 and it is graphed in Figure 4.18. For simplicity, we put the number of variables to be only 10 variables and ignored any extra TCP/IP packet overhead. As shown in Table 4.2 that the amount of data in bytes transferred with Truong Chau et al. [26] is constant with all data change rates, but with our approach it depends on the number of data changes in the OPC data so it is more efficient and effective.

**Table 4.2: Amount of transferred data with data change rates**

| Number of Actual data changes /Minute | Transferred data in Bytes/Minute | |
|---|---|---|
| | Truong Chau et. Al. with T=1 Second | Our approach |
| 0 | 600 | 0 |
| 15 | 600 | 150 |
| 30 | 600 | 300 |
| 60 | 600 | 600 |

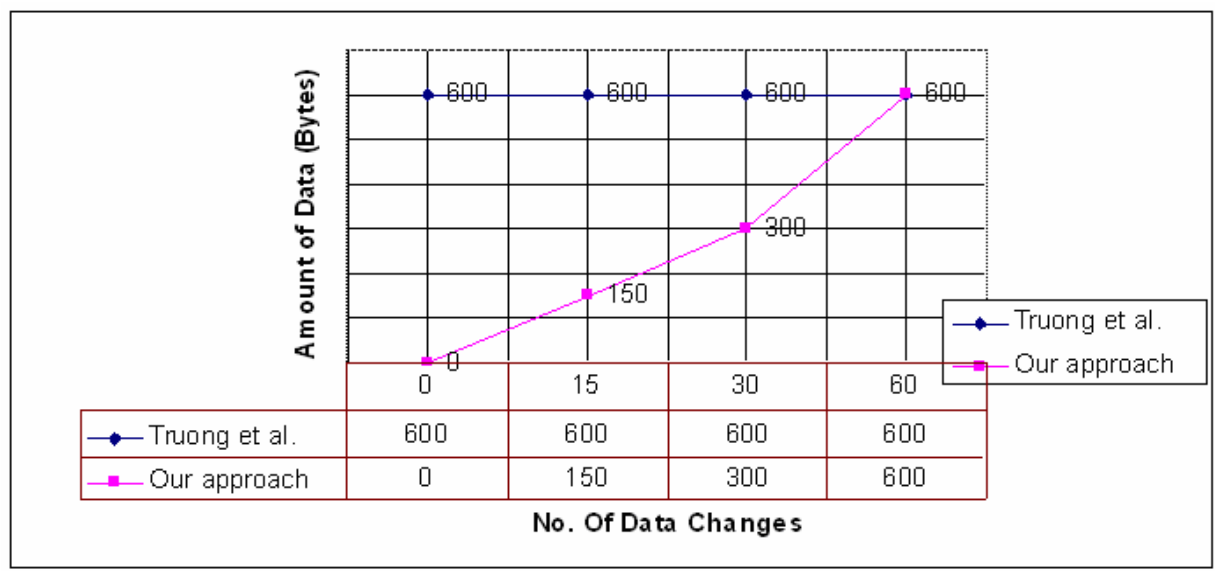

**Figure 4.18: Amount of transferred data with data change rates plot**

Another way to show the difference between our approach and Truong Chau et al. [26] approach is calculating the distribution of inter-arrival time of client requests to the web server. We forced our simulator to change OPC data in random fusion between 1 to 6 seconds and we put the Timer interval for Truong Chau et al [26] approach to 1 second then we got the following results in Table 4.3, which is graphed in Figure 4.19.

**Table 4.3: Client requests inter-arrival time to web server**

| Request number | 1 | 2 | 3 | 4 | 5 | 6 | 7 | 8 | 9 | 10 |
|---|---|---|---|---|---|---|---|---|---|---|
| Delta T for our approach (Seconds) | 2 | 4 | 2 | 5 | 6 | 3 | 1 | 4 | 4 | 5 |
| Delta T for Truong approach (Seconds) | 1 | 1 | 1 | 1 | 1 | 1 | 1 | 1 | 1 | 1 |

As shown with Truong Chau et al. [26] approach the requests inter-arrival rate is constant (one seconds), But with our approach it depends on the OPC data change rate.

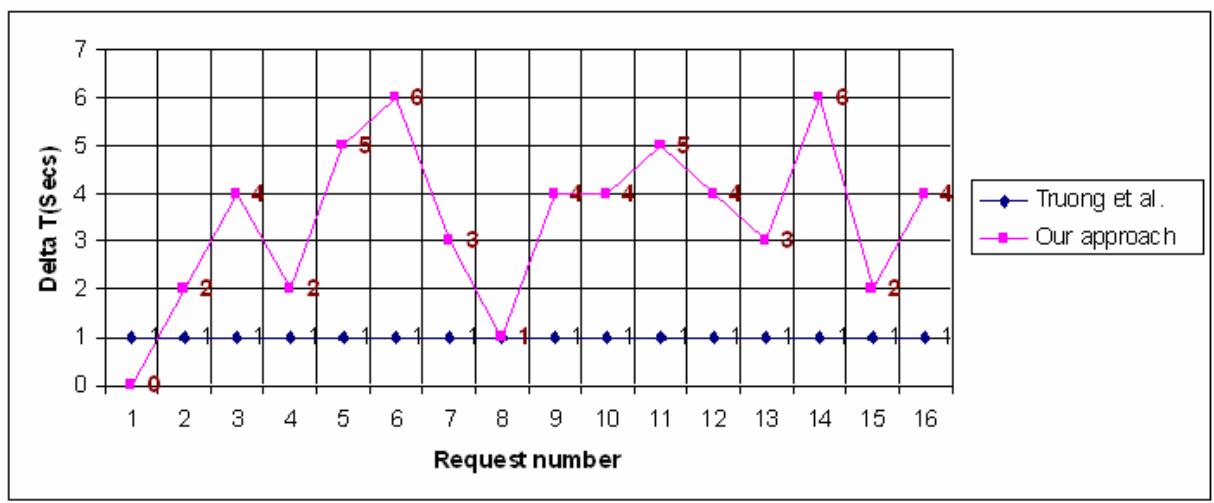

**Figure 4.19: Client requests inter-arrival time to web server plot**

As we mentioned in section 4.4.3 that we achieved a better real time behavior than Truong Chau et al. [26], now we will prove that practically using the simulation software. In our application, the response time can be defined as the time delay for the just changed group of data in the OPC DA server to be delivered to the client (web browser). We will calculate this response time using

Truong Chau et al. [26] approach and using our approach, the results of this experiment are shown in Table 4.4 and Table 4.5 respectively. As shown that with Truong Chau et al. [26] approach the average response time equals 567.17 ms (approximately 0.5 second). But using our approach the average response time is 40.6 ms. Not only this, but also with Truong Chau et al. [26] the number of total client requests which were done to get these 10 data changes is 22 requests which means that there are 12 unuseful client requests. On the other hand, with our approach the number of client requests is equal to the number of data changes, which are 10 in this experiment.

**Table 4.4: Response Time for Truong Chau et al. [26] approach**

| Data Change No. | 1 | 2 | 3 | 4 | 5 | 6 | 7 | 8 | 9 | 10 |
|---|---|---|---|---|---|---|---|---|---|---|
| Response Time (ms) | 265.6 | 609.39 | 203.09 | 906.30 | 328.10 | 718.70 | 328.10 | 1062.5 | 765.59 | 484.39 |

**Table 4.5: Response Time for our approach**

| Data Change No. | 1 | 2 | 3 | 4 | 5 | 6 | 7 | 8 | 9 | 10 |
|---|---|---|---|---|---|---|---|---|---|---|
| Response Time (ms) | 31 | 47 | 47 | 31 | 47 | 47 | 47 | 31 | 31 | 47 |

**Conclusions**

As shown that with our approach for any client the number of requests sent depends on the time delay we put in the web server, this time delay is inversely proportional to the data change rate in the OPC server data. Also only effective data will be red and sent to the client, by effective data we mean that the data is new and different from the data the client got from the previous request. But with Truong Chau et al. [26] approach the number of client requests is fixed according to the client side Timer interval without any attention to the OPC data change rate and according to that the probability that the new data it gets from the web server is the same as the data it already has is high. Also we succeeded to achieve a better real time behavior than Truong Chau et al. [26] because our response time to a data change is in average 40.6 ms, but his response time for a data change is in average .5 second if his Timer Interval is 1 second as discussed earlier.

**4.5.2 Case Study**

As a practical project to apply our approach we selected Qena Paper mill which is the biggest paper mill in Egypt and maybe in Middle East and we selected the Winder station to monitor and control through web using our approach. Winder station takes a paper spool that is about 6 m width as it's input and transfers is to small width rolls as shown in Figure 4.20. This is a complex control process and there are many process variables, which should be monitored continuously for example:

- Speed, which is the speed in (m/min) of both the unwind station, drum1 and drum2.

- Tension, which is the paper tension in (N/m)

- Unwind diameter, which is the input jumbo roll diameter in (mm) of the winder station

- <u>Rewind diameter</u>, which is the output rolls diameter in (mm) of the winder station

- <u>Line force,</u> which is the line force in (N/m), applied on the rewind rolls by drum1, drum2 and rider roll.

As shown in the original winder station desktop application shown in Figure 4.21. To make "webtization" for this project we will apply the methodology, which we have suggested earlier. As shown in Figure 4.22, our gateway to the plant floor will be a web server containing both the ASP.NET AJAX enabled web application and the VB.NET webservice, this web server connected in the same LAN with the OPC DA server and the PLC. We assigned an IP address for each machine for example the PLC unit IP address is 192.168.16.4, the OPC DA server machine IP is 192.168.16.3 and the web server machine IP address is 192.168.16.2. The local LAN is connected to the Internet using a router (REPOTEC in our case) which has two IP addresses, a local one, which is 192.168.16.1, and an international (real) one, which in our case study is 41.238.102.10.

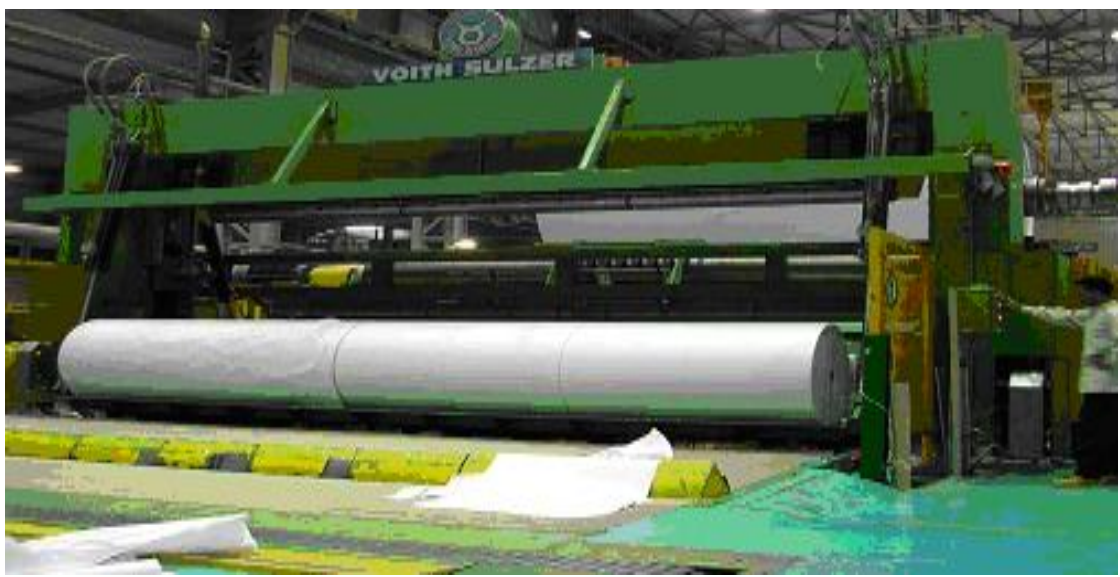

**Figure 4.20: Qena Paper Winder station**

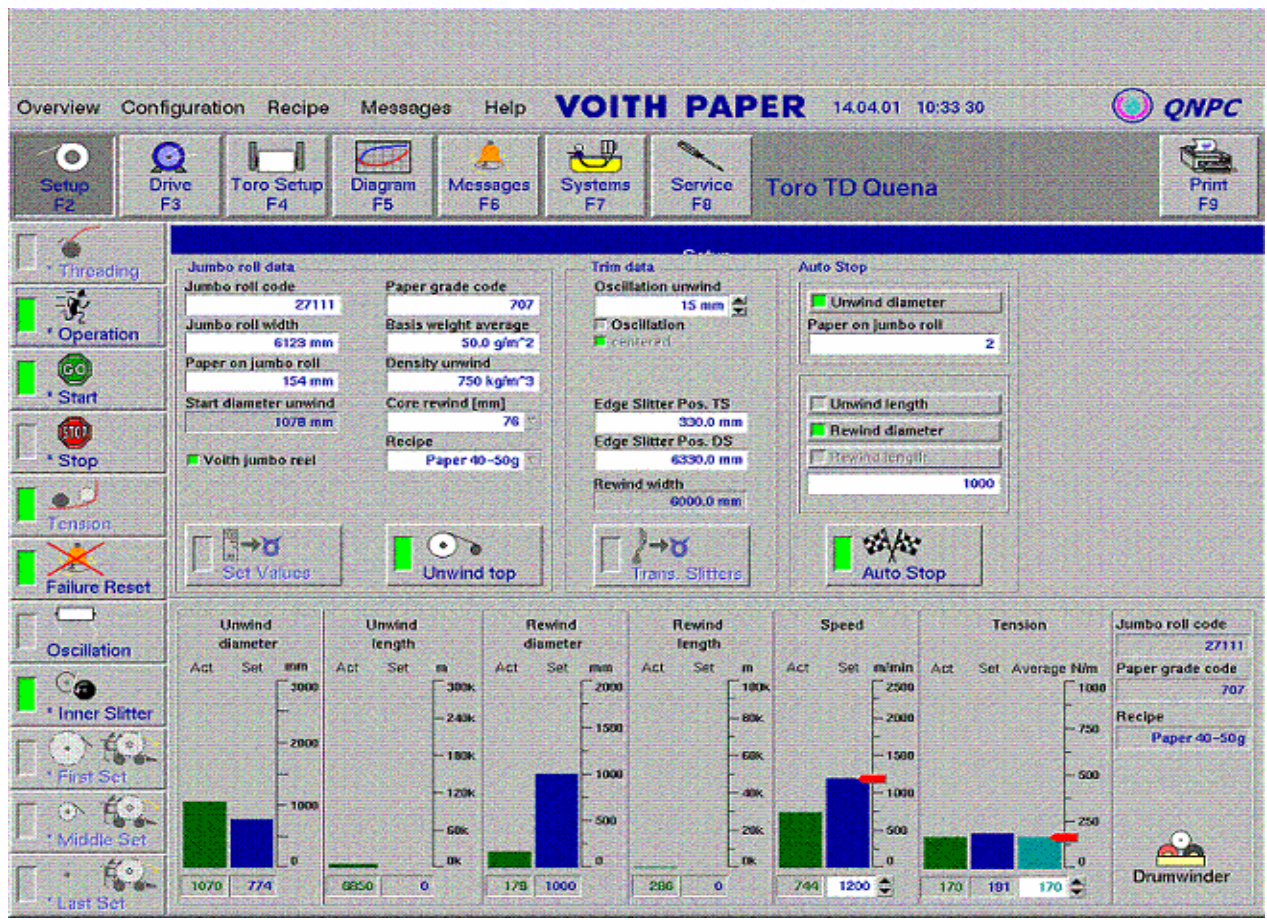

**Figure 4.21: Desktop application for winder station**

Actually, it is preferred to have a static international IP address for the router instead of dynamic IP addresses which are the common case these days, to guarantee that the IP will not change even if the router is powered off and turned on again. As shown in Figure 4.22 the OPC DA server, which we will use, is "OPC.SimaticNet" which belongs to Siemens Company, Also the PLC is Siemens S7-400. The web server we used is Microsoft IIS6.0, which will contain the ASP.NET AJAX, enabled web application that designed for this control process in addition to the VB.NET webservice. As we mentioned that nowadays, most of internet connections use an ADSL router to establish a connection to an ISP to get a dynamic IP number. A dynamic IP address means that each time the router powered on it will get a different randomly selected IP address; this will be the international IP address for the router (see section 4.4.9).

In addition to this IP address, the router has another internal or local IP address to transfer the Internet traffic to the local LAN. The web server installed on the local LAN as shown in Figure 4.22 has the IP address "192.168.16.2" this IP is not visible from the Internet therefore to access this web server from Internet we will use the router NAT utility (see section 4.4.8).

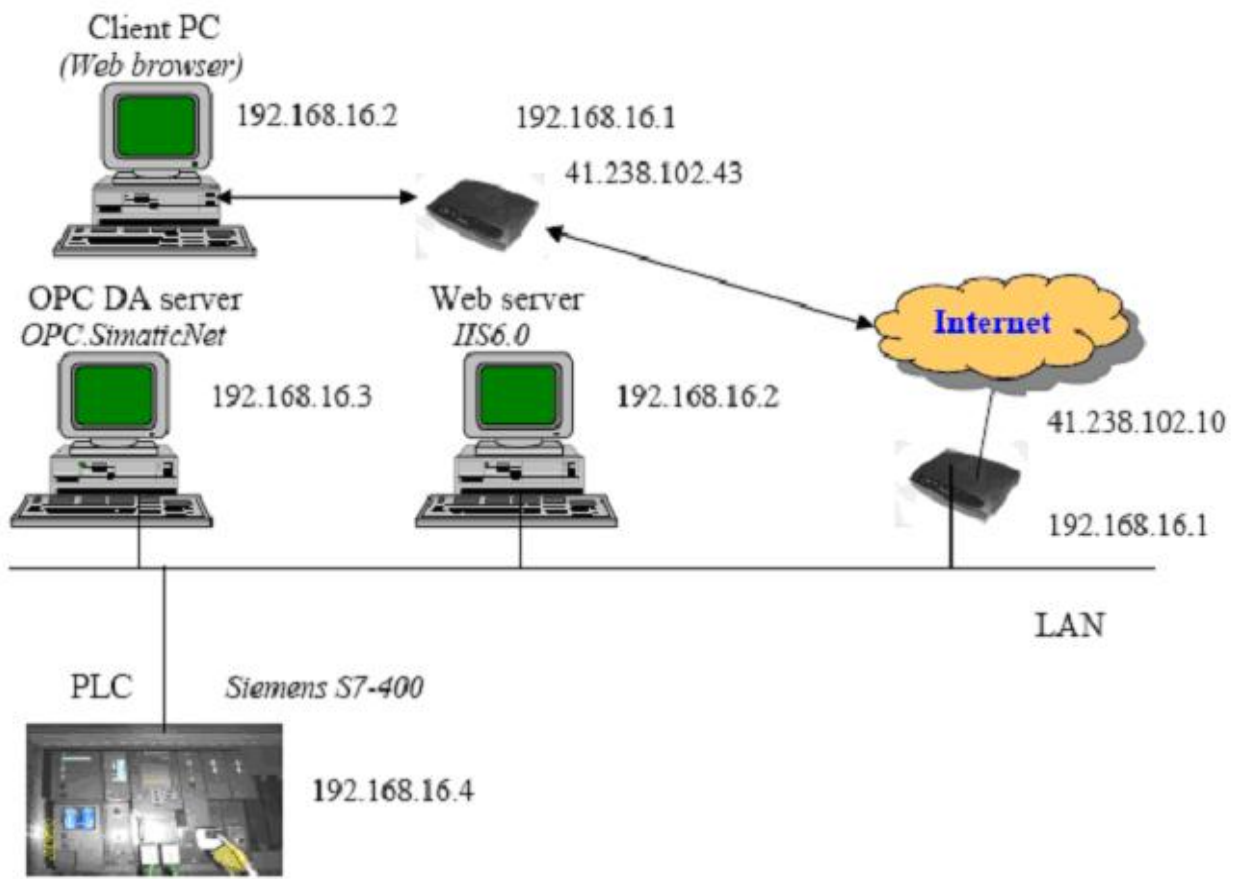

**Figure 4.22: Case Study Overview**

Finally, the implemented user interface is shown in Figure 4.23. In addition, the CPU load and Network traffic are shown in Figure 4.24 and Figure 4.25 respectively.

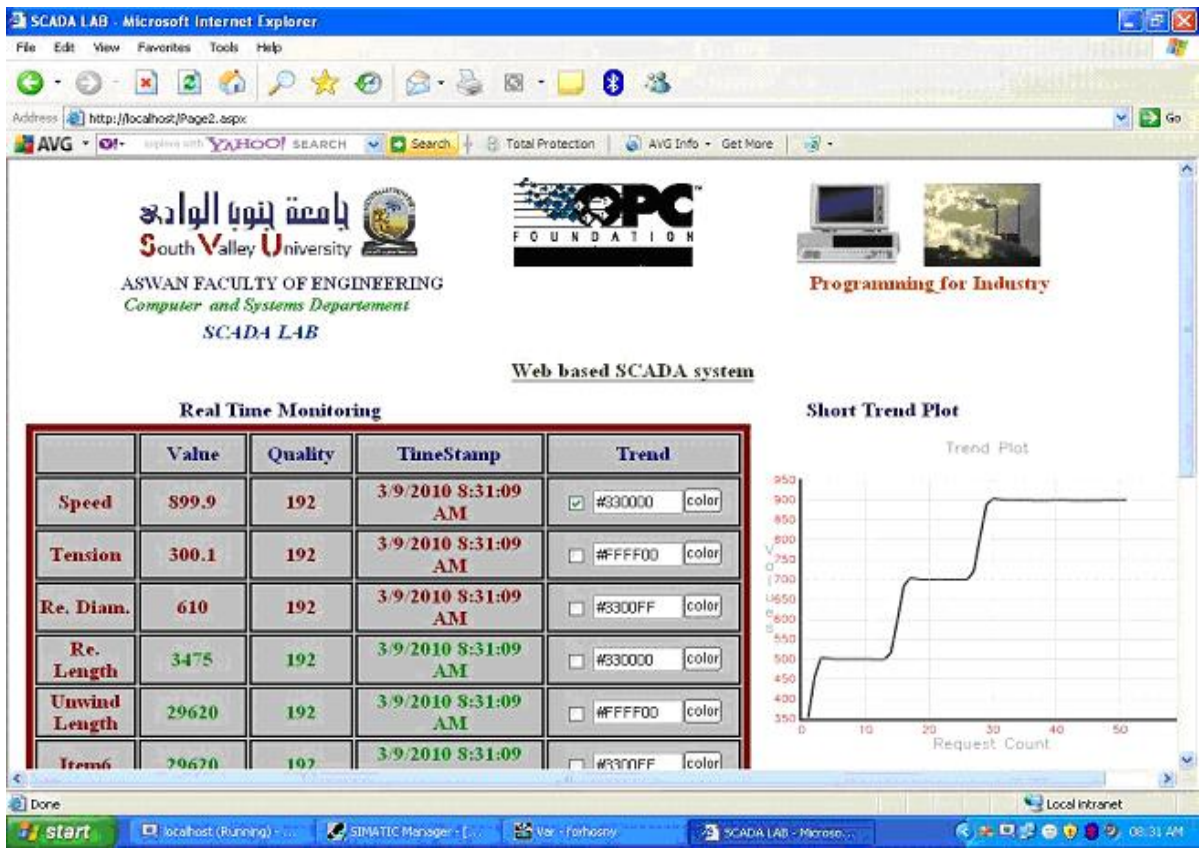

**Figure 4.23: Case Study Presentation**

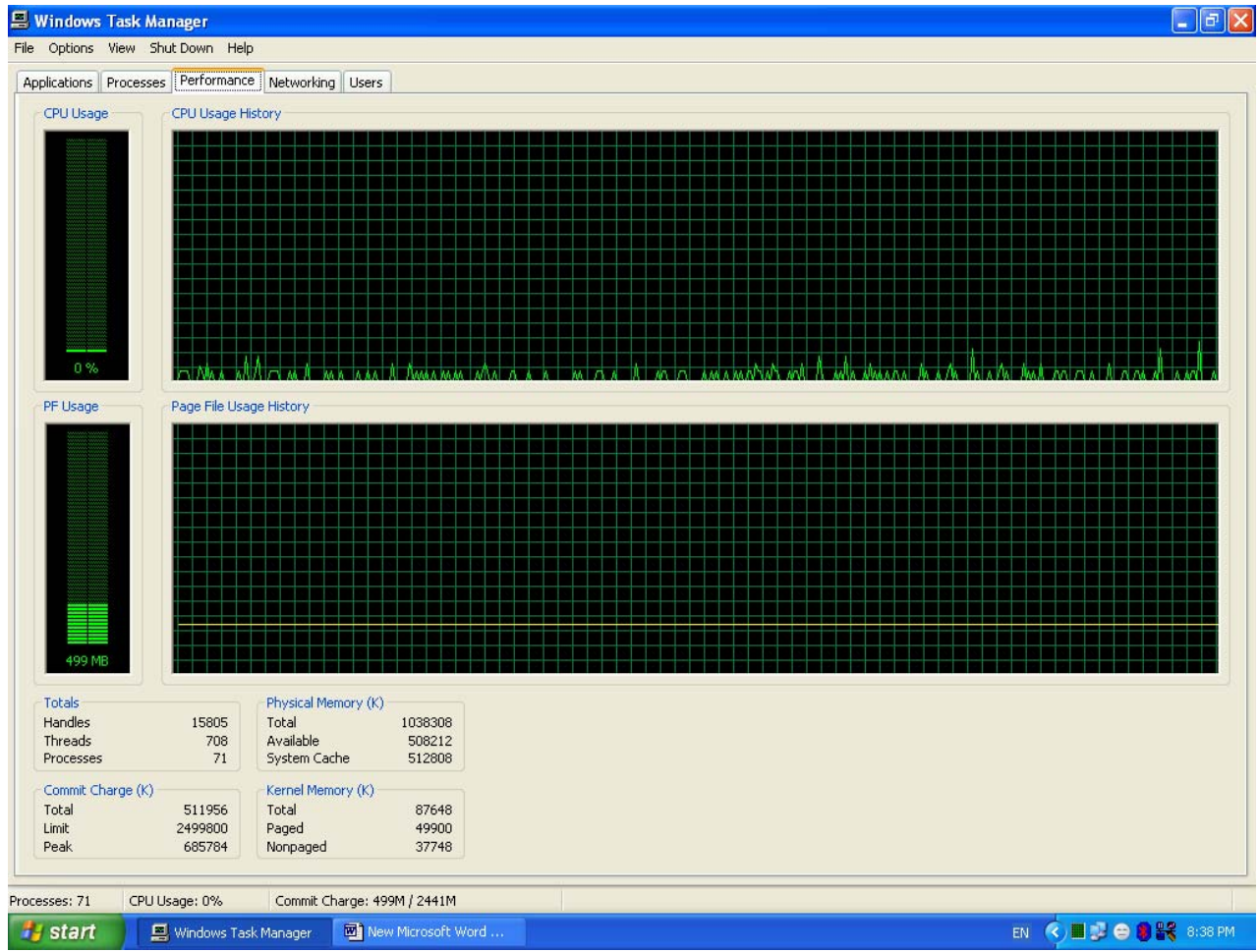

**Figure 4.24: Case Study test (CPU load).**

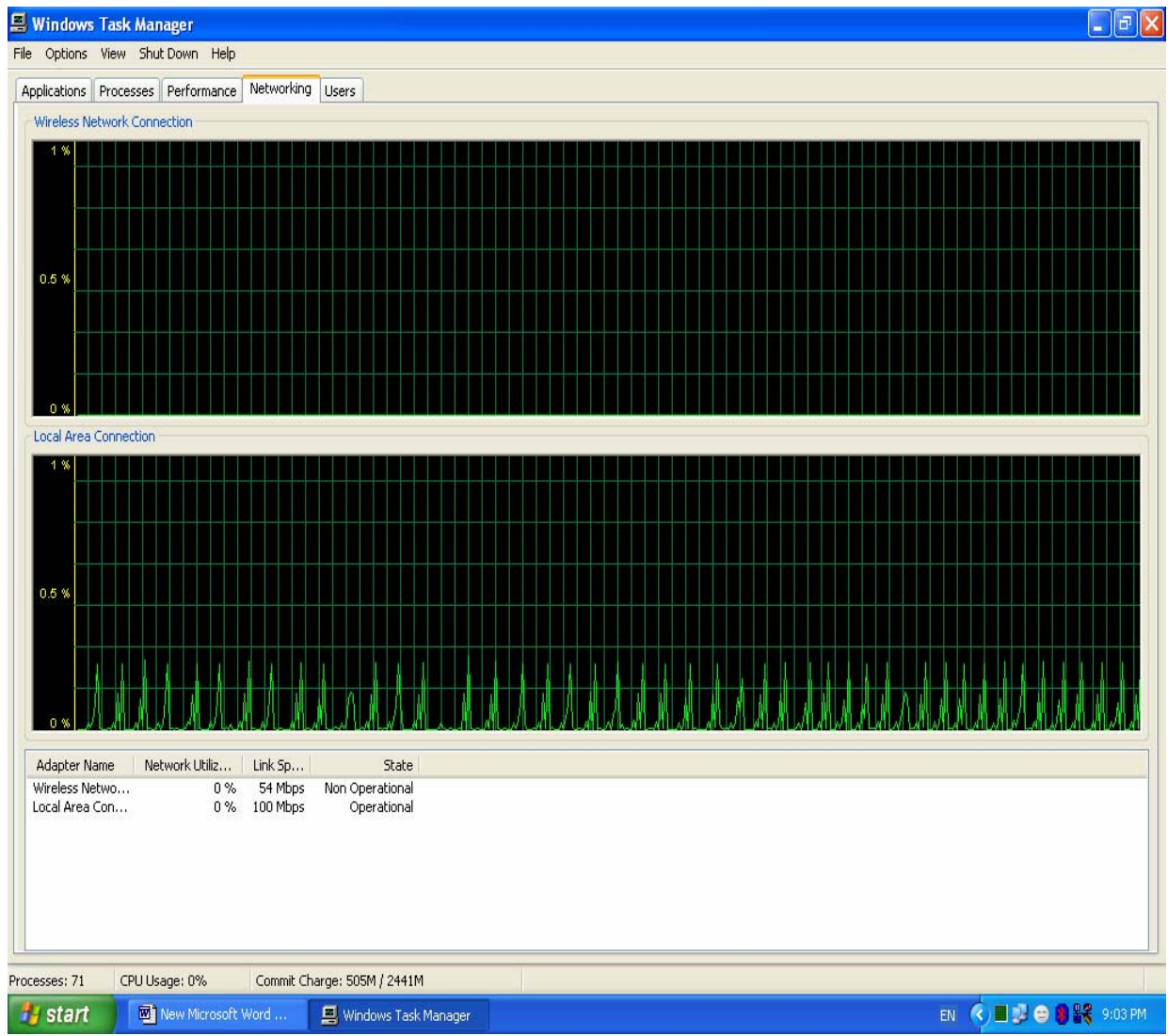

**Figure 4.25: Case Study Test (Network traffic)**

# CHAPTER 5

# WEB-BASED SCADA
# SECURITY

# CHAPTER 5

# WEB-BASED SCADA SECURITY

## Background

The continuous growth of cyber security threats and attacks including the increasing sophistication of malware is impacting the security of critical infrastructure, industrial control systems, and Supervisory Control and Data Acquisition (SCADA) control systems. The reliable operation of modern infrastructures depends on computerized systems and SCADA systems. Since the emergence of Internet and World Wide Web technologies, these systems were integrated with business systems and became more exposed to cyber threats. There is a growing concern about the security and safety of the SCADA control systems. Internet and global e-business application requirements demand that companies increasingly implement computing infrastructures specifically designed for at least 99.999 percent availability. This is the equivalent of less than 5.3 minutes of downtime a year. This is also a requirement for the SCADA networks. In response to these trends, SCADA owners need to address increased security and support for high availability [8]. Since these systems are required to run in a deterministic environment, any change to the SCADA systems that could slow the systems down, induce latency in communications, or bring the systems offline is not permissible [18].

As we mentioned before that for reasons of efficiency, maintenance, and economics, data acquisition and control platforms have migrated from isolated in-plant networks using proprietary hardware and software to PC-based systems using standard software, network protocols, and the Internet. The downside of this transition has been to expose SCADA systems to the same vulnerabilities and threats that plague Windows-based PCs and their associated networks. Some typical attacks that might be mounted against SCADA systems that employ standard hardware and software are listed here:

- Malicious code such as viruses, Trojan horses, and worms
- Unauthorized disclosure of critical data
- Unauthorized modification and manipulation of critical data
- Denial of service
- Unauthorized access to audit logs and modification of audit logs

As we see that the integration of SCADA systems with technologies such as the Internet and wireless means that security is an important issue for these systems. A large body of information-system security knowledge has accumulated concerning the protection of various types of computer systems and networks. The fundamental principles inherent in this knowledge provide a solid foundation for application to SCADA systems. However, some of the characteristics, performance requirements, and protocols of SCADA system components require adapting information-system security methods in industrial settings [18].

Conventional information technology (IT) systems are concerned with providing for internal and external connectivity, productivity, extensive security mechanisms for authentication and authorization, and the three major information security principles of confidentiality, availability, and integrity. Conversely, SCADA systems emphasize reliability, real-time response, tolerance of emergency situations where passwords might be incorrectly entered, personnel safety, product quality, and plant safety. A successful, unauthorized penetration of a SCADA system could result in an intruder taking control of a master or slave unit, disrupting critical processes, falsifying data, and even initiating actions that could result in the loss of human life and destruction of the plant under control. While most organizations claim that their SCADA system is not connected to their enterprise network it has been estimated that, in reality, 80 to 90 percent of SCADA systems are in fact connected to the enterprise network. It is that connection to the enterprise network that opens the door for Internet hackers to attack SCADA systems.

Many believe that it is the interconnection of SCADA and business systems across the enterprise network that poses the greatest risk to SCADA. In other words, SCADA systems were not initially intended to operate within the enterprise environment. Another issue is the inability within SCADA components to deal with the exposure to viruses, worms, and malware that are commonplace today within the enterprise network.

A network protocol alone can only provide security for information as it transits a network. No network protocol protects data before it is sent or after it arrives at its destination. This is the only known weakness in Web security that has been successfully exploited in an actual commercial setting. Unfortunately, it has been exploited more than once [22]. "Security by obscurity" is no longer an option for SCADA security. SCADA systems are a network presence and face significant threats and vulnerabilities. This requires a paradigm change for many personnel working in the critical infrastructure business [4]. Typical high-level weaknesses found in SCADA systems today include:

- Does not require any authentication.
- Does not require any authorization.
- Does not use encryption.
- Does not properly handle errors and exceptions.

Figure 5.1 shows and example for an unsecured communication system which contains no authentication and authorization.

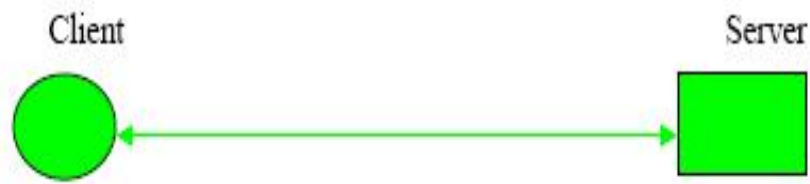

**Figure 5.1: Unsecured communication system, no authentication or authorization**

# 5.1 Design of a Security Policy

In our application we designed a security policy to protect our SCADA system against hackers, unauthorized persons and cyber attacks. In this policy we avoided the high level weakness found in most of today's SCADA systems. So a good security policy may have the following components:

*1. Authentication*

*2. Authorization*

*3. Encryption*

*4. Validation*

And now we will describe each of them.

Authentication (A1)

Authentication is any process by which you verify that someone is who they claim they are. This usually involves a username and a password, but can include any other method of demonstrating identity, such as a smart card, retina scan, special information, voice recognition, or fingerprints. Authentication is equivalent to showing your drivers license at the ticket counter at the airport.

Authorization (A2)

Authorization is finding out if the person, once identified, is permitted to have the resource. This is usually determined by finding out if that person is a part of a particular group, if that person has paid admission, or has a particular level of security clearance. Authorization is equivalent to checking the guest list at an exclusive party, or checking for your ticket when you go to the opera.

Encryption (E)

In cryptography, encryption is the process of transforming information (referred to as plaintext) Using an algorithm (called cipher) to make it unreadable to anyone except those possessing special knowledge, usually referred to as a key. The result of the process is encrypted information (in cryptography, referred to as Ciphertext). In many contexts, the word encryption also implicitly refers to the reverse process, decryption (e.g. "software for encryption" can typically also perform decryption), to make the encrypted information readable again (i.e. to make it unencrypted).

Validation (V)

In common usage, validation is the process of checking if something satisfies a certain criterion. Examples would include checking if a statement is true (validity), if an appliance works as intended, if a computer system is secure, or if computer data are compliant with an open standard. Validation implies one is able to document that a solution or process is correct or is suited for its intended use. Also Validation can mean to declare or make legally valid or to prove valid or confirm the validity of data, information, or processes.

### 5.1.1 Security Policy Overview

As we see in Figure 5.2 that our security policy may consists of four stages, authentication, authorization, optional encryption, and optional validation.

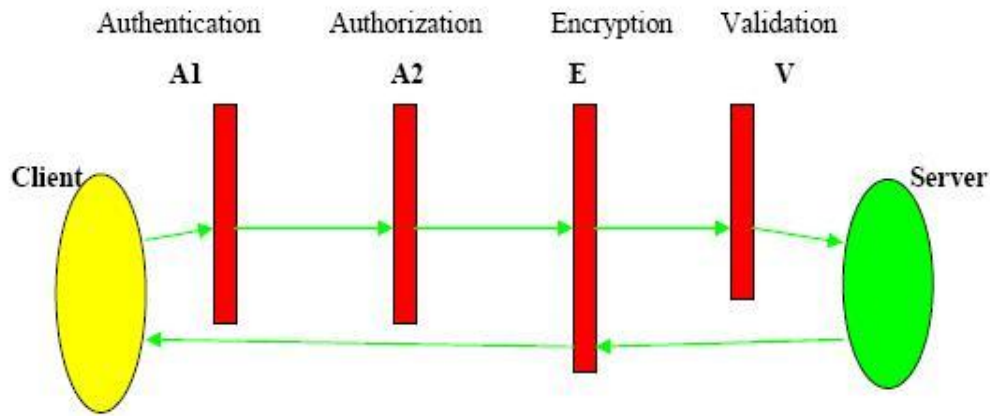

**Figure 5.2: Secured communication system**

## 5.1.2 Security Policy Rules

For any policy there should be rules and for our security policy we suggested the following rules:

1. Multiple logins are not allowed, No more than one client can login with the same username in the same time.

2. If a user tried to login with correct username/password pair and there is already a logged user with the same pair then both of them will disconnected and their machines will be marked as Untrusted (their IPs will be got and stored in a database).

3. Usernames/passwords pairs must be unique, which is the job of system administrator.

4. The user has only three trials to login in the first authentication phase with his username and password and if he failed to login his machine will be marked as Untrusted.

5. If the user succeeded in the first authentication phase he will be asked to give his company secret code which is unique for every employee and if he failed to give it he will be redirected to the login page again after marked as Untrusted.

6. If the user succeeded to pass the authentication phases he will be redirected to the authorization stage in which the system will determine the user's rights.

7. If a user succeeded to login after authentication and authorization his status field in the users' database will be changed to 1 automatically to indicate a logged user.

8. The optional encryption phase can be used to encrypt the transmitted data between the server and the client in both directions.

9. The optional validation phase can be used against authorized people mistakes or low experience.

These are our security policy rules which we will implement in the following section.

## 5.2 Security Policy implementation

We will create a database especially for security let its name be "SecureData.mdb" and we will add a new table into it to store usernames, passwords, roles, status and remote machine IP as shown in Table 5.1.

**Table 5.1: Authentication phase 1 login data**

| Username | Password | Role | Status | IP |
|----------|----------|------|--------|-----|
| hosny | 123456 | admin | 1 | 41.123.42.108 |
| Mohammed | 1234 | user | 0 | |
| Ali | 98765 | operator | 1 | 41.123.234.70 |

Another table for Untrusted IPs can be added which will look like Table 5.2. Again, another table for storing employees company secret codes, where every one in the company has a unique code as shown in Figure 5.3.

**Table 5.2: Untrusted IPs**

| IP |
|----|
| 62.234.45.34 |
| 125.134.87.36 |
| 222.254.61.67 |

**Table 5.3: Company secret Codes**

| Username | secret code |
|----------|-------------|
| Hosny | 2345 |
| Ali | 6657 |
| Rami | 2498 |

When the user opens the web site the first web page he will face is the authentication phase 1 page as shown in Figure 5.3.

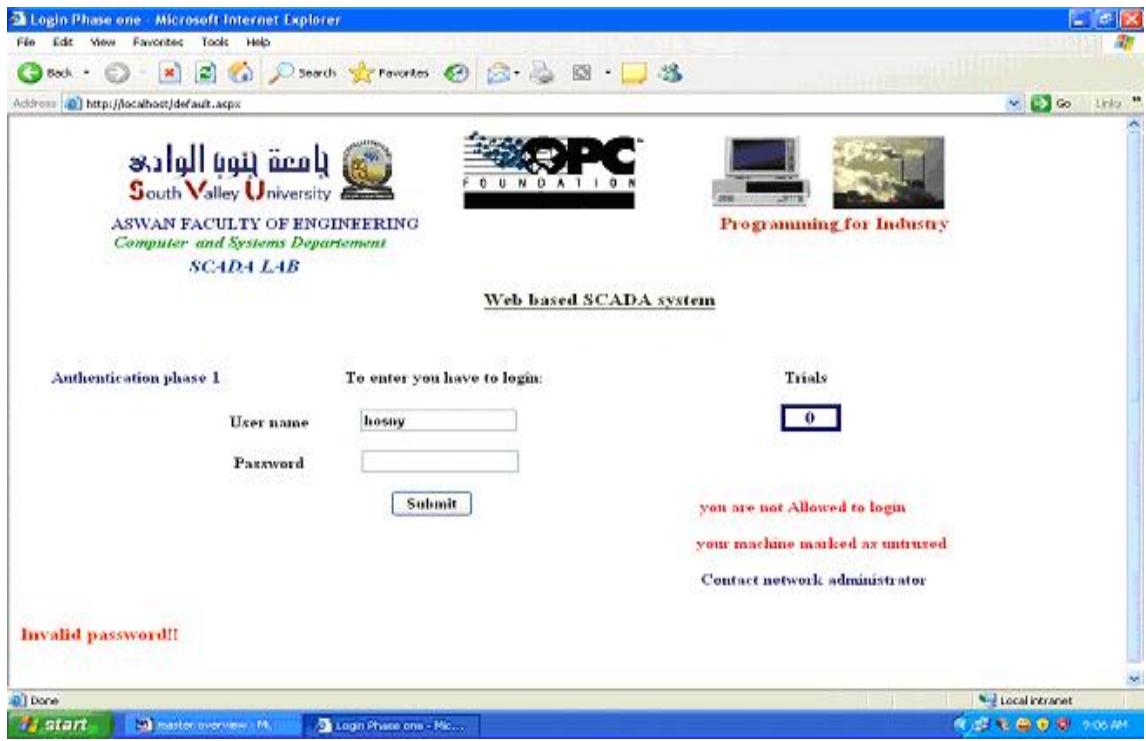

**Figure 5.3: Authentication phase 1**

The username and password will be checked by the server and if the user failed for three times to login to the system, the system will get the IP number for this user machine and marks it as *Untrusted* machine and will not permit this user to login even if he put the correct username and password until he contacts the system administrator. Also if there is another user logged with same username we can force him to logout for safety. On the other hand, if the user is not already logged and he succeeded to pass the first authentication phase, he will be asked to give his company secret code to continue to the next stage which is authorization. In the authorization phase the user rights will be determined for example if the user has the right to monitor or monitor and send setpoints according to his role. If the encryption stage is used (and it's recommended) there will be an encryption algorithm which can be carried out in server side.

Encryption can be in one direction or both directions according to the design. If the validation stage used its better to put it on the server side and there will be a mechanism to check the user entered setpoints with each variable range and required tolerance for change.

There will be a special page for the administrators in which they can manage the users accounts and problems related to login, it maybe similar Figure 5.5. As shown in the figure, the administrator can scan the system to check if a certain user is logging and he can get the user machine IP number and its role. In addition to this the administrator can add a new user and specify a username, password and Company secrete Code (CSC) for him. Also the system administrator can check the system database for untrusted IPs and he can remove any IP number to enable the user who uses this machine to login again to the SCADA system. The implementation code for SCADA security is found in appendix D.

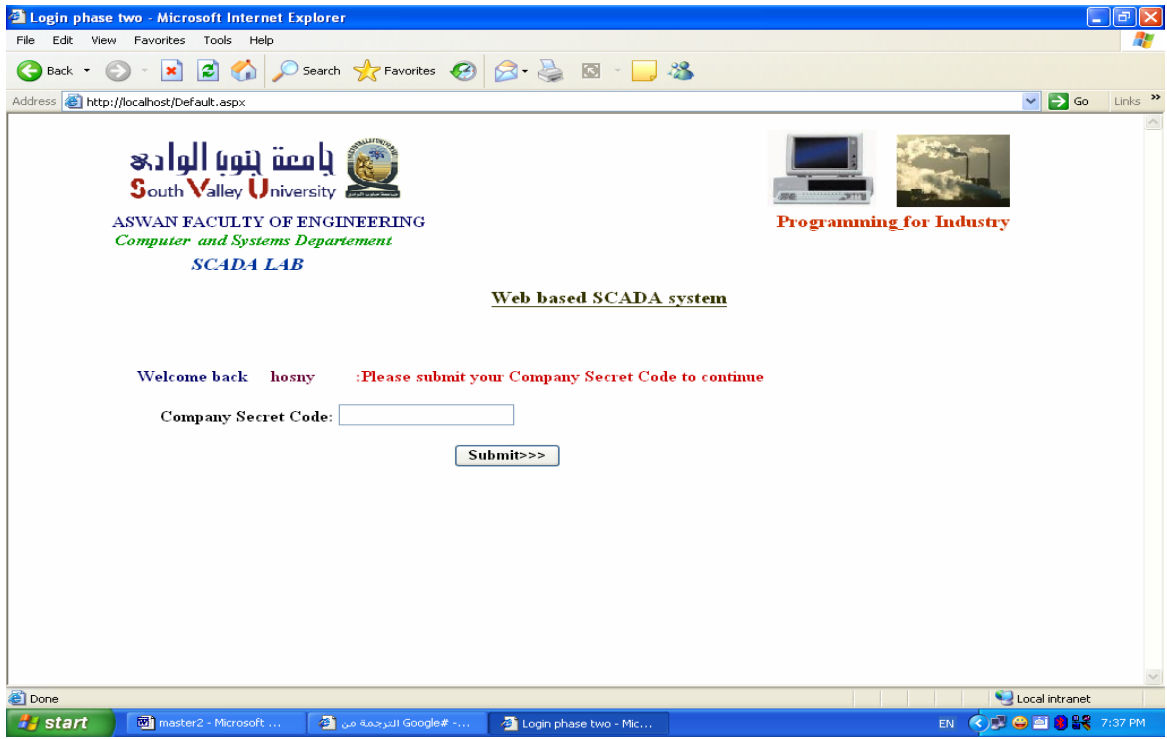

**Figure 5.4: Company Secret Code Page**

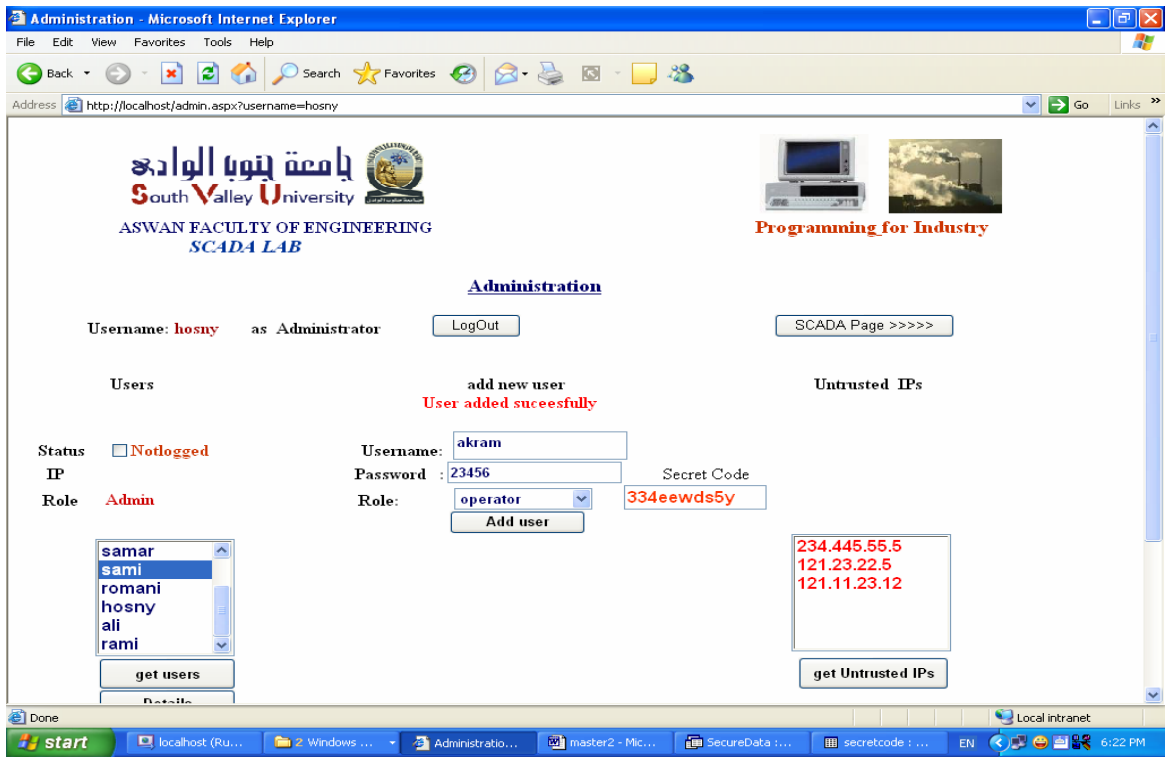

**Fig 5.5: Administration page**

## 5.3 Implementation problems

1. Although we have set the startup page of our project to be Default.aspx which is the login page, there is nothing to prevent a user, malicious or otherwise, from entering the URL of a specific page, such as SCADA page, into the address box of a browser. To avoid all this, it is good practice to check the login status of the user in the Page Load event of every page in the web site. If the user is not logged in (or is not in the correct role), we can then redirect the user to the appropriate page (often the login page).

*Private Sub Page_Load(ByVal sender As Object, ByVal e As System.EventArgs) Handles Me.Load*

*If Page.Request.UrlReferrer = Nothing Then Page.Response.Redirect("default.aspx")*

*Dim s As String*
*s = Request.QueryString("username")*
*username.Text = s*

*opdb("select * from Login where UserName='" + s.Trim() + "'")*
*If myrec.Fields(3).Value = False Then Page.Response.Redirect("default.aspx")*

*End Sub*

2. How to transfer the user details like username through the system pages. Actually there are many ways to do that and we used two of them, the first one is to use a query string (GET request) to pass information. In the source page when we specify the URL of the target page, include the information that we want to pass in the form of key-value pairs at the end of the URL. The first pair is preceded by a question mark (?) and subsequent pairs are preceded by ampersands (&), as shown in the following example:

> *Response.Redirect("page3.aspx?username=" + username.Text)*

In the target page, we can access query string values by using the QueryString property of the HttpRequest object, as shown in the following example:

```
Dim s As String
s = Request.QueryString("username")
username.Text = s
```

The second way is to use a Public Property of the source page. On the source page, we can create one or more public properties like:

```
Public ReadOnly Property CurrentCity () As String
    Get
        Return username.Text
    End Get
End Property
```

In target page code, we use strongly typed members of the PreviousPage property to read the source code properties. The following code example reads the value of the CurrentCity property that is defined in the source page

```
If Not Page.IsPostBack Then
    sourcepage = CType(Context.Handler, ASP.default_aspx)
    username.Text = sourcepage.CurrentCity
End If
```

3. How to get the client machine IP number. This is the function which we used to do that:

```
Public Function IpAddress() As String
    Dim strIpAddress As String
```

```
strIpAddress = Request.ServerVariables("HTTP_X_FORWARDED_FOR")
   If strIpAddress = "" Then
     strIpAddress = Request.ServerVariables("REMOTE_ADDR")
End If
     IpAddress = strIpAddress
End Function
```

4. Getting rid of the Back Button Problem, if a user logout from any web page in our application and he press the back button of the browser he will redirected again to this page without logging in, to solve this problem we added the following code segment in *Page_load()* event

```
Response.Cache.SetCacheability(HttpCacheability.NoCache)
Response.Cache.SetExpires(Now.AddSeconds(-1))
Response.Cache.SetNoStore()
Response.AppendHeader("Pragma", "no-cache")
```

This code segment basically directs the page in concern to expire immediately once it is posted and set the page to cache none of its content.

## 5.4 Security policy analysis

In our design of the security policy we intended to make it sensitive to user mistakes to make sure that if there is any doubt that a user who is not authorized to login to the system tries to login, the system should take an action quickly, that is because unauthorized user can make catastrophes if he succeeded to login to the SCADA system. That is why we restrict the user trials to login to 3 trials in the first authentication phase.

Also, our system can detect bad situations, for example if an unauthorized user succeeded to login and an authorized user tried to login, the system will know that there is something wrong and according to that it will block both, the already logged user and the failed user because the system doesn't know who is the authorized one as both of them used the same username/password pair. Also, the system will get the IP of the failed user machine and add it to the Untrusted

IPs table to prevent him from trying again to login until he contacts with the network administrator. If a user succeeded to give the username/password pair then he will be asked to give his Company Secret Code (CSC) to continue to the SCADA page we can make a Timer with interval (i.e. 30 seconds) in this stage in which he should give his CSC and if he failed his machine marked as Untrusted. With this policy we aim to make the mission too difficult for any attacker to login to the SCADA system.

In the designed policy we concentrate on authentication and authorization because of their importance, but encryption and validation can be used to complete the security system. But there will be some constraints for SCADA security because if complex security approaches used in SCADA systems it may impact the availability and reliability of the SCADA system, so there should be a type of balancing between SCADA security and availability.

There are many other security related issues which can be used in conjunction with our policy such as firewalls which can be used to screen message traffic between a corporate IT network like Internet and a SCADA network. Other security techniques can be used with our security policy but it out of the scope of this thesis for example:

- Audit and monitoring logs
- Biometrics
- Intrusion detection systems
- Malicious code detection and elimination

# CHAPTER 6

# CONCLUSIONS AND FUTURE WORKS

# CHAPTER 6
# CONCLUSIONS AND FUTURE WORKS

## 6.1 Conclusions

As IT technologies progress, web applications will replace desktop applications in most of computer applications fields especially industrial applications like SCADA systems which are now very important in process control. AJAX finally enables us to make an efficient web based SCADA systems because of the features it offers like asynchronous client-server communication and partial page update. Also Webservices can be used as server side scripting. Webservices provide a language-neutral, environment-neutral programming model that accelerates application integration inside and outside the enterprise. Application integration through Webservices yields flexible loosely coupled business systems.

We have seen a plenty of approaches [29, 26, 27, 20, 21, 31, 23, 3, 24, 10] that aimed to achieve a real time web based SCADA system, but the problem that they ignored is server resources and network bandwidth bad performance. Using AJAX in a proper way can give us a very good performance in the server and network as in this research we used AJAX to implement a persistent communication between client and server. Also the threading mechanism used in the Webservice was the key concept to achieve this persistent web communication.

Because of the Internet infrastructure, we concluded that data polling is the only viable solution to get new data, Troung Chau et el. [26] used a simple data polling mechanism by using an AJAX Timer to request new data set every timer tick and as we have explained before that this mechanism gives bad performance, bad effectiveness and bad efficiency. Therefore, we improved this approach by using an advanced data polling method, which depends on OPC DA server data change events, by this way we got an effective, efficient web based SCADA system with a reasonable real time behavior.

In the evaluation section we tested our approach against Troung Chau et al. [26] as an example of the approaches that used fixed client side data polling and concluded that with our approach we got a better web server resources consumption and a better usage of the network bandwidth, also we got an effective data transfer from server to client.

## 6.2 Future work

As we mentioned in chapter 1 that desktop applications have many advantages over web applications so we need to add these advantages to web applications to be competitive to desktop applications and to be feasible and suitable to be used for real time SCADA systems. AJAX is the latest technology in this area and it is now an attractive area for most automation researchers. Special AJAX techniques can be used to build a more efficient, secure, reliable, robust and safe SCADA systems.

Also as we have mentioned before that security and validation are very important in the new generation of web based SCADA systems, the reason for that is that the convergence between SCADA systems and internet make it vulnerable to cyber attacks ,hackers hands, worms, viruses, ...etc. Early SCADA systems benefited from the "Security by obscurity" that was afforded by their specialized operating systems and communications, but now the situation is deferent because the entire world now walks to standard systems. So, in future we will need to concentrate our efforts to secure SCADA systems and there are many researchers and companies working on that now. Also there are many features belong to SCADA system that need to be considered in future such as safety, reliability. Also, the GUI can be improved ,in our research we just used a text form to display data but it can be a graphical and animated pictures of tanks, valves, piping, pumps,…etc, that is because we concentrated on the main ideas in the approach design. We can summaries the future work in the following items:

1. Enhance the system to have alarming and history mechanisms.

2. Solving the problem of concurrent user access to web based SCADA.

3. Strengthening system security by adding encryption and validation techniques.

4. Analyzing the effect of complex security approaches on web based SCADA availability and reliability.

5. Checking the performance of the designed system with multi-Clients applications.

# APPENDICES

# APPENDIX A

## Flow charts for the system modules

# APPENDIX A

## Flow charts for client side

## AJAX scripts

Channel 1(Real Time Monitoring)

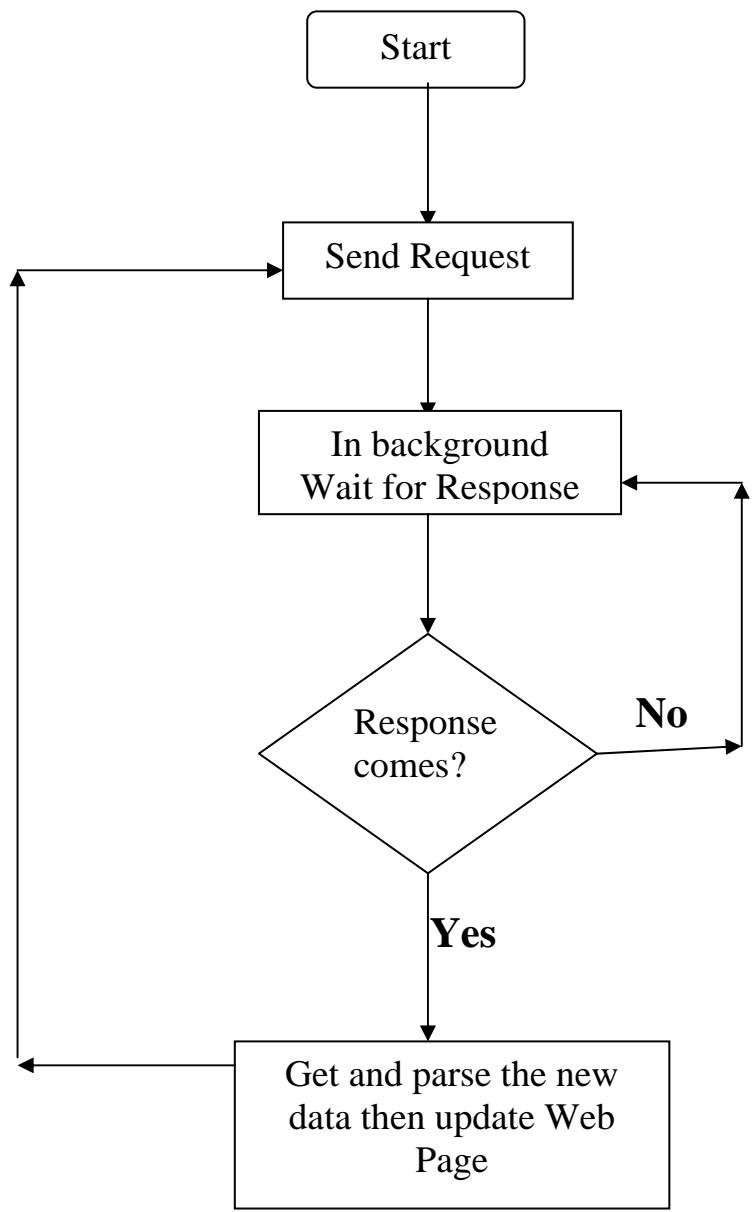

Channel 2 (Direct writing to server)

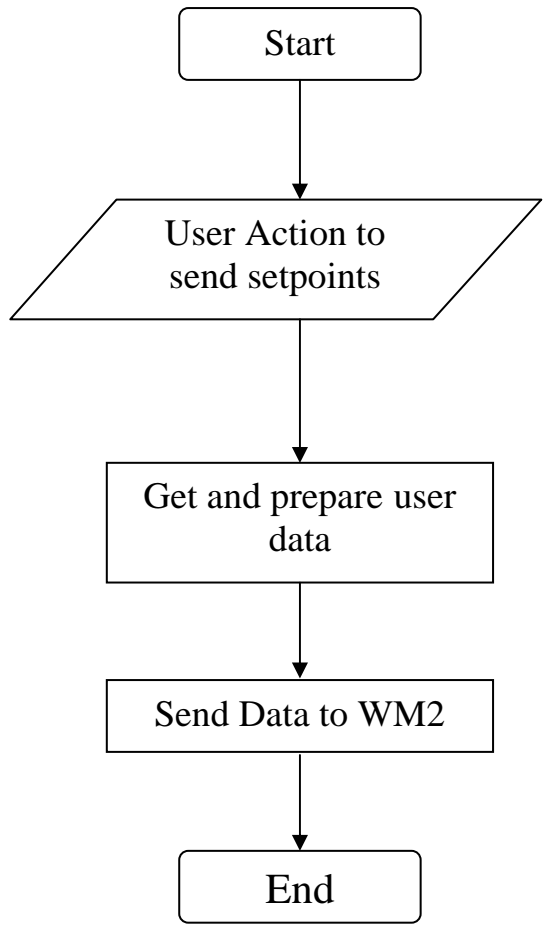

## Flow charts for Server side

WM1 (For Real Time Monitoring)

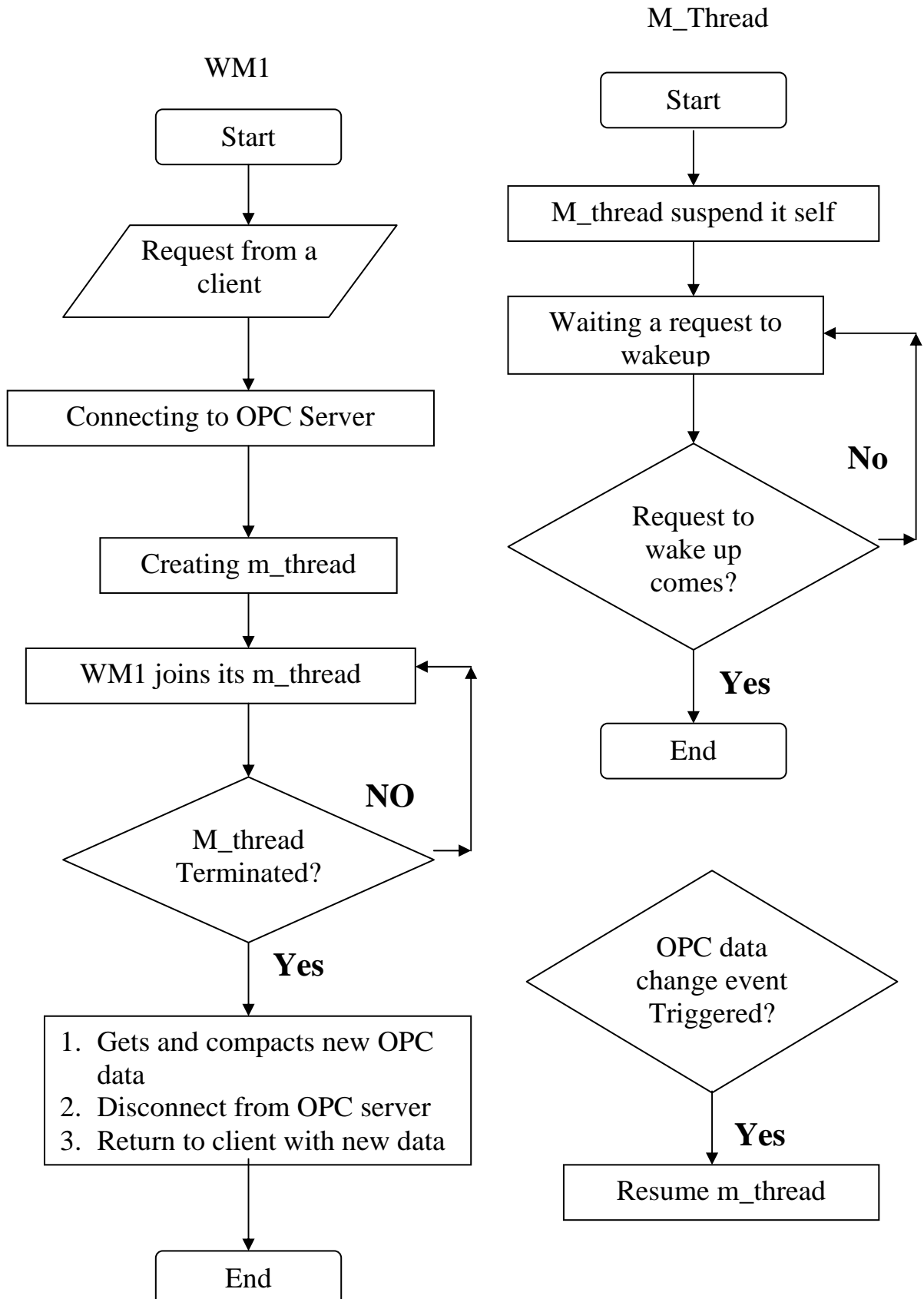

WM2 for direct writing to OPC DA server

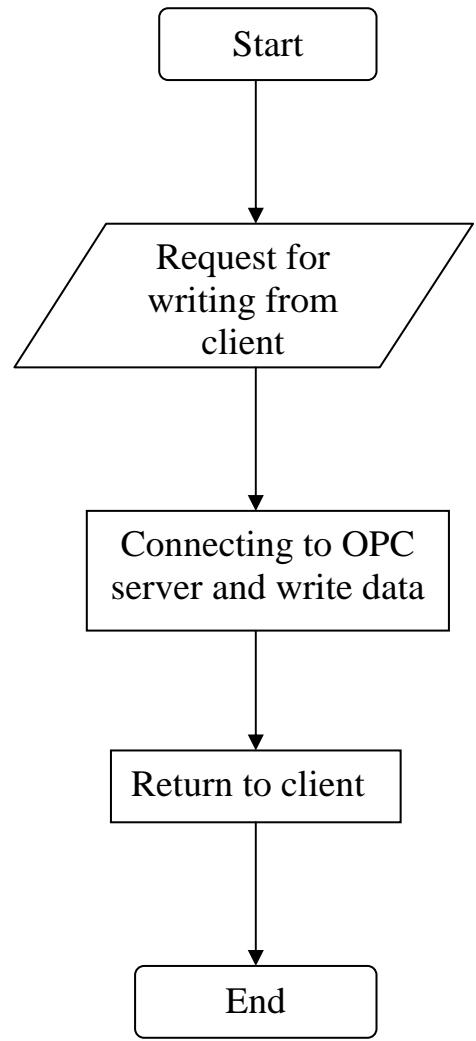

# APPENDIX B
## Client side implementation (JavaScript)

# APPENDIX B

## Client side implementation (JavaScript)

### File name: "Page3.aspx"

```
<script type="text/javascript" language="JavaScript" src="progress.js">
</script>
<script type="text/javascript" src="jscharts.js"></script>
<script type="text/javascript" src="rgbcolor.js"></script>
<link rel="stylesheet" href="js_color_picker_v2.css" media="screen">
<script type="text/javascript" src="color_functions.js"></script>
<script type="text/javascript" src="js_color_picker_v2.js"></script>

<script type="text/javascript" language="JavaScript" >
var DelayInSeconds=500;
var x=0;
var mydata1=[];
var mydata2=[];
var mydata3=[];
var permision=new Boolean;
permision=true;
var req1=new String("");
    function InitConstantCall()
    {
    $get('Text1').value="#330000"
    $get('Text2').value="#FFFF00"
    $get('Text3').value="#3300FF"
    setTimeout('ConstantCall()',DelayInSeconds);
    }

    function senddata()
    {
    if (($get('TextBox2').value > 0 & $get('TextBox2').value < 100) &
($get('TextBox3').value > 0 & $get('TextBox3').value<100)&
($get('TextBox4').value > 0 & $get('TextBox4').value<100))
    {
    var mystr=new String("");
    permision=false;
    mystr=$get('TextBox2').value + ";" + $get('TextBox3').value + ";" +
$get('TextBox4').value;
    MyWebService.senddata(mystr,onReturn);
    }
    else
    alert("Invalid value!Check setpoints values");
    }

    function onReturn()
    {
    permision=true;
    setTimeout('ConstantCall()',DelayInSeconds);

    }

    function ConstantCall()
    {
    if(permision)
        MyWebService.GetUpdate(OnComplete);

    }
```

```
function OnComplete(data){
    x+=1;

 req1=data.toString().split(';');

 req2=req1[0].split('~');
 $get('mytable').rows[1].cells[1].innerHTML=req2[0];
 $get('mytable').rows[1].cells[2].innerHTML=req2[1];
 $get('mytable').rows[1].cells[3].innerHTML=req2[2];

 req3=req1[1].split('~');
 $get('mytable').rows[2].cells[1].innerHTML=req3[0];
 $get('mytable').rows[2].cells[2].innerHTML=req3[1];
 $get('mytable').rows[2].cells[3].innerHTML=req3[2];

 req4=req1[2].split('~');
 $get('mytable').rows[3].cells[1].innerHTML=req4[0];
 $get('mytable').rows[3].cells[2].innerHTML=req4[1];
 $get('mytable').rows[3].cells[3].innerHTML=req4[2];

 req5=req1[3].toString().split('~');
 $get('mytable').rows[4].cells[1].innerHTML=req5[0];
 $get('mytable').rows[4].cells[2].innerHTML=req5[1];
 $get('mytable').rows[4].cells[3].innerHTML=req5[2];

 req6=req1[4].split('~');
 $get('mytable').rows[5].cells[1].innerHTML=req6[0];
 $get('mytable').rows[5].cells[2].innerHTML=req6[1];
 $get('mytable').rows[5].cells[3].innerHTML=req6[2];

 req7=req1[5].split('~');
 $get('mytable').rows[6].cells[1].innerHTML=req7[0];
 $get('mytable').rows[6].cells[2].innerHTML=req7[1];
 $get('mytable').rows[6].cells[3].innerHTML=req7[2];

 mydata1.push([x,parseFloat(req2[0])]);
 mydata2.push([x,parseFloat(req3[0])]);
 mydata3.push([x,parseFloat(req4[0])]);
 if(mydata1.length >=(parseInt($get('TextBox1').value)))
    for(i=0;i<=(parseInt($get('TextBox1').value));i++)
    {
    mydata1.pop();
    mydata2.pop();
    mydata3.pop();
    }

 if($get('CheckBox1').checked ||$get('CheckBox2').checked
||$get('CheckBox3').checked)
 if(mydata1.length >1 & mydata2.length >1 & mydata3.length >1)
 showtrend();
 $get('mytable').rows[7].cells[1].innerHTML=x;
 if (permission)
 setTimeout('ConstantCall()',DelayInSeconds);

    }

 function showtrend()
 {
```

```
    var myChart = new JSChart('chartcontainer', 'line');
    myChart.setSize(400,300);
    myChart.setLineColor('#ff0f0f');

    if($get('CheckBox1').checked==true)
    {
    myChart.setDataArray(mydata1,'l1');
    myChart.setLineColor($get('Text1').value, 'l1');
    }
     if($get('CheckBox2').checked==true)
    {
    myChart.setDataArray(mydata2,'l2');
    myChart.setLineColor($get('Text2').value, 'l2');
    }
     if($get('CheckBox3').checked==true)
     {
    myChart.setDataArray(mydata3,'l3');
    myChart.setLineColor($get('Text3').value, 'l3');
    }
    myChart.setTitle('Trend Plot');
    myChart.setAxisNameX('Request Count');
    myChart.setAxisNameY('Values');
    myChart.setAxisColor('#H00000');
    myChart.setAxisValuesColor('#ff0f0f');

    myChart.draw();
    }
function Button1_onclick() {

}

function Button4_onclick() {

}

</script>
```

# APPENDIX C

**Server side implementation (Webservice in VB.NET)**

# APPENDIX C

## Server side implementation (Webservice in VB.NET)

### File name: "myWebService.vb"

```
Imports System.Web
Imports System.Web.Services
Imports System.Web.Services.Protocols
Imports System
Imports System.Threading
' To allow this Web Service to be called from script, using ASP.NET AJAX,
uncomment the following line.
<System.Web.Script.Services.ScriptService()> _
<WebService(Namespace:="http://tempuri.org/")> _
<WebServiceBinding(ConformsTo:=WsiProfiles.BasicProfile1_1)> _
<Global.Microsoft.VisualBasic.CompilerServices.DesignerGenerated()> _
Public Class MyWebService
    Inherits System.Web.Services.WebService
    Dim OPCMyServer As OPCAutomation.OPCServer
    Dim OPCMyGroups As OPCAutomation.OPCGroups
    Dim WithEvents OPCMyGroup As OPCAutomation.OPCGroup
    Dim OPCMyItems As OPCAutomation.OPCItems
    Dim OPCMyItem As OPCAutomation.OPCItem

    Dim sItemName(8) As String
    Dim cH(8) As Integer
    Dim sH(8) As Integer
    Dim oVal(8) As Object
    Dim dTime(8) As Date
    Dim wQuality(8) As Short
    Dim ServerHandles As Array
    Dim Errors As Array
    Dim x1 As Integer
    Dim mycon As ADODB.Connection
    Dim myrec As ADODB.Recordset
    Dim mystr As String
    Private m_Thread As Thread

    <WebMethod()> _
        Public Function GetUpdate() As String
        opcconnect()        ' this function connects to the OPC DA server
        Thread.Sleep(50) 'for stability
        m_Thread = New Thread(AddressOf mythread) 'child thread
        m_Thread.IsBackground = True
        m_Thread.Start() 'start the child thread
        m_Thread.Join()    ' parent thread join its child thread
        getdata()          ' control comes to parent thread
        m_Thread = Nothing
        opcdisconnect()
        Return mystr        'return to client (AJAX)
    End Function
    <WebMethod()> _
      Public Function senddata(ByVal data As String) As String
        Dim connected As Boolean
        If OPCMyServer Is Nothing Then
            opcconnect()
            connected = True
        End If
        Dim mydata As Array
        mydata = data.Split(";")
        OPCMyGroup.OPCItems.Item(4).Write(Val(mydata(0)))
```

```
    OPCMyGroup.OPCItems.Item(5).Write(Val(mydata(1)))
    OPCMyGroup.OPCItems.Item(6).Write(Val(mydata(2)))
    If connected Then opcdisconnect()
    Return "Done"
End Function
Sub getdata()
    Dim anItem As OPCAutomation.OPCItem
    mystr = ""
    For Each anItem In OPCMyGroup.OPCItems
        anItem.Read(OPCAutomation.OPCDataSource.OPCDevice)  ', value,
qual, time    ' If subscribed, don't do this!
        oVal(anItem.ClientHandle) = anItem.Value
        dTime(anItem.ClientHandle) = anItem.TimeStamp
        wQuality(anItem.ClientHandle) = anItem.Quality
        mystr = mystr + Str(oVal(anItem.ClientHandle)) + "~" +
Str(wQuality(anItem.ClientHandle)) + "~" +
dTime(anItem.ClientHandle).ToLocalTime.ToString + ";"
    Next anItem
    mystr = mystr.Remove(mystr.Length - 1)
    anItem = Nothing
End Sub
Sub mythread()
    m_Thread.Suspend()
End Sub
    Sub opcconnect()
    OPCMyServer = New OPCAutomation.OPCServer
    OPCMyServer.Connect("opc.simaticnet", "192.168.16.3")
    OPCMyGroups = OPCMyServer.OPCGroups
    OPCMyGroup = OPCMyGroups.Add("Group1")
    OPCMyGroup.UpdateRate = 300
    OPCMyGroup.IsActive = True
    OPCMyGroup.IsSubscribed = True
    OPCMyItems = OPCMyGroup.OPCItems
    x1 = 0
    For i = 1 To 8
        sItemName(i) = "s7:[@LOCALSERVER]db1,w" + Format$(x1)
        x1 = x1 + 2
        cH(i) = i
    Next i
    OPCMyItems.AddItems(8, sItemName, cH, ServerHandles, Errors)
End Sub
Sub opcdisconnect()
    OPCMyGroup.IsActive = False
    OPCMyItems = Nothing
    OPCMyItem = Nothing
    OPCMyGroups = Nothing
    OPCMyGroup = Nothing

    OPCMyServer.Disconnect()
    OPCMyServer = Nothing
    System.GC.Collect() 'Execute Garbage Collection compulsorily.
End Sub
    Private Sub OPCMyGroup_DataChange(ByVal TransactionID As Integer, ByVal
NumItems As Integer, ByRef ClientHandles As System.Array, ByRef ItemValues
As System.Array, ByRef Qualities As System.Array, ByRef TimeStamps As
System.Array) Handles OPCMyGroup.DataChange
        m_Thread.Resume()
    End Sub
End Class
```

# APPENDIX D

## SCADA security code (VB.NET)

# APPENDIX D

## SCADA security code (VB.NET)

Login page (authentication phase 1), **File name: "default.aspx.vb"**

```vbnet
Partial Class login
    Inherits System.Web.UI.Page
    Dim mycon As ADODB.Connection
    Dim myrec As ADODB.Recordset
    Dim mystr As String
    Dim IP1 As String

    Protected Sub Page_Load(ByVal sender As Object, ByVal e As
System.EventArgs) Handles Me.Load
        checkIPadress()
    End Sub

    Protected Sub btnSubmit_Click(ByVal sender As Object, ByVal e As
System.EventArgs) Handles btnSubmit.Click
        if checkIPadress() then

            Exit Sub
        End If

        opdb("select * from Login where UserName='" +
txtUserName.Text.Trim() + "'")
        If (myrec.RecordCount <= 0) Then
            lblMessage.Text = "Invalid user name"
            counter.Text = Val(counter.Text) - 1
            If Val(counter.Text) = 0 Then
                untrusteduser()
            End If

        Else
            If myrec.Fields(1).Value = txtPassword.Text.Trim() Then
                If myrec.Fields(3).Value = True Then
                    msg0.Text = "This username is already logged"
                    myrec.Fields(3).Value = False
                    IP1 = myrec.Fields(4).Value
                    myrec.Fields(4).Value = ""
                    myrec.Update()
                    untrusteduser()
                    Exit Sub
                End If
                lblMessage.Text = "Welcome back " + txtUserName.Text.Trim()
                txtPassword.Text = String.Empty
                opdb("select * from Login where UserName='" +
txtUserName.Text.Trim() + "'")
                myrec.Fields(3).Value = True
                myrec.Update()
                Server.Transfer("Page2.aspx?IP=" + IP1)

            Else
                lblMessage.Text = "Invalid password!!"
                counter.Text = Val(counter.Text) - 1
                If Val(counter.Text) = 0 Then
                    untrusteduser()
                End If
```

```vbnet
        End If
        End If
    End Sub
    Public Sub opdb(ByVal mystr As String)
        mycon = New ADODB.Connection
        myrec = New ADODB.Recordset
        Dim cnstr = "provider=microsoft.jet.oledb.4.0;data source=" &
"C:\SecureData.mdb" & ";persist security info=false"
        mycon.Open(cnstr)
        myrec.Open(mystr, mycon, 3, 3)
    End Sub

    Public Function IpAddress() As String
        Dim strIpAddress As String
        strIpAddress = Request.ServerVariables("HTTP_X_FORWARDED_FOR")
        If strIpAddress = "" Then
            strIpAddress = Request.ServerVariables("REMOTE_ADDR")
        End If
        IpAddress = strIpAddress
    End Function

    Public ReadOnly Property CurrentCity() As String
        Get
            Return txtUserName.Text
        End Get
    End Property

Function checkIPaddress() as boolean
        opdb("select * from untrusted where IP='" + IpAddress() + "'")
        If (myrec.RecordCount > 0) Then
            txtUserName.BackColor = Drawing.Color.Red
            txtPassword.BackColor = Drawing.Color.Red
            txtUserName.Enabled = 0
            txtPassword.Enabled = 0
            counter.Text = 0
            msg1.Text = "you are not Allowed to login"
            msg2.Text = " your machine marked as untrused"
            msg3.Text = "Contact network administrator"

         checkIPaddress=True

         endif

        checkIPaddress=False

 End Function
```

Login page (authentication phase 2), File name: "Page2.aspx.vb"

```vb
Partial Class login
    Inherits System.Web.UI.Page
    Dim sourcepage As ASP.default_aspx
    Dim mycon As ADODB.Connection
    Dim myrec As ADODB.Recordset
    Dim mystr As String
    Dim IP2 As String

    Public Sub Page_Load(ByVal sender As Object, ByVal e As
System.EventArgs) Handles Me.Load
        Response.Cache.SetCacheability(HttpCacheability.NoCache)
        Response.Cache.SetExpires(Now.AddSeconds(-1))
        Response.Cache.SetNoStore()
        Response.AppendHeader("Pragma", "no-cache")
        If Page.Request.UrlReferrer = Nothing Then
Page.Response.Redirect("default.aspx")

        If Not Page.IsPostBack Then
            sourcepage = CType(Context.Handler, ASP.default_aspx)
            labusername.Text = sourcepage.CurrentCity
        End If
        IP2 = Request.QueryString("IP")
    End Sub

    Public Function IpAddress() As String
        Dim strIpAddress As String
        strIpAddress = Request.ServerVariables("HTTP_X_FORWARDED_FOR")
        If strIpAddress = "" Then
            strIpAddress = Request.ServerVariables("REMOTE_ADDR")
        End If
        IpAddress = strIpAddress
    End Function

    Public Sub opdb(ByVal mystr As String)
        mycon = New ADODB.Connection
        myrec = New ADODB.Recordset
        Dim cnstr = "provider=microsoft.jet.oledb.4.0;data source=" &
"C:\SecureData.mdb" & ";persist security info=false"
        mycon.Open(cnstr)
        myrec.Open(mystr, mycon, 3, 3)
    End Sub

    Public ReadOnly Property CurrentCity2() As String
        Get
            Return labusername.Text
        End Get
    End Property

    Protected Sub btnsubmit_Click(ByVal sender As Object, ByVal e As
System.EventArgs) Handles btnsubmit.Click
        opdb("select * from secretcode where username='" +
labusername.Text.Trim() + "'")
        If myrec.Fields(1).Value = txtsecretcode.Text.Trim() Then
            opdb("select * from login where UserName='" +
labusername.Text.Trim() + "'")
            myrec.Fields(4).Value = IpAddress()
            myrec.Update()
            If myrec.Fields(2).Value = "admin" Or myrec.Fields(2).Value =
"Admin" Then
```

```
                Response.Redirect("admin.aspx?username=" +
labusername.Text)
            Else
                Response.Redirect("page3.aspx?username=" +
labusername.Text)
            End If
        Else
            untrusteduser()
            logout()
        End If
    End Sub

    Public Sub untrusteduser()
        opdb("untrusted")
        myrec.AddNew()
        myrec.Fields(0).Value = IpAddress()
        myrec.Update()
        myrec.AddNew()
        myrec.Fields(0).Value = IP2
        myrec.Update()
    End Sub
    Sub logout()
        opdb("select * from Login where UserName='" +
labusername.Text.Trim() + "'")
        myrec.Fields(3).Value = False
        myrec.Fields(4).Value = ""
        myrec.Update()
        Page.Response.Redirect("default.aspx")
    End Sub
End Class
```

Administration Page, File name: "admin.aspx.vb"

```vb
Partial Class analysis
    Inherits System.Web.UI.Page
    Dim mycon As ADODB.Connection
    Dim myrec As ADODB.Recordset
    Dim mystr As String
    Dim selecteduser As String

    Protected Sub Page_Load(ByVal sender As Object, ByVal e As
System.EventArgs) Handles Me.Load
        If Page.Request.UrlReferrer = Nothing Then
            Response.Redirect("default.aspx")
        End If
        Response.Cache.SetCacheability(HttpCacheability.NoCache)
        Response.Cache.SetExpires(Now.AddSeconds(-1))
        Response.Cache.SetNoStore()
        Response.AppendHeader("Pragma", "no-cache")

        Dim s As String
        s = Request.QueryString("username")
        labusername.Text = s
        opdb("select * from Login where UserName='" + s.Trim() + "'")
        If myrec.Fields(3).Value = False Then
Response.Redirect("default.aspx")

    End Sub

    Public Sub opdb(ByVal mystr As String)
        mycon = New ADODB.Connection
        myrec = New ADODB.Recordset
        Dim cnstr = "provider=microsoft.jet.oledb.4.0;data source=" &
"C:\SecureData.mdb" & ";persist security info=false"
        mycon.Open(cnstr)
        myrec.Open(mystr, mycon, 3, 3)
    End Sub
    Sub logout()
        opdb("select * from Login where UserName='" +
labusername.Text.Trim() + "'")
        myrec.Fields(3).Value = False
        myrec.Fields(4).Value = ""
        myrec.Update()
        Page.Response.Redirect("default.aspx")
    End Sub

    Protected Sub btnlogout_Click(ByVal sender As Object, ByVal e As
System.EventArgs) Handles btnlogout.Click
        logout()
    End Sub

    Protected Sub Button1_Click(ByVal sender As Object, ByVal e As
System.EventArgs) Handles Button1.Click
        Response.Redirect("page3.aspx?username=" + labusername.Text.Trim())
    End Sub

    Protected Sub Button2_Click(ByVal sender As Object, ByVal e As
System.EventArgs) Handles Button2.Click
        lst1.Items.Clear()
        labIP.Text = ""
```

```
        labrole.Text = ""
        chkstatus.Text = ""
        opdb("select * from login")
        myrec.MoveFirst()
        Do While Not myrec.EOF
            lst1.Items.Add(myrec.Fields(0).Value)
            myrec.MoveNext()
        Loop
    End Sub

    Protected Sub Button3_Click(ByVal sender As Object, ByVal e As
System.EventArgs) Handles Button3.Click
        opdb("select * from untrusted where IP='" + lst2.SelectedItem.Value
+ "'")
        myrec.Delete()
        lst2.Items.Remove(lst2.SelectedItem)
    End Sub

    Protected Sub Button4_Click(ByVal sender As Object, ByVal e As
System.EventArgs) Handles Button4.Click
        lst2.Items.Clear()
        opdb("select * from untrusted")
        If myrec.RecordCount > 0 Then myrec.MoveFirst()
        Do While Not myrec.EOF
            If myrec.Fields(0).Value <> "" Then
                lst2.Items.Add(myrec.Fields(0).Value)
            End If
            myrec.MoveNext()
        Loop
    End Sub

    Protected Sub Button5_Click(ByVal sender As Object, ByVal e As
System.EventArgs) Handles Button5.Click
        If txtusername.Text = "" Or txtpassword.Text = "" Or lst3.Text = ""
Then
            labshow.Text = "Complete user data"
            Exit Sub
        End If
        opdb("select * from login where username='" &
txtusername.Text.Trim() & "'")
        If myrec.RecordCount > 0 Then
            labshow.Text = "This username is already used,try another one"
            Exit Sub
        End If
        myrec.AddNew()
        myrec.Fields(0).Value = txtusername.Text.Trim
        myrec.Fields(1).Value = txtpassword.Text.Trim
        myrec.Fields(2).Value = lst3.Text.Trim
        myrec.Fields(3).Value = False
        myrec.Fields(4).Value = ""
        myrec.Update()
        opdb("secretcode")
        myrec.AddNew()
        myrec.Fields(0).Value = txtusername.Text.Trim
        myrec.Fields(1).Value = txtsecretcode.Text.Trim
        myrec.Update()
        labshow.Text = "User added suceesfully"
    End Sub
```

```vb
    Protected Sub btngetusers_Click(ByVal sender As Object, ByVal e As
System.EventArgs) Handles btngetusers.Click
        On Error Resume Next
        opdb("select * from Login where UserName='" +
lst1.SelectedItem.Value + "'")
        chkstatus.Checked = myrec.Fields(3).Value = True
        If chkstatus.Checked = True Then
            chkstatus.Text = "Logged"
        Else
            chkstatus.Text = "Notlogged"
        End If
        labIP.Text = myrec.Fields(4).Value
        labrole.Text = myrec.Fields(2).Value
    End Sub

    Protected Sub chkstatus_CheckedChanged(ByVal sender As Object, ByVal e
As System.EventArgs) Handles chkstatus.CheckedChanged
        If chkstatus.Checked = False Then
            chkstatus.Text = "Not Logged"
            labIP.Text = ""
            If lst1.SelectedItem.Value <> "" Then
                opdb("select * from Login where UserName='" +
lst1.SelectedItem.Value.Trim + "'")
                If myrec.Fields(3).Value = True Then
                    myrec.Fields(3).Value = False
                    myrec.Fields(4).Value = ""
                    myrec.Update()
                End If
            End If
        Else
            chkstatus.Text = "Logged"
        End If
    End Sub
End Class
```

# APPENDIX E

**Simulation**
**VB.NET 2008**

# APPENDIX E

## Simulation

## VB.NET 2008

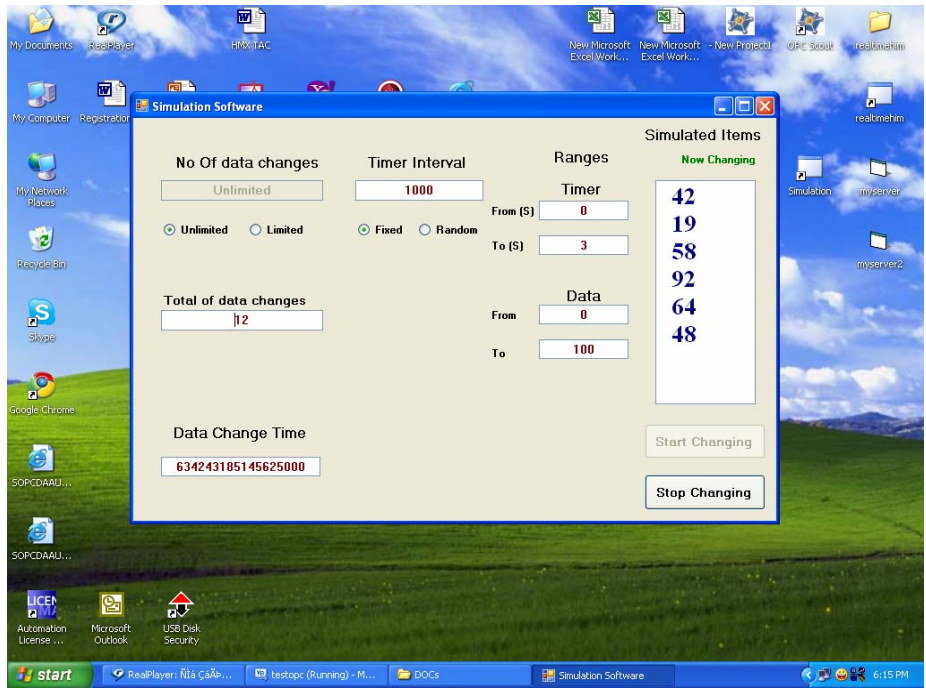

Simulation program

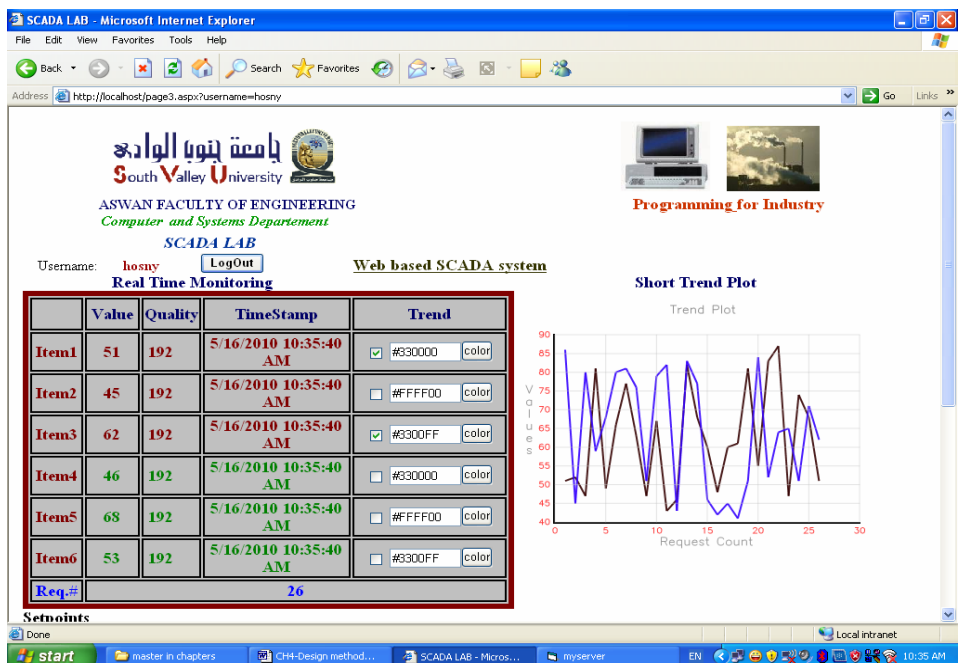

Project output

```vbnet
Imports System
Public Class Form1

    Dim WithEvents OPCMyServer As OPCAutomation.OPCServer
    Dim OPCMyGroups As OPCAutomation.OPCGroups
    Dim WithEvents OPCMyGroup As OPCAutomation.OPCGroup
    Dim OPCMyItems As OPCAutomation.OPCItems
    Dim OPCMyItem As OPCAutomation.OPCItem

    Dim itemid(6) As String
    Dim cH(6) As Integer
    Dim sH(6) As Integer
    Dim oVal(6) As Object
    Dim dTime(6) As Date
    Dim wQuality(6) As Short
    Dim ServerHandles As Array
    Dim Errors As Array
    Dim x1 As Integer
    Dim mycon As ADODB.Connection
    Dim myrec As ADODB.Recordset
    Dim mystr As String
    Dim startTick As Long
    Dim endTick As Long
    Dim tick As Long
    Dim seconds As Long
    Dim milliseconds As Long
    Dim microseconds As Long
    Public count As Integer
    Dim Totaldatachanges As Integer
    Dim RandomClass As New Random()

    Public Sub opdb(ByVal mystr As String)
        mycon = New ADODB.Connection
        myrec = New ADODB.Recordset
        Dim cnstr = "provider=microsoft.jet.oledb.4.0;data source=" & _
"D:\realtimehim.mdb" & ";persist security info=false"
        mycon.Open(cnstr)
        myrec.Open(mystr, mycon, 3, 3)
    End Sub

    Sub opcconnect()
        OPCMyServer = New OPCAutomation.OPCServer
        OPCMyServer.Connect("opc.simaticnet", "")
        OPCMyGroups = OPCMyServer.OPCGroups
        OPCMyGroup = OPCMyGroups.Add("Group1")
        OPCMyGroup.UpdateRate = 400
        OPCMyGroup.IsActive = True
        OPCMyGroup.IsSubscribed = True
        OPCMyItems = OPCMyGroup.OPCItems

        Dim constr = "s7:[@LOCALSERVER]db1,w"
        x1 = 0
        For i = 1 To 6
            itemid(i) = constr + Format$(x1)
            x1 = x1 + 2
            cH(i) = i
        Next
        OPCMyItems.AddItems(6, itemid, cH, ServerHandles, Errors)
    End Sub
    Sub opcdisconnect()
        OPCMyGroup.IsActive = False
```

```vbnet
        OPCMyItems = Nothing
        OPCMyItem = Nothing
        OPCMyGroups = Nothing
        OPCMyGroup = Nothing

        OPCMyServer.Disconnect()
        OPCMyServer = Nothing
        System.GC.Collect() 'Execute Garbage Collection compulsorily.
    End Sub

    Private Sub Form1_Load(ByVal sender As System.Object, ByVal e As
System.EventArgs) Handles MyBase.Load
        opcconnect()
        labstatus.ForeColor = Color.Red
        labstatus.Text = "Stopped"
        btnstopwriting.Enabled = 0
        txtnoofdatachanges.Text = "Unlimited"
        txtnoofdatachanges.Enabled = 0
        Totaldatachanges = 0
        txtInterval.Text = 1000
        rdofixed.Checked = 1
    End Sub

    Private Sub OPCMyGroup_DataChange(ByVal TransactionID As Integer, ByVal
NumItems As Integer, ByRef ClientHandles As System.Array, ByRef ItemValues
As System.Array, ByRef Qualities As System.Array, ByRef TimeStamps As
System.Array) Handles OPCMyGroup.DataChange

        startTick = DateTime.Now.Ticks
        txtstart.Text = startTick
        opdb("tbl")
        myrec.AddNew()
        myrec.Fields(0).Value = startTick
        myrec.Update()

        lstsimulatedItems.Items.Clear()
        For i = 1 To 6
            lstsimulatedItems.Items.Add("  " & ItemValues(i))
        Next
        Totaldatachanges = Totaldatachanges + 1
        txttotaldatachanges.Text = Totaldatachanges
        If rdolimited.Checked And Totaldatachanges >=
Val(txtnoofdatachanges.Text) And txtnoofdatachanges.Text <> "" Then
            OPCMyGroup.IsActive = 0
            Timer1.Enabled = 0
            labstatus.ForeColor = Color.Red
            labstatus.Text = "Stopped"
            btnstopwriting.Enabled = 0
            btnstartwriting.Enabled = 1
        End If
    End Sub

    Private Sub Timer1_Tick(ByVal sender As System.Object, ByVal e As
System.EventArgs) Handles Timer1.Tick
        Timer2.Enabled = 1
    End Sub

    Private Sub Timer2_Tick(ByVal sender As System.Object, ByVal e As
System.EventArgs) Handles Timer2.Tick
        Dim RandomNumber As Double
        For i = 1 To 6
```

```vb
        oVal(i) = RandomClass.Next(Val(txtDataFrom.Text),
Val(txtDataTo.Text))
        Next
        OPCMyGroup.SyncWrite(6, ServerHandles, oVal, Errors)
        Timer2.Enabled = 0

        If rdorandom.Checked Then
            RandomNumber = RandomClass.NextDouble()
            Timer1.Interval = 1000 * Val(txtRTimerTo.Text) * RandomNumber
            txtInterval.Text = Timer1.Interval
        End If
    End Sub

    Private Sub btnstartwriting_Click(ByVal sender As System.Object, ByVal
e As System.EventArgs) Handles btnstartwriting.Click
        Timer1.Enabled = 1
        OPCMyGroup.IsActive = 1
        labstatus.ForeColor = Color.Green
        labstatus.Text = "Now Changing"
        btnstopwriting.Enabled = 1
        btnstartwriting.Enabled = 0
        If rdolimited.Checked Then
            Totaldatachanges = 0
            txttotaldatachanges.Text = 0
            opdb("tb1")
            Do While Not myrec.EOF
                myrec.Delete()
                myrec.Update()
                myrec.MoveNext()
            Loop
        End If
    End Sub

    Private Sub btnstopwriting_Click(ByVal sender As System.Object, ByVal e
As System.EventArgs) Handles btnstopwriting.Click
        Timer1.Enabled = 0
        OPCMyGroup.IsActive = 0
        labstatus.ForeColor = Color.Red
        labstatus.Text = "Stopped"
        btnstartwriting.Enabled = 1
        btnstopwriting.Enabled = 0
    End Sub

    Private Sub rdofixed_CheckedChanged(ByVal sender As System.Object,
ByVal e As System.EventArgs) Handles rdofixed.CheckedChanged

        txtInterval.Text = 1000
        Timer1.Interval = 1000

    End Sub

    Private Sub rdolimited_CheckedChanged(ByVal sender As System.Object,
ByVal e As System.EventArgs) Handles rdolimited.CheckedChanged

    End Sub

    Private Sub rdounlimited_CheckedChanged(ByVal sender As System.Object,
ByVal e As System.EventArgs) Handles rdounlimited.CheckedChanged
        txtnoofdatachanges.Enabled = 0
        txtnoofdatachanges.Text = "unlimited"
    End Sub
```

```vb
    Private Sub rdorandom_CheckedChanged(ByVal sender As System.Object,
ByVal e As System.EventArgs) Handles rdorandom.CheckedChanged

    End Sub

    Private Sub rdolimited_Click(ByVal sender As Object, ByVal e As
System.EventArgs) Handles rdolimited.Click
        If rdolimited.Checked Then
            txtnoofdatachanges.Enabled = 1
            txtnoofdatachanges.Text = "10"
        End If
    End Sub
End Class
```

# LIST OF PUBLICATIONS

## LIST OF PUBLICATIONS

# LIST OF REFERENCES

# GLOSSARY

# GLOSSARY

## AJAX

*Asynchronous JavaScript and XML*

## API

*Application Programming Interface*

## ASP

*Active Server Pages*

## ASP.NET

*Active Server Pages in .NET*

## CAD

*Computer-aided design*

## CAM

*Computer-aided manufacturing*

## CAx

*Computer-aided technologies*

## CGI

*Common Gateway interface, it's a set of standards where a program or script (a series of commands) can send data back to the web server where it can be processed.*

## COM

*Component Object Model*

## ControlNet

*An Allen Bradley communications protocol applied to control systems*

## CORBA

*The Common Object Request Broker Architecture (CORBA) is a standard defined by the Object Management Group (OMG) that enables software components written in multiple computer languages and running on multiple computers to work together, i.e. it supports multiple platforms.*

## CSS

*Class Style Sheet*

## DCOM

*Distributed Component Object Model*

## DeviceNet

*An Allen Bradley control network protocol that is used to connect PLCs and local controllers*

## DHTML

*Dynamic Hyper Text Mark up Language*

## DOM

*The Document Object Model is a platform- and language-neutral interface that will allow programs and scripts to dynamically access and update the content, structure and style of documents.*

## Echo2

*Echo is an open-source framework for developing rich web applications. From the developer's perspective, Echo behaves as a user interface toolkit--like Swing or Eclipse SWT.*

## Fieldbus

*Communication protocols that facilitate interchange of messages among field devices. Some examples of Fieldbus protocols are Foundation Fieldbus, Modbus, DeviceNet, and Profibus.*

## FTP

*File Transfer Protocol*

## HTML

*Hyper text and mark up language*

## HTTP

*Hyper Text Transfer Protocol*

## IDE

*An integrated development environment (IDE) also known as integrated design environment or integrated debugging environment is a software application that provides comprehensive facilities to computer programmers for software development.*

## IT

*Information Technology*

## JRE

*Java Runtime Environment*

## JSON

*JSON (JavaScript Object Notation) is a lightweight data-interchange format. It is easy for humans to read and write. It is easy for machines to parse and generate.*

## JVM

*Java Virtual Machine*

## LAMP

*LAMP is an acronym for a solution stack of free, open source software, originally coined from the first letters of Linux (operating system), Apache HTTP Server, MySQL (database software), and PHP, Python or Perl (scripting language), principal components to build a viable general purpose web server.*

## NAT

*Network Address Translation*

## OPC

*Before: OLE for Process Control, Now: Open Process Control*

## Real-Time

*An action that occurs at the same rate as actual time; no lag time, no processing time*

## RMI

*Java Remote Method Invocation (Java RMI) enables the programmer to create distributed Java technology-based to Java technology-based applications, in which the methods of remote Java objects can be invoked from other Java virtual machines\*, possibly on different hosts. RMI uses object serialization to marshal and unmarshal parameters and does not truncate types, supporting true object-oriented polymorphism*

## RTOS

*A computer operating system that implements process and services in a deterministic manner*

## SCADA

*Supervisory Control and Data Acquisition*

## SOAP

*Simple Object Access Protocol*

## TCP/IP

*Transmission control protocol/Internet packet*

## URL

*Uniform Resource Locator*

## W3C

*World Wide Web Consortium*

## WSDL

*Web Service description language*

## XHR

*XMLHttpRequest (XHR) is a DOM API that can be used inside a web browser scripting language, such as JavaScript, to send an HTTP or an HTTPS request directly to a web server and load the server response data directly back into the scripting language*

## XHTML

*Extensible Hyper Text Mark up Language*

## XML

*Extensible mark up language*

## XSLT

*XSLT (XSL Transformations) is a declarative, XML-based language used for the transformation of XML documents into other XML documents.*

## ZK

*ZK is an open-source Ajax Web application framework, written in Java, that enables creation of rich graphical user interfaces for Web applications with no JavaScript and little programming knowledge.*

صناعية حقيقية (Simulation) لاستخدامه فى عملية المقارنة والاختبار وكذلك تم تنفيذ النظام المقترح على عملية صناعية فعلية حقيقية فى شركة قنا لصناعة الورق.

فى الفصل الخامس تم التعرض لمشكلة الأمن والحماية لأنظمة التحكم والمراقبة من خلال شبكة المعلومات حيث أن هذه المشكلة قد ظهرت أهميتها عندما تم ربط هذه الأنظمة بالانترنت حيث أصبحت أكثر عرضة للاختراق والتخريب, لذلك فقد تم تصميم سياسة حماية وتأمين للنظام المقترح.

فى الفصل السادس تم استعراض ما تم الحصول عليه من نتائج وخلاصة ما تم عمله فى هذه الرسالة مع اقتراح ما يمكن عمله كامتداد لهذا العمل.

# الملخص العربى

تطورت نظم السيطرة الرقابية المعتمدة على الحاسب والحصول على البيانات (سكادا- SCADA) على مدى الأربع عقود الماضية ، من قائمة بذاتها ، وعمليات مجزأة الى بنية شبكية تتواصل عبر مسافات شاسعة. وبالإضافة إلى ذلك ، فقد انتقلت التطبيقات الخاصة بهم من الأجهزة والبرامج المخصصة الى الأجهزة والبرمجيات القياسية. وقد أدت هذه التغيرات الى خفض تكلفة التنفيذ والتشغيل والصيانة ، فضلا عن توفير المعلومات للإدارة التنفيذية في وقت حدوثها والتي يمكن استخدامها لدعم التخطيط والإشراف واتخاذ القرارات. ولأسباب تتعلق بالكفاءة ، والصيانة ، والاقتصاد ، انتقلت هذه الأنظمة من شبكات معزولة داخل المصنع باستخدام الأجهزة والبرمجيات المسجلة للنظم القائمة على الحاسب الشخصى باستخدام برمجيات قياسية ، وبروتوكولات شبكة قياسية ، والإنترنت. وهناك اتجاه ناشئ في العديد من المنظمات التي تضم سكادا ووحدات تكنولوجيا المعلومات نحو توحيد بعض الأنشطة المتداخلة. على سبيل المثال ، يمكن أن يتم استيعاب هندسة التحكم وتكاملها تماما مع قسم تكنولوجيا المعلومات. والدافع وراء هذا الاتجاه هو تحقيق توفير في التكاليف عن طريق توحيد المناهج المتباينة ، والشبكات ، والبرمجيات ، وأدوات الصيانة. وبالإضافة إلى ذلك ، دمج جميع أنظمة سكادا مع الشركات المالية وإدارة بيانات العملاء يؤدى الى زيادة القدرة على تشغيل المنظمة بكفاءة وفعالية أكبر.

**الهدف الرئيسى** من هذه الرسالة هو نظام سكادا فعال على شبكة الإنترنت عن طريق الوصول إلى مخدم ويب من خلال شبكة الانترنت (Web Server). ولكى نفعل ذلك علينا تجميع بعض من التكنولوجيات الحديثة مثل تكنولوجيا المعلومات وخدمات الويب أجاكس (AJAX) مع بروتوكولات صناعية قياسية مثل بروتوكول " أو بى سى داتا اكسس" (OPC DA). هدفنا هو نظام سكادا فعال وآمن والذى يستهلك أقل قدر ممكن من موارد الخادم مثل حمل وحدة المعالجة والذاكرة وكذلك الشبكة. هذا بالإضافة الى توفير أساليب الحماية ضد عمليات الاختراق والتخريب من خلال شبكة الانترنت. ولتحقيق هذا الهدف سوف نضطر لمواجهة وحل الكثير من التحديات المرتبطة بالانترنت.

وتتكون الرسالة من ستة فصول ومجموعة من الملاحق والمراجع كالتالى:-

فى الفصل الأول مقدمة تم من خلالها تعريف كلا من الميكنة (Automation) ونظم سكادا (SCADA) وموقعها فى هيكلية الميكنة وكذلك تم التعرف على بعض البروتوكلات الصناعية ومنها برتوكول "أو بى سى داتا أكسس" (OPC DA). كذلك تم استعراض بعض المفاهيم فى نظم تكنولوجيا المعلومات (IT) مثل HTTP و Web Servers و Web Applications.

فى الفصل الثانى تم استعراض أحدث تكنولوجيات وأدوات نظم المعلومات بشئ من التفصيل مع ذكر المزايا والعيوب لكلا منها.

فى الفصل الثالث تم استعراض الطرق والحلول السابقة لعمل نظم سكادا من خلال شبكة المعلومات حيث تم عرض مجموعة من الأبحاث التى استخدم أصحابها نظم المعلومات الحديثة (Modern IT) والتقليدية لتحقيق هذا الهدف , وقد تم تحليل كل بحث وذكر مميزاته وعيوبه.

فى الفصل الرابع استعراض وتصميم وتنفيذ الطرق المقترحة لعمل نظام تحكم ومراقبة من خلال شبكة المعلومات , وقد تم اختيار احد الأبحاث المعروضة فى الفصل الثالث حيث تم حل الكثير من المشاكل التى وقع فيها الباحث وتحسين أداء النظام المقترح. أيضا تم مقارنة النتائج التى حصلنا عليها مع البحث المختار لاظهار صور التحسين فى الأداء والكفاءة كما تم تصميم وتنفيذ برنامج يعمل كمحاكاة لعملية

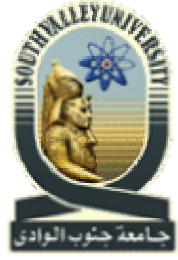

**كلية الهندسة بأسوان**
**قسم الهندسة الكهربية**

# نظام سكادا الفعال بواسطة شبكة المعلومات الدولية

رسالة من احدى متطلبات الحصول على درجة الماجستير
فى الهندسة الكهربية

مقدمة من

## المهندس/ حسنى احمد عباس احمد

مهندس تحكم بشركة قنا لصناعة الورق
بكالوريوس الهندسة الكهربية- جامعة أسيوط ١٩٩٨

### قسم الهندسة الكهربية – كلية الهندسة بأسوان – جامعة جنوب الوادى

| لجنة المناقشة | لجنة الاشراف |
|---|---|
| **أ.د/ محمد حسين أمين** | **أ.د/ عبد المجيد محمد على** |
| أستاذ بكلية الهندسة- جامعة أسيوط | أستاذ بكلية الهندسة بأسوان- جامعة جنوب الوادى |
| **أ.د/ يوسف بسيونى مهدى** | **د/ احمد مصطفى عبد الرحمن** |
| وكيل كلية الحاسبات والمعلومات للدراسات العليا والبحوث – جامعة أسيوط | مدرس بكلية الهندسة بأسوان- جامعة جنوب الوادى |
| **أ.د/ عبد المجيد محمد على** | |
| أستاذ بكلية الهندسة بأسوان- جامعة جنوب الوادى | |

يونيو ٢٠١١

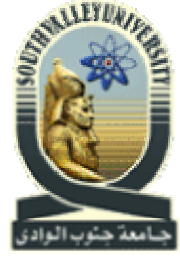

**كلية الهندسة بأسوان**
**قسم الهندسة الكهربية**

# نظام سكادا الفعال بواسطة شبكة المعلومات الدولية

## رسالة ماجستير

مقدمة من

## المهندس/ حسنى احمد عباس احمد

مهندس تحكم بشركة قنا لصناعة الورق

**قسم الهندسة الكهربية**
**كلية الهندسة**
**جامعة جنوب الوادى**

يونيو ٢٠١١